\documentclass[12pt,preprint]{emulateapj}

\slugcomment{Astrophysical Journal}

\shorttitle{Infrared AO imaging of merging galaxies}
\shortauthors{Imanishi et al.}

\begin{document}

\title{Subaru Adaptive-optics High-spatial-resolution Infrared $K$- and
$L'$-band Imaging Search for Deeply Buried Dual AGNs in Merging Galaxies} 


\author{Masatoshi Imanishi\altaffilmark{1} and Yuriko Saito\altaffilmark{1}}
\affil{Subaru Telescope, 650 North A'ohoku Place, Hilo, Hawaii 96720,  U.S.A.}
\email{masa.imanishi@nao.ac.jp}

\altaffiltext{1}{Department of Astronomy, School of Science, Graduate
University for Advanced Studies (SOKENDAI), Mitaka, Tokyo 181-8588}

\begin{abstract}
We present the results of infrared $K$- (2.2 $\mu$m) and $L'$-band 
(3.8 $\mu$m) high-spatial-resolution ($<$0$\farcs$2) imaging
observations of nearby gas- and dust-rich infrared luminous merging
galaxies, assisted by the adaptive optics (AO) system on the Subaru 8.2-m
telescope.   
We investigate the presence and frequency of red $K-L'$ compact sources,
which are sensitive indicators of active galactic nuclei (AGNs),
including AGNs that are deeply buried in gas and dust. 
We observed 29 merging systems and confirmed at least one AGN
in all but one system.
However, luminous dual AGNs were detected in only four of the 29 systems 
($\sim$14\%), despite our method's being sensitive to buried AGNs.
For multiple nuclei sources, we compared the estimated AGN luminosities
with supermassive black hole (SMBH) masses inferred from large aperture
$K$-band stellar emission photometry in individual nuclei. 
We found that mass accretion rates onto SMBHs are significantly
different among multiple SMBHs, such that larger-mass SMBHs
generally show higher mass accretion rates when normalized to SMBH mass.
Our results suggest that non-synchronous mass accretion onto SMBHs in
gas- and dust-rich infrared luminous merging galaxies hampers the
observational detection of kiloparsec-scale multiple active SMBHs.
This could explain the significantly smaller detection fraction of 
kiloparsec-scale dual AGNs when compared with the number expected from
simple theoretical predictions.   
Our results also indicate that mass accretion onto SMBHs is dominated by
local conditions, rather than by global galaxy properties, 
reinforcing the importance of observations to our understanding of how
multiple SMBHs are activated and acquire mass in gas- and dust-rich
merging galaxies. 
\end{abstract}


\section{Introduction}

Recent observations have revealed that supermassive black holes (SMBHs)
are ubiquitously present at the center of galaxy spheroidal components, 
and that the masses of SMBHs and spheroidal stars are correlated
\citep{mag98,fer00,gul09,mcc13}, suggesting that SMBHs are an important
galaxy ingredient whose formation is closely related to galaxy formation.
The widely accepted cold dark matter-based galaxy formation theories
postulate that small gas-rich galaxies merge and grow into massive
galaxies, as seen in the present-day universe \citep{whi78}.
If SMBHs are present in the progenitor gas-rich small galaxies, 
then the merging galaxies should have multiple SMBHs. 
In this case, kiloparsec-scale dual active galactic nuclei (AGNs) are 
expected to be common if the mass accretion onto both SMBHs is
sufficiently high to create luminous observable AGNs \citep{col11}. 

Optical spectroscopic searches for AGNs with double-peaked emission
lines \citep{wan09,liu10,smi10,liu11,pil12,ge12,bar13} and subsequent
follow-up observations at other wavelengths have been extensively
performed in the search for kiloparsec-scale dual AGNs
\citep{fu11,ros11,she11,tin11,com12,fu12,liu13}. 
These studies have provided some examples of kiloparsec-scale dual AGNs,
but the detected fraction of optically identifiable kiloparsec-scale
dual AGNs ($<$a few \%) is significantly smaller than the number derived
from the simple theoretical prediction that the majority of galaxy mergers 
should have multiple SMBHs and become dual 
AGNs if both SMBHs are sufficiently mass-accreting \citep{ros11,yu11}.   
Several scenarios have been proposed to reconcile this discrepancy between 
theory and observations, but it is still unclear which scenario
is most likely. 
First, double-peaked emission is expected only in dual active SMBHs whose
orbital planes are aligned relatively edge-on along our line
of sight. Those systems whose orbital planes are aligned face-on 
are overlooked with these methods \citep{ros11,wan12}. 
Although this scenario may be able to explain the discrepancy by a
factor of a few, it is probably difficult to account for the difference 
by more than an order of magnitude, provided that the alignment of the
orbital planes of two active SMBHs is random in terms of our
line-of-sight.  
A second, more plausible explanation is that most AGNs in gas-rich
galaxy mergers are deeply buried in gas and dust along virtually 
all directions \citep{hop05,hop06} and can become optically elusive
\citep{mai03,ros11}. 
Third, it is also possible that even though multiple SMBHs are present, 
only one of them has sufficient mass accretion to be observationally
detectable as an AGN (i.e., non-synchronous SMBH mass accretion) during
galaxy mergers \citep{van12}.   
If we are to unveil observationally the true fraction of 
kiloparsec-scale dual AGNs in gas- and dust-rich merging galaxies, 
it is of particular importance to apply observational methods that are 
sensitive to AGNs whose SMBHs orbital motion is relatively face-on 
along our line-of-sight and to deeply buried AGNs.  
 
High-spatial-resolution imaging observations at the wavelengths of 
strong dust penetration are a powerful tool for this purpose because 
imaging observations can, in principle, preferentially detect AGNs
with face-on multiple SMBH orbit geometry, and buried AGNs are
detectable at such wavelengths.
X-rays in the 2--10 keV range have higher dust-penetrating power than
the optical in the Galactic interstellar medium \citep{ryt96}, and it is
well known that an AGN is a much stronger X-ray emitter (relative to the 
bolometric luminosity) than a starburst \citep{ran03,sha11}, 
so X-ray observations are expected to be sensitive to AGNs, including 
obscured ones. 
In fact, strong dual AGN candidates were discovered 
from 2--10 keV X-ray observations of several nearby gas- and dust-rich 
infrared luminous merging galaxies
\citep{kom03,bal04,bia08,pic10,kos11,fab11}. 
Although the presence of dual AGNs is strongly suggested, 
the observed 2--10 keV X-ray emission is, in most cases, only a
scattered component behind Compton-thick (N$_{\rm H}$ $>$ 10$^{24}$
cm$^{-2}$) obscuring material rather than a directly transmitted
component.
Thus, the intrinsic AGN X-ray luminosity
is difficult to estimate because the scattering efficiency is unknown.
To extend the 2--10 keV X-ray dual AGN survey systematically to
gas- and dust-rich merging galaxies, the Chandra X-ray Observatory, with
its spatial resolution of $\sim$0$\farcs$5, is a particularly
powerful tool for spatially resolving closely separated multiple AGNs.
In fact, the Chandra X-ray Observatory has been used to find obscured
dual AGNs in further gas- and dust-rich merging galaxies
\citep{ten05,iwa11b,ten12,kos12,liu13}.   
However, the detected X-ray fluxes are, in many cases, so faint that
detailed spectral analysis is hampered.  
Both the scattered X-ray component of Compton-thick obscured AGNs
and the emission originating from stars can reproduce the observed faint 
X-ray fluxes, often making it difficult to interpret the detected
X-ray emission as solid evidence for an AGN \citep{ten09,iwa11b,liu13}. 
X-ray observations at $>$10 keV could directly detect transmitted X-ray
emission from some mildly Compton-thick AGNs \citep{ito08}, but
the spatial resolution at $>$10 keV is still insufficient ($>$15$''$)
\citep{har13} to spatially resolve many interesting closely separated
dual AGNs.     

High-spatial-resolution radio observation using the VLBI technique 
is another powerful tool to detect closely separated dual AGNs, as
radio wavelengths are less susceptible to dust extinction \citep{rod06}.
However, the radio VLBI technique is sensitive only to a small fraction
of radio-loud AGN population, and it is not sensitive to the radio-quiet 
AGNs that comprise the majority of the AGN population \citep{gol99,whi00}.
The very small detectable fraction ($<$0.05\%) of dual AGNs discovered using 
this radio VLBI technique \citep{bur11} is difficult to interpret
directly in relation to the actual fraction of dual AGNs. 

Infrared observations at $>$2 $\mu$m are potentially another
effective (even improved) method for investigating buried dual AGNs in
gas- and dust-rich merging galaxies.  
First, compared with the optical, dust extinction is substantially reduced 
($<$0.06 $\times$ A$_{\rm V}$) \citep{nis08,nis09}.
Additionally, AGNs, including both radio-loud and radio-quiet ones, are 
observationally distinguishable from starbursts. 
Strong polycyclic aromatic hydrocarbons (PAH) emission features seen at
3--20 $\mu$m are usually observed in starbursts, but not in pure AGNs
\citep{moo86,imd00}, due to the destruction of PAHs by strong AGN
X-ray emission \citep{voi92}. 
In a pure AGN, a PAH-free continuum due to AGN-heated, large 
($\sim$0.1 $\mu$m) hot dust grains is observed. 
Thus, infrared spectroscopy can be a unique means to find obscured
AGNs by separating them from starburst activity, as demonstrated by the
successful detection of many buried AGNs in the brightest main nuclei of
gas- and dust-rich merging galaxies
\citep{idm06,ima06,ima07a,arm07,ima08,nar08,vei09,ima09,nar09,ima10a,ima10b,nar10}. 
Such infrared spectroscopy with high-spatial-resolution is, in
principle, useful for dual AGN searches, but in practice, 
its application to faint AGNs in secondary galaxy nuclei 
is limited by spectroscopic sensitivity.

However, these infrared spectroscopic observations have clearly shown
that the infrared 2--5 $\mu$m continuum emission in AGNs is
systematically redder than the emission from starbursts
\citep{ima08,san08,ris10,ima10b}. 
The radiative energy generation efficiency of an AGN (= mass accreting
SMBH; 6--42\% of Mc$^{2}$) \citep{bar70,tho74} is much higher than that
of a starburst (= nuclear fusion reaction inside stars; 
$\sim$0.7\% of Mc$^{2}$).
Thus, high luminosity can be generated from a very compact region in an AGN.   
A larger amount of dust in the close vicinity ($<$10 pc) of an
AGN can be heated to high temperatures with several 100 K \citep{bar87},
producing stronger infrared $L'$-band (3.8 $\mu$m) radiation 
than a starburst whose infrared 2--5 $\mu$m  
flux is usually dominated by stellar photospheric blue emission.  
Observationally, pure AGNs are known to display strong $L'$-band flux
excess relative to the $K$-band (2.2 $\mu$m) ($K-L' >$ 2.0)
\citep{iva00,alo03,vid13}, 
when compared to starburst activity ($K-L' \sim$ 0.5) \citep{hun02}. 
Hence, we can scrutinize dual AGNs, including deeply buried ones,
through the detection of spatially compact red $K-L'$ sources. 
An important point is that while simple high-spatial-resolution imaging 
observations at wavelengths shorter than the $K$-band often cannot 
easily differentiate AGNs from compact starbursts for spatially compact 
emission at the distance of interesting gas- and dust-rich merging
galaxies \citep{sco00}, 
high-spatial-resolution imaging at both $K$ and $L'$ can 
better distinguish between AGNs and compact starbursts by combining 
morphological and color information.  
More importantly, this infrared {\it imaging} method is more sensitive 
than the infrared {\it spectroscopic} method and thus is applicable to
a larger number of fainter secondary nuclei of gas- and dust-rich merging
galaxies, which is crucial for dual AGN studies. 
Given that the ratio of infrared dust extinction to X-ray absorption
(dust + gas) toward obscured AGNs is empirically known to be much smaller
than the Galactic value \citep{alo97,gra97,fad98,geo11}, the infrared
$K$ and $L'$-band observations could be sensitive even to Compton-thick
buried AGNs, as was demonstrated in some sources
\citep{imd00,idm06,ten05,ima08,ten09}, making the infrared $K$- and
$L'$-band imaging method particularly promising.

The recent availability of laser-guide-star adaptive optics (LGS-AO) on 
ground-based 8--10-m telescopes has enabled us to routinely achieve 
spatial resolutions of $<$0$\farcs$2 in infrared $K$- and $L'$-band images
for the bulk of extra-galactic sources. 
Theoretical simulations of gas- and dust-rich galaxy mergers
\citep{hop05,hop06,van12} predict that dual AGNs become luminous, 
particularly at the late merging stage, at separations of less than a
few kiloparsecs.
This physical separation corresponds to $<$1$''$ at z $>$ 0.1, so 
sub-arcsecond resolution observations are crucial. 
The AO-assisted infrared $K$- and $L'$-band imaging method has several
advantages: 
(1) it has even better spatial resolution ($<$0$\farcs$2) than the Chandra
X-ray Observatory has ($\sim$0$\farcs$5); (2) it is sensitive to buried
AGNs with face-on SMBH orbits, which are missed in the previously employed
optically based searches; and (3) it can trace radio-quiet AGNs. All of
these qualities make this an excellent and unique method  
for a systematic search for dual AGNs in gas- and dust-rich merging
galaxies.    
We have thus embarked on AO-assisted infrared $K$- and $L'$-band imaging
observations of nearby gas- and dust-rich galaxy mergers using the Subaru
8.2-m telescope atop Mauna Kea, Hawaii.  
This is one of the best sites
in the world for conducting highly sensitive $L'$-band observations, 
due to high elevation ($\sim$4200 m) and the resulting very low
precipitable water vapor value.
 
Throughout this paper, we adopt H$_{0}$ $=$ 71 km s$^{-1}$ Mpc$^{-1}$, 
$\Omega_{\rm M}$ = 0.27, and $\Omega_{\rm \Lambda}$ = 0.73
\citep{kom09}.
The luminosity distance is calculated using the cosmological calculator
created by \citet{wri06}. 
In Section 2, we describe our sample population. In Section 3, we
present our observations and data analysis, followed by results in
Section 4. We discuss the implications of our results in Section 5, and
conclude with a summary of our work in Section 6.  

\section{Target Selection}

We primarily target luminous infrared galaxies (LIRGs) whose infrared
luminosities exceed L$_{\rm IR}$ $>$ 10$^{11}$L$_{\odot}$ because they
are a representative sample of gas- and dust-rich galaxy mergers in the
local universe \citep{sam96}.
In gas-rich LIRGs, it is expected that many SMBHs are mass accreting and
so the detection rate of multiple SMBHs through luminous AGN
searches is higher than in gas-poor galaxy mergers.

Previous systematic infrared spectroscopy revealed that 
buried AGNs play an increasing energetic role with increasing galaxy 
infrared luminosity and become particularly important in ultraluminous 
infrared galaxies (ULIRGs) with L$_{\rm IR}$ $>$ 10$^{12}$L$_{\odot}$ 
\citep{ima09,vei09,ima10a,ima10b,nar10}.
Thus, ULIRGs are of particular interest.
Among AGN-hosting ULIRGs at the brightest nuclei, those with
spatially resolved multiple nuclei in seeing-limited images \citep{kim02}
are our first targets because investigating the presence of AGNs in the
fainter secondary nuclei is a straightforward way to search for dual
AGNs.  
In addition to these, we also include apparently single-nucleus ULIRGs 
in seeing-limited images because it may be possible to detect
closely separated dual AGNs in our high-spatial-resolution
AO images.
Multiple-nuclei merging ULIRGs without strong AGN signatures in the
brightest main nuclei are also added to see whether AGNs exist in the
fainter secondary nuclei.

In addition to ULIRGs, we include five LIRGs for which previous 2--10
keV X-ray observations have suggested the presence of dual AGNs
\citep{kom03,bia08,pic10,kos11,fab11}.  
Our aim in including these targets in our sample is to confirm
that dual AGN signatures could be found using our infrared
$K$- and $L'$-band imaging method. 

In total, 29 merging systems were observed, and their basic properties
are summarized in Table 1. 
Our sample is neither homogeneous nor complete in a statistical sense. 
Our goal is to determine whether our infrared method, sensitive to
buried AGNs, results in a substantially higher fraction of dual AGN
detections than previous optically based methods, and to investigate
whether the observed dual AGN fraction approaches unity, as predicted by
theories, in gas- and dust-rich merging galaxies in the local universe.

\section{Observation and Data Analysis}

We used the IRCS infrared camera and spectrograph \citep{kob00} at the  
Nasmyth focus of the Subaru 8.2-m telescope atop Mauna Kea, Hawaii
\citep{iye04}, together with the 188-element adaptive optics (AO)
system, using laser-guide stars (LGS) or natural-guide stars (NGS)
\citep{hay08,hay10}.    
For the LGS-AO to work properly, we need to find a star or compact
object brighter than 18--19 mag in the optical $R$-band within $\sim$90$''$
from the target to use as a guide star for good tip-tilt correction. 
The AO correction itself is made using a laser spot with an optical
$R$-band magnitude of $\sim$10.5 mag.
Because most extra-galactic sources have such guide stars, 
we can now routinely obtain high-quality LGS-AO images with significantly
improved image sizes compared with seeing-limited observations.
When laser launch was possible, we performed LGS-AO
observations. 
However, when it was not, we adopted NGS-AO observations. 
To achieve good NGS-AO performance, we need a star or
compact source brighter than 16.5 mag in the optical $R$-band within
$\sim$30$''$ from the target to serve a guide star for AO correction. 
The number of external galaxies with such suitable NGS-AO guide stars is 
limited.   
Table 2 tabulates our observations, including information on 
guide stars used to achieve good AO performance and standard stars 
for photometric calibration.
In summary, NGS-AO was used during the observing runs in 2011 June and
2013 May, and LGS-AO was adopted for all the remaining observing runs.

During our Subaru AO observations, the sky was clear.
The seeing in the $K$-band (2.2 $\mu$m), measured in images before
starting actual AO exposures, was 0$\farcs$3--0$\farcs$7 
in full width at half maximum (FWHM).  
The precipitable water vapor value was low at $<$1 mm for the 2012 May 
run,  
$<$2 mm for 2011 June and July and 2012 October runs, and $\sim$3 mm for
2012 March and April and 2013 May runs; however, it was high at 4--5
mm for the 2011 August run. 

In the $K$-band, we employed the 52 mas (52.77 mas pixel$^{-1}$) imaging
mode to observe simultaneously some widely separated multiple galaxy
nuclei and as many stars as possible to probe the point-spread function,
within the field of view of 54$\farcs$04 $\times$ 54$\farcs$04 (1024
pixels $\times$ 1024 pixels). 
The only exception is Mrk 231, which has very bright nuclear emission, 
for which we employed the 20 mas (20.57 mas pixel$^{-1}$) imaging
mode to scatter the bright nuclear glare into many pixels and avoid
array saturation. 
In the $L'$-band, as Earth's atmospheric background emission is high,
we used the 20 mas imaging mode, whose field of view is 21$\farcs$06
$\times$ 21$\farcs$06 in the full-array mode.  
Even using this 20 mas mode, we had to use a sub-array mode to avoid
saturating the array for the $L'$-band observations of some
sources.

For the $K$-band observations of target merging (U)LIRGs, the exposure
times were 0.26--30 sec, and 2--120 coadds were applied. 
In the $L'$-band, the exposure times were
0.076--0.12 sec, with 250--300 coadds.
The individual exposure time was set so that signal levels at the object
positions were below the linearity level of the IRCS imaging  
array ($<$4000 ADU).
Exposure times were dependent on the condition of Earth's atmosphere and
on the ambient background emission level at the time of observations, as
well as on the brightness of the object nuclei.  

For target merging (U)LIRGs, we adopted nine-point 
dithering patterns to observe objects at nine different array
positions and to overcome the effects of bad pixels.
One dataset consisted of the nine dithered frames. 
For faint (U)LIRGs, this nine-point dithering pattern was repeated 
multiple times.  
The only exception was the IRAS 16474$+$3430 $K$-band data taken 
in 2011 July, for which a five-point dithering pattern was employed.

For all observing runs, photometric $K$- and $L'$-bands standard stars
(Table 2)  
were observed, with a mean airmass difference of $<$0.2 in
directions similar to the individual (U)LIRGs to correct for the
transmission of Earth's atmosphere at the time of observations and to
provide a flux calibration.  
For standard star observations, NGS-AO was used whenever possible,
using the standard stars themselves as AO guide stars to 
confirm the performance of the Subaru AO using point sources and to observe
the stability and variation in the point-spread function under
varying Earth atmospheric conditions.  
For a few very bright standard stars, we performed non-AO observations
so that the signals did not exceed the linearity level of the IRCS array.
For a few standard stars, we did not have enough time to optimize the AO 
parameters and so used AO without the best tuning.
These data are useful for photometry but not for investigating
the point-spread function.

Standard data analysis procedures were employed using IRAF   
\footnote{IRAF is distributed by the National Optical Astronomy
Observatories, which are operated by the Association of Universities
for Research in Astronomy, Inc. (AURA), under cooperative agreement
with the National Science Foundation.}.
We first inspected the individual frames by eye. 
A very small fraction of frames showed strange patterns compared with 
the majority of the remaining normal frames.
These strange frames were discarded from our analysis. 
We created median-combined sky frames to make a sky flat image 
for each dataset.
In this procedure, the positions of bright objects and bad pixels 
were masked.
Individual frames were sky-subtracted and divided by the 
sky flat frames for flat fielding.
When the effects of cosmic ray hits and bad pixels remained 
for a few pixels at this stage, we made 
corrections manually by replacing the signals of these pixels with 
the interpolated values of the surrounding pixels.  
Then, we shifted the sky-subtracted, flat-fielded images so that the
peak position of each target object landed on the same array pixel. 
For images that contained sufficiently bright compact sources inside the 
field of view of individual frames, the pixel shift was determined using 
these bright compact sources. 
The $K$-band frames often contained such sources. 
For a small fraction of the $K$-band data and the majority of the
$L'$-band data, however, no such bright compact sources were seen in
individual frames.  
In those cases, we used the offset values calculated from the dithering
amplitude and pixel scale. 
We then average-combined these shifted frames to obtain final images.
Based on a comparison of the FWHM values of compact objects 
between the resulting combined images and individual frames, we found
this offset estimate to be very accurate for our AO data,  
due to the good quality of the AO guiding of the Subaru telescope. 

\section{Results}

\subsection{$K$- and $L'$-band images}

Figure 1 presents the infrared $K$- and $L'$-band AO images of
the observed (U)LIRG nuclei. 
The achieved image sizes for stars and compact sources are usually
0$\farcs$1--0$\farcs$2 in FWHM both in the $K$- and $L'$-bands.
Due to the higher atmospheric background emission, the sensitivity 
in the $L'$-band is much lower (worse) than that in the $K$-band. 
Consequently, the detection significance in the $L'$-band should be 
much lower than that in the $K$-band, unless the sources are very red 
in $K-L'$. 
In Figure 1, the detection rate of merging (U)LIRG nuclei is smaller in
the $L'$-band than in the $K$-band, and yet a significant fraction of
the observed merging nuclei are clearly detected in our highly sensitive
$L'$-band AO images.  

In the $L'$-band, the background emission for space-based infrared
satellites is much lower than that for ground-based telescopes.
So, as far as the detection of a single source is concerned, 
observations using space-based infrared
satellites with small apertures could also be useful.
WISE (40 cm) and Spitzer (75 cm) infrared satellites have imaging
capabilities at 3.4 $\mu$m \citep{wri10} and 3.6 $\mu$m \citep{faz04},
respectively.
These wavelengths are similar to our $L'$-band (3.8 $\mu$m). 
However, the spatial resolution of WISE at 3.4 $\mu$m and Spitzer at 3.6
$\mu$m are $\sim$6$\farcs$1 \citep{wri10} and $\sim$1$\farcs$7
\citep{faz04}, respectively. 
As our primary scientific goal is to locate closely-spaced dual AGNs
and unveil their properties, higher-spatial-resolution
ground-based AO images are more effective.
As illustrative examples, Figure 2 compares our AO images of two
ULIRGs at $L'$ (Arp 220 and IRAS
16474$+$3430) with Spitzer IRAC camera images at 3.6 $\mu$m analyzed
from archival data.   
Whereas multiple nuclei are clearly resolved in our AO images, the
nuclei are not resolved in the Spitzer IRAC data.  
It is clear that our ground-based AO images are better for
our scientific purpose. 

\subsection{Photometric aperture size}

$K$- and $L'$-band emission originating from buried AGNs are dominated by
AGN-heated hot dust located in the innermost part 
($<$10 pc in physical scale or $<$0$\farcs$1 at $z>$ 0.01) of the
surrounding dusty material, and so are expected to be almost point sources.
Higher-spatial-resolution photometry can minimize the 
contamination from spatially extended stellar-origin emission in host 
galaxies.
In this regard, AO images have an advantage over seeing-sized non-AO
images; the higher fraction of signal from compact sources
is concentrated on a smaller central region.
In our AO data, the measured image sizes of stars and compact galaxy 
nuclei are usually 0$\farcs$1--0$\farcs$2 in FWHM.
However, we should note that in ground-based AO imaging data, 
even though the peak signal level of a compact source is higher than
that in non-AO data, a considerable fraction of emission from a compact
source spreads over a seeing-sized halo outside the AO core.
The signal fraction within the AO core (0$\farcs$1--0$\farcs$2 in FWHM)
is usually significantly less than unity at $<$4 $\mu$m
\citep{min10,min12} and varies depending on the brightness of guide
stars used for the LGS-AO tip-tilt correction or NGS-AO correction,
their separation from the target objects, and Earth's atmospheric
turbulence at the time of observations.   
Hence, if we extract signal only from the AO-core emission component
(0$\farcs$1--0$\farcs$2 in FWHM), we will recover, for example, 50\% of
the spatially unresolved compact source flux in some cases and 75\% in
other cases. 
This difference will introduce a large uncertainty in the photometry of
the nuclear compact sources among different galaxies.
Larger aperture photometry can obviously cover a larger signal fraction 
of compact sources, 
but at the same time, the contamination from spatially extended stellar 
emission will increase. 
We must find the optimum aperture size such that the bulk of the compact 
source signal is covered, and yet the contamination from stellar
emission is minimal. 
 
Figure 3 plots the growth of the curves of the encircled signal in the 
$L'$-band using standard stars and a compact ULIRG observed with NGS-AO.
Figure 4 displays the same plot for any compact sources found in the
science target data, observed with LGS-AO. 
Although AO performance varies slightly among the different observing
runs, 85--93\% of total signal is usually recovered using a 
$\sim$0$\farcs$5 (25 pixels $\times$ 20.57 mas pixel$^{-1}$)-radius
aperture.  
Even in the data taken in 2012 October, when seeing was most unstable 
among our observing runs in Table 2, $>$83\% of the total signal is 
included within the selected aperture size. 
Thus, with the $\sim$0$\farcs$5 radius aperture, we can consistently
recover 83--93\% of the spatially unresolved compact source flux in our
$L'$-band LGS- and NGS-AO data.    

Figure 5 shows the growth of the curves of the encircled signal in the
$K$-band for standard stars and one ULIRG observed with NGS-AO.
Figure 6 displays the same plot for any compact sources detected inside
the science target frames taken with LGS-AO.
In the $K$-band, Earth's atmospheric turbulence effects are larger than
in the $L'$-band.  
Also, the selected standard stars are generally fainter in the optical 
$R$-band than the $L'$-band standard stars.
Thus, poorer growth of curves may be anticipated, and yet 75--90\% of
signals are usually recovered with the $\sim$0$\farcs$5-radius aperture
in the $K$-band.  

For $L'$-band data, a standard star was observed with and without 
NGS-AO on the same night (2011 August 22 UT) (Figure 3, upper left). 
The AO data provide higher encircled signals at small radii than non-AO
data, demonstrating the merit of AO for spatially unresolved
compact source photometry using a small aperture.     

From the fact that 75--90\% and 83--93\% of the spatially unresolved
compact source signals are consistently recovered with a
$\sim$0$\farcs$5-radius aperture in $K$- and $L'$-band data,
respectively, taken on different nights, under different Earth's
atmospheric conditions, and with different guide star properties, we 
can safely conclude that Subaru AO, including both NGS-AO and LGS-AO,
can provide such stable data, as long as the AO correction performs 
sufficiently well.  
We thus apply the same aperture size ($\sim$0$\farcs$5 radius) for the
photometry of nuclear compact sources (= hot dust emission heated by the
putative AGNs) in the target (U)LIRGs
and consider that 75--90\% and 83--93\% of compact source signal is
recovered in the $K$- and $L'$-band, respectively. 
Table 3 summarizes the photometric measurements of individual merging
nuclei using a $\sim$0$\farcs$5-radius aperture. 
The measurements are performed by varying the regions selected for sky
subtraction, 
but are usually found to agree within much better than 0.2 mag.

The more significant source of photometric uncertainty is the slight 
variation in compact source signal fraction within the
$\sim$0$\farcs$5-radius aperture.   
For example, if two different galaxies included 75\% and 90\% of the 
compact source signal, respectively, for the selected $\sim$0$\farcs$5-
radius aperture, then the measured photometry could differ 
by $\sim$0.2 mag between these sources, even if the intrinsic 
compact source flux were the same.  
Also, for the same galaxy nucleus,
if 75\% and 93\% of compact source signal were included with the 
$\sim$0$\farcs$5-radius aperture in the $K$-band and $L'$-band,
respectively, the derived $K-L'$ color could differ from the actual
color by $\sim$0.2 mag. 
Therefore, we must be aware that the $\sim$0.2 mag photometric 
uncertainty among different galaxy nuclei or between the $K$- 
and $L'$-bands for the same galaxy nucleus is unavoidable in 
our AO photometry of spatially unresolved compact emission. 
In general, since the signal fraction of compact emission within the
$\sim$0$\farcs$5-radius aperture is higher in the $L'$- than in the
$K$-band (Figures 3--6), the estimated $K-L'$ colors can be at most
$\sim$0.2 mag redder than the true colors of the compact sources.
The uncertainty in the $K-L'$ colors of compact source emission
in a redder sense is reduced somewhat by the fact that
the spatially extended stellar contribution to an observed flux 
(within the $\sim$0$\farcs$5-radius aperture) is higher in the $K$-
than in the $L'$-band.  
This partially compensates for the slightly higher missing flux of
compact source emission in the $K$- than in the $L'$-band.  
In summary, the measured $K-L'$ colors of spatially compact source
emission should agree with the true colors within no redder than 0.2 mag.  

As to the absolute flux of the compact source emission, the
uncertainty is different from that of the $K-L'$ color.  
While our $\sim$0$\farcs$5-radius aperture recovers 75--90\% and 
83--93\% of spatially unresolved compact source signals in the $K$- and
$L'$-bands, respectively, a $\sim$2$''$-radius aperture is basically
used for the photometry of AO-corrected standard stars (point sources
with virtually no spatially extended emission) to recover as much
point source signal as possible ($>$95\%).  
Thus, when the $\sim$0$\farcs$5-radius aperture photometry of merging
nuclei is compared with the $\sim$2$''$-radius aperture photometry of
standard stars,  
the $K$- and $L'$-band compact source fluxes of merging galaxy nuclei could 
be underestimated by as much as $\sim$0.3 mag (the maximum 
difference between 75--90\% and 100\%) in the $K$-band and 
$\sim$0.2 mag (the maximum difference between 83--93\% and 100\%)
in the $L'$-band. 
Table 4 compares our nuclear 0$\farcs$5-radius (1$\farcs$0 diameter) 
$K$-band photometry, with 1$\farcs$1 diameter 2.2 $\mu$m photometry by
\citet{sco00} for (U)LIRG nuclei observed by both groups.
Our photometry tends to be fainter by a few tenths of a magnitude, most
likely because our aperture is slightly smaller, and 
some fraction of the seeing-sized AO halo signal of compact source
emission is not covered.  
Hence, our AO photometry should not miss the spatially compact emission
from the putative AGNs with more than $\sim$0.3 mag in the $K$-band and
$\sim$0.2 mag in the $L'$-band, which will not affect our main
discussions and conclusions. 

\section{Discussion}

\subsection{Galaxy nuclei with luminous detectable AGNs}

As mentioned in the introduction, AGNs should have redder $K-L'$ colors
than starbursts due to the larger amount of hot (several 100 K)
dust emission in the former. 
Although the intrinsic $K-L'$ colors could have some scatter for 
individual starbursts and AGNs, we adopt the values for 
intrinsic $K-L'$ color = 0.5 mag for starburst activity \citep{hun02} and 
$K-L'$ = 2.0 mag for AGNs \citep{iva00,alo03,vid13}.
The observed $K-L'$ colors vary as a function of the AGN contribution to
the observed flux, increasing (reddening) with increasing AGN contribution.
Our calculation shows that the $K-L'$ color becomes $>$1.0 mag when the AGN
contribution to the $L'$-band flux exceeds $\sim$50\%. 
Hence, for the purposes of this analysis, we consider that galaxy 
nuclei with $K-L'\gtrsim$ 1.0 mag contain luminous recognizable AGNs
rather than an intrinsic color scatter of starbursts.  
Based on Table 3 (column 4), at least one AGN ($K-L' \gtrsim$ 1.0 mag) 
is detected with our AO-assisted infrared imaging method in all
sources except IRAS 21208$-$0519, demonstrating that our method is very
effective at detecting AGNs in many gas- and dust-rich merging (U)LIRG
nuclei. 
In Table 3 (column 5), the estimated AGN contribution to the observed
$L'$-band flux is shown for (U)LIRG nuclei with $K-L'$ = 0.5--2.0 mag.
(U)LIRG nuclei with $K-L'$ $<$ 0.5 mag and $>$ 2.0 mag are taken 
as 100\% dominated by starbursts and AGNs, respectively. 

We now comment on some strengths of our method and discuss some caveats. 
First, the infrared $L'$-band is very sensitive to an AGN. 
Assuming the typical spectral energy distribution of an AGN and a
starburst, the $L'$-band-to-bolometric luminosity ratio of an AGN
($\sim$0.2) is two orders of magnitude higher than that of a starburst
($\sim$0.002) \citep{ris10}.
This means that for an AGN whose bolometric contribution is only 10\%
(5\%), as much as 91\% (84\%) of the observed infrared $L'$-band flux
comes from the AGN (no dust extinction case).  
Even if the dust extinction of a buried AGN is larger by A$_{\rm V}$
$\sim$ 35 mag than the surrounding starbursts, 52\% (34\%) of the
infrared $L'$-band flux originates in the buried AGN, if we adopt the
dust extinction curve derived by \citet{nis08,nis09}. 
In short, we should be able to detect moderately luminous 
buried AGNs even in galaxies with coexisting strong starbursts.
This is likely to be the primary reason that our method allows the
detection of many AGNs in the gas- and dust-rich merging (U)LIRGs that
usually accompany strong starburst activity.  

Next, we discuss an alternative possible mechanism for increasing $K-L'$ 
colors.
Dust extinction can redden the colors of obscured starbursts.
If the $K-L'$ colors become $\gtrsim$1.0 mag by this mechanism without
invoking AGN-heated hot dust emission, some galaxy nuclei with 
$K-L'>$ 1.0 mag may not contain luminous AGNs.
However, recent observations have shown that the dust extinction curve is
relatively flat in the infrared $K$- and $L'$-band wavelength range,
with dust extinction at $L'$ only 0.5--0.7 times the extinction at $K$ 
(A$_{\rm L'}$ = 0.5--0.7 $\times$ A$_{\rm K}$) \citep{nis08,nis09,gao09}.
Adopting A$_{\rm K}$/A$_{\rm V}$ = 0.062 \citep{nis08}, even in the case
of dust extinction as large as A$_{\rm V}$ = 10 mag (foreground screen
dust absorption case), the
$K-L'$ colors change by only $\sim$0.3 mag. 
In fact, in normal starburst galaxies, stellar energy sources and 
dust and gas are spatially well mixed \citep{pux91,mcl93,for01}, 
and the actual flux attenuation 
should be much lower than the foreground screen dust model for the same 
dust extinction toward the most dust-obscured region at the other side. 
Hence, we anticipate that a color reddening of starbursts to 
$K-L'\gtrsim$ 1.0 mag by dust extinction is not very common, and therefore 
that $K-L'\gtrsim$ 1.0 mag nuclei are dominated by luminous 
hot-dust-emitting AGNs, 
with a very small contribution from highly obscured starburst nuclei. 

Dust extinction for buried-AGN-heated hot dust emission could also alter  
the observed $K-L'$ colors.  
Given that the dust extinction of buried-AGN-heated $K$- and $L'$-band
emitting hot dust in the inner part of the dusty envelope is most likely
to be substantially larger than that in the surrounding starburst
regions and that a foreground screen dust model is applied to buried
AGNs \citep{ima07a}, this effect may not be neglected.
We now assume the dust extinction curve of 
A$_{\rm K}$/A$_{\rm V}$ = 0.062 and A$_{\rm L'}$/A$_{\rm V}$ = 0.031 
derived by \citet{nis08,nis09}, and consider the two examples of 
A$_{\rm V}$ = 16 mag and A$_{\rm V}$ =32 mag as dust extinction for
the buried-AGN-heated hot dust emission. 
Flux attenuation of starburst-origin $K$- and $L'$-band emission by dust
extinction is assumed to be negligible here.
The buried-AGN-origin $K-L'$ colors change from 2.0 mag to 2.5 mag and
3.0 mag in the case of A$_{\rm V}$ = 16 mag and 32 mag,
respectively, while the $K-L'$ colors of starbursts remain unchanged as
0.5 mag.  
In this case, the observed $K-L'$ colors become $\gtrsim$1.0 mag, when 
the contributions from buried-AGN-origin emission to the observed
$L'$-band fluxes are 44\% and 41\%, respectively, which are smaller than
no dust extinction case for the AGN-heated hot dust emission ($\sim$50\%
AGN contribution is required to make $K-L'$ $\gtrsim$ 1.0 mag).  
However, in the case of A$_{\rm V}$ = 16 mag and 32 mag dust extinction
toward the buried-AGN-heated hot dust region, the intrinsic
AGN-heated hot dust emission luminosities will increase by a factor of
1.6 and 2.5, respectively, after correction for the flux attenuation by
dust extinction.  
When this correction is applied, the fraction of the intrinsic
AGN-origin $L'$-band flux, relative to the observed $L'$-band flux,
becomes 55\% and 64\% for the A$_{\rm V}$ = 16 mag and 32 mag dust
extinction case, respectively. 
These fractions are even larger than that in the case of no dust
extinction for AGN-heated hot dust emission. Therefore, our
argument that important AGN contributions are required to reproduce the
observed colors of $K-L'$ $\gtrsim$ 1.0 mag will not change, or even be
strengthened.     

Now that the AGN contribution to the observed $L'$-band flux has been  
estimated (Table 3, column 5; no dust extinction case for the 
AGN-heated hot dust emission), we can derive the luminosity of the
AGN-heated hot dust emission in the close vicinity ($<$10 pc) 
of an AGN, which dominates the AGN-originated $L'$-band flux.  
As our main targets are optically elusive buried AGNs surrounded by
dust and gas with a covering factor close to unity, we here assume a
simple spherical dust distribution. 
In this geometry, dust has a strong temperature gradient.
Inner (outer) dust has a higher (lower) temperature, and dust emission
luminosity is conserved at each temperature from the hot inner regions to
the cool outer regions \citep{ima07a,ima09}.  
The inner hot dust should generate most of the $L'$-band (3.8 $\mu$m)
emission, and the intrinsic AGN-origin $L'$-band luminosity
($\nu$L$_{\nu}$ or $\lambda$L$_{\lambda}$), after subtracting the
stellar contamination, should be as large as the intrinsic AGN
UV--optical energetic radiation luminosity at the very center. 

The estimated AGN luminosity is tabulated in Table 3 (column 6).
This AGN luminosity is derived from the observed AGN-origin 
$L'$-band luminosity, with no dust extinction correction.  
Correction for possible dust extinction of the AGN-origin $L'$-band 
emission will increase the intrinsic 
AGN UV--optical energetic radiation luminosity.
Additionally, if the dust covering factor around an AGN is substantially 
below unity, the AGN-heated hot dust emission luminosity 
underestimates the intrinsic AGN UV--optical energetic radiation
luminosity.
For these reasons, the derived AGN luminosity in our method should be
taken as a lower limit.

\subsection{Infrared properties of X-ray dual AGNs}

Among the 29 observed infrared luminous merging systems, signatures of
dual AGNs with separations of $>$0$\farcs$5 were previously reported
from X-ray observations for the following five sources:  
Mrk 463 \citep{bia08}, Mrk 739 \citep{kos11}, NGC 3393 \citep{fab11}, 
NGC 6240 \citep{kom03}, and IRAS 20210$+$1121 \citep{pic10}. 
Using our methods, we detect double nuclear emission in Mrk 463, 
Mrk 739, NGC 6240, and IRAS 20210$+$1121, but not in NGC 3393.
However, only NGC 6240 displays the double red ($K-L'\gtrsim$ 1.0 mag)
nuclei that are indicative of a dual AGN. 
Mrk 463, Mrk 739, and IRAS 20210$+$1121 show only one 
red ($K-L'\gtrsim$ 1.0 mag) nucleus, with a blue $K-L'$ color 
($<$1.0 mag) at the other nucleus. 

In NGC 3393, we see no signature of a secondary nucleus at $\sim$0$\farcs$6
away from the primary nucleus \citep{fab11}, although our AO images 
should be able to resolve both nuclei spatially.
A similar case is found in an AGN at z = 0.16 (SDSS J171544.05+600835.7), 
in which double X-ray emission is found with the Chandra X-ray
Observatory, but the infrared $K$-band image shows a single nuclear 
morphology \citep{com11}. 
However, no X-ray spectra are shown in this source, and the origin of
the detected X-ray emission is less clear than NGC 3393. 

X-ray observations are sensitive to an AGN, irrespective of the presence
of hot dust in the close vicinity of a mass-accreting SMBH, and our 
infrared imaging method requires the presence of hot dust 
for AGN detection. 
If the detected X-ray emission from both nuclei originates in luminous AGNs, 
one nucleus in Mrk 463, Mrk 739, NGC 3393, and IRAS 20210$+$1121 
could be a hot-dust-deficient AGN, in which case the contribution from
AGN-heated hot dust emission to the observed $L'$-band flux would be small,
and the observed infrared $K$- and $L'$-band fluxes would be dominated by
nuclear stellar-origin emission.  

\subsection{Fraction of infrared dual AGNs}

Among the 29 observed merging (U)LIRGs, 
only four systems (Mrk 273, Arp 220, IRAS 16474$+$3430, and NGC 6240) show   
red ($K-L'\gtrsim$ 1.0 mag) colors in both merging nuclei, indicative of dual
AGNs.  
Thus, the fraction of detected dual AGNs is $\sim$14\% (4/29).  
Because the bottom five sources in Table 1 are known dual AGNs from 
previous X-ray observations, their inclusion could bias the fraction of 
detectable dual AGNs. 
However, only one out of the five sources (=20\%) is an infrared dual AGN,
not significantly biasing the total dual AGN fraction. 
The detected dual AGN fraction in our infrared imaging method is apparently
slightly higher than that of previous optical dual AGN surveys 
with $<$5\% \citep{liu10,she11}, 
although the sample size is still small, and the sample selection criteria
are different.  
However, the detected dual AGN fraction is far below unity and 
is much smaller than the value expected from the simple prediction that
the majority of gas- and dust-rich merging galaxies should have multiple
active SMBHs.  

Because our infrared imaging method is sensitive to buried AGNs, it is 
unlikely that the small detectable dual AGN fraction is due to 
the elusiveness of AGNs obscured by dust (see $\S$1). 
Previous optical spectroscopic dual AGN searches and our method 
are sensitive to multiple active SMBHs with edge-on and face-on orbiting 
geometry, respectively, and the detectable dual AGN fraction is still
much less than unity, even after combining these two methods. 
Thus, the most natural explanation is non-simultaneous SMBH 
activation \citep{van12}.
We can observationally identify the presence of SMBHs at the 
distance of the observed merging galaxies, only if the SMBHs are
actively mass accreting.
If only one of the multiple SMBHs is sufficiently active to be
observationally detectable over the long time period of a galaxy merger, 
and if the phase during which both SMBHs are active is
short, then most merging galaxies with multiple SMBHs cannot be 
identified as dual AGNs.
This scenario predicts that the mass accretion rates onto SMBHs should be 
different for multiple SMBHs. 
To test this scenario, we compare the buried AGN luminosity, derived from
our infrared observations, with $K$-band photometry, including 
host galaxy emission \citep{kim02,skr06}, 
for spatially resolved multiple-nuclei merging systems in seeing-limited 
images. 
These $K$-band luminosities should reflect the stellar luminosity, and 
SMBH mass ratios in individual merging nuclei are expected to be 
roughly proportional to the $K$-band stellar luminosity   
ratios \citep{mar03,vik12}.

Figure 7 (Left) compares the $K$-band stellar luminosity ratio 
(i.e., SMBH mass
ratio) and nuclear $L'$-band luminosity ratio (Table 5) between
seeing-based spatially resolved multiple nuclei. 
If both SMBHs in multiple nuclei systems have similar mass accretion
rates when normalized to the SMBH mass, then the sources are expected to
be distributed around the solid line. 
However, most sources are located far from the solid line,
suggesting that the mass accretion rates per SMBH mass (= Eddington
ratio) are significantly different between the two nuclei in the
majority of the observed multiple-nuclei merging systems.  
In Figure 7 (right), we also compare the $K$-band stellar luminosity 
ratio with AGN-origin $L'$-band luminosity ratio after subtracting 
stellar contamination. 
In Figure 7 (left) and (right), some ambiguity remains. 
In the left panel, stellar emission could contribute to the $L'$-band 
flux, whereas in the right panel, the AGN subtraction process could
introduce some uncertainty. 
However, both plots show basically the same behavior: most 
sources largely deviate from the solid line, strongly suggesting that
SMBH activation is non-synchronous.
The bulk of the observed merging (U)LIRGs are distributed along the
upper-left side of the solid line, indicating that larger-mass SMBHs
generally have higher Eddington ratios than smaller-mass SMBHs. 

In this comparison, we need to include some cautionary statements. 
First, the large-aperture $K$-band photometry could include emission
from AGNs, particularly for luminous AGNs that are weakly obscured by dust.
In these systems, AGN-origin nuclear emission could contribute
substantially to the observed $K$-band flux.
This might result in an overestimate of the inferred SMBH mass from the
observed $K$-band luminosity. 
However, since our primary targets are obscured AGNs in gas- and dust-rich 
merging (U)LIRGs, this effect is not expected to be severe in most cases. 
In Figure 7, the bulk of the observed sources are distributed around the
upper-left side of the solid line, a region in which larger SMBHs are
more actively mass accreting (= higher Eddington ratios).  
The AGN contribution to the $K$-band flux can be high in such 
larger-mass SMBHs with high Eddington ratios if they are less
dust obscured. 
If this effect is present in some sources, then the actual mass of the
larger SMBH is smaller than our estimate.
In that case, the true Eddington ratios in larger-mass SMBHs 
would become even higher, increasing the non-simultaneity of the
mass-accretion rates onto multiple SMBHs.  
Thus, our main conclusion does not change. 

Second, our method provides higher AGN luminosity in the 
$L'$-band brighter nucleus than in the $L'$-band fainter nucleus 
in each merging system because the possible dust extinction of AGN-heated
hot dust emission is not taken into account. 
Assuming that the Eddington ratios are similar for SMBHs at two nuclei, if
dust extinction of AGN-heated hot dust is generally smaller in the 
larger-mass SMBHs than in the smaller-mass SMBHs, then many sources
would appear at the upper-left side of the solid line in Figure 7. The
trends observed in Figure 7 could be reproduced without introducing
non-synchronous SMBH mass accretion.
Similar Eddington ratios would mean that larger-mass SMBHs have higher
absolute mass accretion rates, requiring a larger amount of
fuel in the vicinity of the SMBHs, which could obscure the SMBHs.  
In gas- and dust-rich infrared luminous merging galaxies, more luminous
AGNs with higher absolute mass accretion rates are predicted to be more
highly obscured \citep{hop06}, making this second scenario (similar 
Eddington ratios with less dust extinction in the larger-mass SMBHs)
unlikely, and still supporting the suggestion that larger-mass SMBHs
have generally higher Eddington ratios. 

Third, when we convert $K$-band stellar luminosity to SMBH mass, 
we need to mention that younger star formation is likely to have 
higher $K$-band 
luminosity than older star formation for a given galaxy stellar 
mass \citep{bel01}. 
The inferred SMBH mass can be overestimated for younger star formation. 
Actively mass-accreting SMBHs indicate the presence of
dynamically settled nuclear gas, which is likely to cause active young 
nuclear starbursts \citep{ima02,ima03,iw04,oi10,ima11a}.  
Thus, the SMBH mass could be overestimated for active SMBHs with higher
Eddington ratios.
Now, Figure 7 suggests that larger-mass SMBHs have higher
Eddington ratios in general than do smaller-mass SMBHs. 
If the abscissa is changed from $K$-band stellar luminosity to actual
SMBH mass, many sources currently located around the upper-left side of
the solid line in Figure 7 would move leftward, even strengthening the 
scenario that larger mass SMBHs have higher Eddington ratios. 
Thus, this uncertainty also will not change our conclusion. 

Figure 8 plots the comparison of the nuclear $L'$-band to 
galaxy-wide $K$-band
stellar luminosity ratio (Left) and AGN-origin $L'$-band to
$K$-band stellar luminosity ratio (Right), between two nuclei, 
as a function of apparent nuclear separation.
Since the $K$-band stellar luminosity is taken as the indicator
of SMBH mass \citep{mar03,vik12}, the ordinate corresponds to the ratio 
of SMBH-mass-normalized mass accretion rates (= Eddington ratio) between
the two nuclei. 
Sources close to (far away from) the solid horizontal lines mean that 
SMBH-mass-normalized mass accretion rates are similar (largely different) 
between the two nuclei. 
If dual SMBH activation preferentially occurs at a later merging stage
\citep{van12}, it is expected that sources with small apparent nuclear
separation tend to be distributed around the solid horizontal lines. 
No such trend is seen.

Finally, our comparison of SMBH mass and AGN luminosity 
above is limited to (U)LIRGs with
spatially resolved nuclei in seeing-limited images.  
Multiple active closely-separated SMBHs ($<<$1$''$),
which are spatially resolvable only with our AO images 
(e.g., IRAS 16474$+$3430 in Figure 1, $L'$-band image), are very
interesting, in terms of the theoretical prediction that such SMBHs in
the late stages of gas- and dust-rich galaxy mergers become particularly
active and become luminous AGNs \citep{hop05,hop06,van12}.  
However, although AGN luminosity ratios for closely separated systems 
could be derived from our AO-assisted high-spatial-resolution infrared
$K$- and $L'$-band imaging data (Table 3), SMBH mass ratios are
difficult to obtain from stellar emission luminosity in $K$-band
images because host galaxy's stellar emission strongly overlaps between 
multiple nuclei.
Spatially resolved velocity information obtained with AO-assisted
spectroscopy will be useful for inferring SMBH ratios in such
small-separation SMBH systems, but such observations are still limited
to very bright nuclei only \citep{med11,u13}. 
Investigating the properties of these closely separated SMBHs
requires additional future work. 

In summary, our results support the scenario proposed by the theory
\citep{van12} that SMBH mass accretion is not simultaneous among
multiple SMBHs in gas- and dust-rich merging galaxies.
In general, larger-mass SMBHs are more actively mass accreting  
(normalized to SMBH mass) in merging (U)LIRGs with multiple nuclei in 
seeing-limited images.
This non-synchronous SMBH activation may reduce the fraction of
observable dual AGN, compared to the fraction of multiple SMBHs, in
merging galaxies. 
Our results suggest that mass accretion onto SMBHs is dominated by the
local environment on the small scale rather than by global galaxy
properties, even in gas- and dust-rich infrared luminous merging
galaxies.  
In this case, it is not easy to predict SMBH activity through
theoretical simulations of galaxy mergers.
Thus, observations are important for understanding how multiple SMBHs 
are activated during the gas- and dust-rich galaxy merger process. 

\section{Summary}

We conducted infrared $K$- and $L'$-band 
high-spatial-resolution ($<$0$\farcs$2)
imaging observations of nearby infrared luminous merging galaxies 
using Subaru LGS/NGS-AO to search for kiloparsec-scale multiple AGNs
surrounded by dust through the detection of red $K-L'$ compact sources. 
Given the gas- and dust-rich nature of these galaxies, many SMBHs are
expected to be mass accreting and hence to become luminous AGNs, but
these AGNs are deeply buried in gas and dust.
Using our infrared method, which is sensitive to buried AGNs,
the observational detection of multiple SMBHs is expected to be more
feasible than in gas-poor galaxy mergers where many SMBHs may not be
actively mass accreting due to the paucity of surrounding gas.   
We present the following main conclusions.

\begin{enumerate}
\item Among 29 observed merging systems, at least one AGN was 
found in all sources except one, demonstrating the effectiveness of our   
method for the purpose of AGN detection in these gas- and dust-rich 
infrared luminous merging galaxies.

\item Kiloparsec-scale dual AGNs were seen in only four of 29 galaxies, 
even using our powerful method, which is sensitive to deeply buried
AGNs. 
This fraction seems slightly higher than the fraction determined by 
previously published optical spectroscopic dual AGN searches, despite 
the small sample size and the differences in criteria used for sample
selection.  
However, the fraction is still significantly smaller than the value
derived from the simple theoretical prediction that most gas- and
dust-rich merging galaxies are expected to contain multiple active
SMBHs. 

\item The AGN luminosity ratios derived from AGN-origin $L'$-band 
emission between two nuclei are, in most cases, higher than the 
SMBH mass ratios inferred from large-aperture $K$-band photometric
observations.  
When normalized to SMBH mass, larger-mass SMBHs are generally more
highly mass accreting than are smaller-mass SMBHs in most of the
observed infrared luminous merging galaxies with spatially resolved
nuclei in seeing-limited images.  
This trend is independent off the apparent nuclear separation.

\item When combined with previous optically based dual AGN searches,
our observational results suggest that the most likely reason for the
small observed dual AGN fraction in merging galaxies is that mass
accretion onto multiple SMBHs is non-simultaneous rather than the result
of the orbiting geometry of multiple SMBHs or the optical elusiveness of
AGNs deeply buried in gas and dust. 

\item Our results suggest that in gas- and dust-rich infrared luminous
merging galaxies, mass accretion onto SMBHs is primarily determined by
local conditions rather than by global galaxy properties.
This makes  
theoretical prediction difficult and necessitates the inclusion of 
observational constraints when attempting to understand what is happening 
for SMBHs in gas- and dust-rich galaxy mergers.   

\end{enumerate} 

We thank the anonymous referee for his/her useful comment.  
We are grateful to Drs. Minowa and Ishii for their support during 
our observations at the Subaru Telescope and to Sayaka Yamaguchi for 
her English proofreading. 
M.I. is supported by a Grant-in-Aid for Scientific Research
(23540273). 
This research made use of (1) the SIMBAD database, operated at CDS,
Strasbourg, France, and the NASA/IPAC Extragalactic Database (NED)
operated by the Jet Propulsion Laboratory, California Institute of
Technology, under contract with the National Aeronautics and Space
Administration,
(2) data products from the Two Micron All Sky
Survey, which is a joint project of the University of Massachusetts and
the Infrared Processing and Analysis Center/California Institute of
Technology, funded by the National Aeronautics and Space Administration
and the National Science Foundation, and
(3) the NASA/ IPAC Infrared Science Archive, which is operated by the
Jet Propulsion Laboratory, California Institute of Technology, under
contract with the National Aeronautics and Space Administration.

\clearpage


\clearpage

\begin{deluxetable}{lcrrrrccc}
\tabletypesize{\scriptsize}
\tablecaption{
Properties of the Observed Infrared Luminous Merging 
Galaxies \label{tbl-1}}
\tablewidth{0pt}
\tablehead{
\colhead{Object} & \colhead{z} &  
\colhead{f$_{\rm 12}$}  & \colhead{f$_{\rm 25}$}  & 
\colhead{f$_{\rm 60}$}  & \colhead{f$_{\rm 100}$}  & 
\colhead{log L$_{\rm IR}$} & \colhead{Optical} & \colhead{AGN in the} \\ 
 &    & \colhead{(Jy)} & \colhead{(Jy)}  & \colhead{(Jy)} 
& \colhead{(Jy)}  & \colhead{(L$_{\odot}$)}   & \colhead{Class} &
\colhead{main nucleus ?} \\ 
\colhead{(1)} & \colhead{(2)} & \colhead{(3)} & \colhead{(4)} & 
\colhead{(5)} & \colhead{(6)} & \colhead{(7)} & \colhead{(8)} &
\colhead{(9)} 
}
\startdata
IRAS 00091$-$0738 & 0.118 & $<$0.07 & 0.22 & 2.63 & 2.52 & 12.2  & HII$^{a}$(cp$^{b}$) & Y$^{1,2,3}$\\
IRAS 00188$-$0856 & 0.128 & $<$0.12 & 0.37 & 2.59 & 3.40 & 12.4 & LI$^{a}$(Sy2$^{b}$) & Y$^{1,2,3,4,5}$\\  
IRAS 05024$-$1941 & 0.192 & 0.15 & 0.14 & 1.06 & 1.34 & 12.5  & Sy2$^{a,b}$ &Y$^{2,3}$\\
IRAS 05189$-$2524 & 0.042 & 0.73 & 3.44 & 13.67 & 11.36 & 12.1 & Sy2$^{a,b}$  &Y$^{2,3,6,7,8,9,10,11,12,13,14,15}$\\
IRAS 08572$+$3915 & 0.058 & 0.32 & 1.70 & 7.43  & 4.59  & 12.1  & LI$^{a}$(Sy2$^{b}$) &Y$^{1,2,3,4,7,9,12,14,16,17}$\\  
IRAS 12127$-$1412 & 0.133 & $<$0.13 & 0.24 & 1.54 & 1.13 & 12.2 & LI$^{a}$(HII$^{b}$) &Y$^{1,2,3,4,5}$\\  
IRAS 12540+5708 (Mrk 231) & 0.042 & 1.87 & 8.66 & 31.99 & 30.29 & 12.5  & Sy1$^{a,b}$ & Y$^{2,3,5,7,9,12,18,19}$\\  
IRAS 13335$-$2612 & 0.125 & $<$0.13 & $<$0.14 & 1.40 & 2.10 & 12.1 & LI$^{a}$(cp$^{b}$) &Y$^{2}$\\
IRAS 13428$+$5608 (Mrk 273) & 0.038 & 0.24 & 2.28 & 21.74 & 21.38 & 12.1  & Sy2$^{a,b}$ & Y$^{2,3,6,7,8,12,13,15,20,21,22,23}$\\
IRAS 13443$+$0802 & 0.135 & $<$0.12 & $<$0.11 & 1.50 & 1.99 & 12.2  & HII+Sy2$^{a}$(cp+Sy2$^{b}$) & N\\ 
IRAS 13451$+$1232 (PKS 1345+12) & 0.122 & 0.14 & 0.67 & 1.92 & 2.06 & 12.3 & Sy2$^{a,b}$ &Y$^{2,3,4,5,11,13,15,24}$\\ 
IRAS 14348$-$1447 & 0.083 & 0.07 & 0.49 & 6.87 & 7.07 & 12.3 & LI$^{a}$(cp$^{b}$) &Y$^{1,2,3,4,5,12}$\\  
IRAS 15327$+$2340 (Arp 220) & 0.018 & 0.48 & 7.92 & 103.33 & 112.40 & 12.1 & LI$^{a,b}$ &Y$^{2,3,12,15,25}$\\  
IRAS 16468$+$5200 & 0.150 & $<$0.06 & 0.10 & 1.01 & 1.04 & 12.1 & LI$^{a}$(cp$^{b}$) &Y$^{1,3,14}$\\  
IRAS 16474$+$3430  & 0.111 & $<$0.13 & 0.20 & 2.27 & 2.88 & 12.2 & HII$^{a}$(cp$^{b}$) &Y$^{4}$\\ 
IRAS 16487$+$5447 & 0.104 & $<$0.07 & 0.20 & 2.88 & 3.07 & 12.2  & LI$^{a}$(cp$^{b}$) &Y$^{3,4}$\\ 
IRAS 17044$+$6720   & 0.135 & $<$0.07 & 0.36 & 1.28 & 0.98 & 12.2 & LI$^{a}$(Sy2$^{b}$) &Y$^{1,3,4,14}$\\  
IRAS 21208$-$0519 & 0.130 & $<$0.09 & $<$0.15 & 1.17 & 1.66 & 12.0 & HII$^{a}$(cp$^{b}$) &Y$^{2}$\\ 
IRAS 23233$+$2817 & 0.114 & $<$0.13 & 0.28 & 1.26 & 2.11 & 12.1 & Sy2$^{a,b}$ &Y$^{5}$\\
IRAS 23234$+$0946 & 0.128 & $<$0.06 & 0.08 & 1.56 & 2.11 & 12.1 & LI$^{a}$(cp$^{b}$) &Y$^{2,3}$\\ 
IRAS 23327$+$2913 & 0.107 & $<$0.06 & 0.22 & 2.10 & 2.81 & 12.1 & LI$^{a}$(Sy2$^{b}$) &Y$^{1,3}$\\ 
IRAS 23389$+$0300 & 0.145 & $<$0.09 & $<$0.35 & 1.23 & 1.17 & 12.2  & Sy2$^{a,b}$ &N\\ 
IRAS 23498$+$2423 & 0.212 & $<$0.10 & 0.12 & 1.02 & 1.45 & 12.5 & Sy2$^{a,b}$ &Y$^{2,3,13,14,24}$\\
UGC 5101 & 0.040 & 0.25 & 1.03 & 11.54 & 20.23 & 12.0 & LI$^{c}$(Sy2$^{b}$) &Y$^{2,12,13,14,26,27}$\\  
Mrk 463 &  0.051 &  0.51 & 1.58 & 2.18 &  1.92 & 11.8 & Sy2$^{d,e,f}$ & Y$^{11,12,13,24,28,29}$\\  
Mrk 739 &  0.030 &  0.16 &  0.31 & 1.26 & 2.41 & 10.9 & Sy1+HII$^{g}$ &Y$^{30}$\\  
NGC 3393 &  0.013 & 0.13 &  0.75 & 2.25 & 3.87 & 10.4 & Sy2$^{h}$ &Y$^{31}$\\  
NGC 6240 &  0.024 &  0.56 & 3.42 & 22.68 & 27.78 & 11.8 & LI$^{b,c}$ & Y$^{2,12,13,32}$\\  
IRAS 20210$+$1121 &  0.056 &  0.29 &  1.40 & 3.39 & 2.68 & 11.9 & Sy2+LI$^{i}$& Y$^{33}$\\  
\enddata

\tablecomments{
Col.(1): Object name.
Col.(2): Redshift. 
Cols.(3)--(6): f$_{12}$, f$_{25}$, f$_{60}$, and f$_{100}$ are 
{\it IRAS} fluxes at 12 $\mu$m, 25 $\mu$m, 60 $\mu$m, and 100 $\mu$m.
For the first 23 galaxies, the flux is derived from \citet{kim98}. 
For the last six galaxies, we use the IRAS Faint Source Catalog.
Col.(7): Decimal logarithm of infrared (8$-$1000 $\mu$m) luminosity
in units of solar luminosity (L$_{\odot}$), calculated with
$L_{\rm IR} = 2.1 \times 10^{39} \times$ D(Mpc)$^{2}$
$\times$ (13.48 $\times$ $f_{12}$ + 5.16 $\times$ $f_{25}$ +
$2.58 \times f_{60} + f_{100}$) [ergs s$^{-1}$] \citep{sam96}, where 
we adopt H$_{0}$ $=$ 71 km s$^{-1}$ Mpc$^{-1}$, $\Omega_{\rm M}$ = 0.27,
and $\Omega_{\rm \Lambda}$ = 0.73 \citep{kom09}, to estimate  
the luminosity distance D (Mpc) from the redshift.
Col.(8): Optical spectral classification and references. 
Sy1, Sy2, LI, HII, and cp mean Seyfert 1, Seyfert 2, LINER, HII-region,
and starburst+AGN composite type, respectively.
$^{a}$: \citet{vei99a}. 
$^{b}$: \citet{yua10}.
$^{c}$: \citet{vei95}.
$^{d}$: \citet{shu81}.
$^{e}$: \citet{hut89}.
$^{f}$: \citet{san88}.
$^{g}$: \citet{kos11}.
$^{h}$: \citet{dia88}.
$^{i}$: \citet{per90}.
Col.(9): The presence of AGN signatures in the brightest main nucleus 
(Y = yes, N = no), and several selected representative references.
$^{1}$: \citet{ima07a}.
$^{2}$: \citet{vei09}. 
$^{3}$: \citet{nar10}.
$^{4}$: \citet{idm06}. 
$^{5}$: \citet{ima10b}. 
$^{6}$: \citet{vei99b}.
$^{7}$: \citet{imd00}.
$^{8}$: \citet{ris00}.
$^{9}$: \citet{soi00}.
$^{10}$: \citet{sev01}.
$^{11}$: \citet{ima04}.
$^{12}$: \citet{arm07}.
$^{13}$: \citet{far07}.
$^{14}$: \citet{ima08}.
$^{15}$: \citet{ten09}.
$^{16}$: \citet{dud97}.
$^{17}$: \citet{ima11b}.  
$^{18}$: \citet{mal00}.  
$^{19}$: \citet{bra04}.  
$^{20}$: \citet{xia02}. 
$^{21}$: \citet{bal05}. 
$^{22}$: \citet{iwa11a}.
$^{23}$: \citet{u13}.
$^{24}$: \citet{vei97}.
$^{25}$: \citet{dow07}.
$^{26}$: \citet{idm01}. 
$^{27}$: \citet{ima03}. 
$^{28}$: \citet{uen96}. 
$^{29}$: \citet{bia08}.
$^{30}$: \citet{kos11}.
$^{31}$: \citet{fab11}. 
$^{32}$: \citet{kom03}. 
$^{33}$: \citet{pic10}.
}

\end{deluxetable}

\begin{deluxetable}{lclc|lr|lrr}
\tabletypesize{\scriptsize}
\tablecaption{Observation Log \label{tbl-2}}
\tablewidth{0pt}
\tablehead{
\colhead{Object} & \colhead{Band} & \colhead{Date} & \colhead{Exposure} 
& \multicolumn{2}{c}{Standard Star} & \multicolumn{3}{c}{LGS-AO or NGS-AO guide star} \\
\colhead{} & \colhead{} & \colhead{(UT)} & \colhead{(min)} &
\colhead{Name} & \colhead{mag} & \colhead{Name} & \colhead{$R$-band} &
\colhead{separation}  \\ 
\colhead{} & \colhead{} & \colhead{} & \colhead{} & \colhead{} & \colhead{} & 
\colhead{USNO} & \colhead{mag} & \colhead{(arcsec)}  \\
\colhead{(1)} & \colhead{(2)} & \colhead{(3)} & \colhead{(4)} &
\colhead{(5)} & \colhead{(6)} & \colhead{(7)} & \colhead{(8)} & \colhead{(9)} 
}
\startdata
IRAS 00091$-$0738 & K  & 2011 August 23 & 11.3 & FS2 & 10.5 & 0826-0002500 & 17 & 57 \\
                  & L' & 2011 August 22, 23 & 27 & HD 1160 & 7.1 & 0826-0002500 & 17 & 57 \\
IRAS 00188$-$0856 & K  & 2011 August 23 & 11.7 & FS2 & 10.5 & 0813-0003721 & 16 & 24 \\
                  & L' & 2011 August 22 & 9 & HD1160 & 7.1 & 0813-0003721 & 16 & 24 \\
IRAS 05024$-$1941 & K  & 2012 October 16 & 9 & FS13 & 10.1 & 0703-0054437 & 15 & 10 \\
                  & L' & 2012 October 16 & 27 & HD22686 & 7.2 & 0703-0054437 & 15 & 10 \\
IRAS 05189$-$2524 & K  & 2012 October 16 & 9 & FS13 & 10.1 & nucleus & 11 & 0 \\
                  & L' & 2012 October 16 & 9 & HD22686 & 7.2 & nucleus & 11 & 0 \\
IRAS 08572$+$3915 & K  & 2012 April 15 & 4.5 & FS125 & 10.4 & 1290-0192776 & 14 & 58 \\ 
                  & L' & 2012 April 15 & 13.5 & HD84800 & 7.5 & 1290-0192776 & 14 & 58 \\
IRAS 12127$-$1412 & K  & 2012 May 20 & 9 & FS132 & 11.8 & 0755-0260793 & 13 & 20 \\ 
                  & L' & 2012 May 20 & 9 & HD106965 & 7.3 & 0755-0260793 & 13 & 20 \\
Mrk 231           & K  & 2013 May 8 & 15.6 & FS133 & 11.9 & nucleus \tablenotemark{a} & 9 & 0 \\
                  & L' & 2013 May 8 & 13.7 & HD129653 & 6.9 & nucleus \tablenotemark{a} & 9 & 0 \\
IRAS 13335$-$2612 & K  & 2012 May 20 & 9 & S791-C & 11.2 & 0635-0309957 & 14 & 43 \\ 
                  & L' & 2012 May 20 & 18 & HD106965 & 7.3 & 0635-0309957 & 14 & 43 \\
Mrk 273           & K  & 2013 May 8 & 9 & FS133 & 11.9 & 1458-0231011 \tablenotemark{a} & 16 & 34 \\
                  & L' & 2013 May 8 & 22.5 & HD129653 & 6.9 & 1458-0231011 \tablenotemark{a} & 16 & 34 \\
IRAS 13443$+$0802 & K  & 2012 May 20 & 9 & S791-C & 11.2 & 0977-0294304 & 16 & 57 \\ 
                  & L' & 2012 May 20 & 18 & HD106965 & 7.3 & 0977-0294304 & 16 & 57 \\
PKS 1345+12       & K  & 2012 March 23 & 4 & P272D & 11.2 & western nucleus & 12 & 0 \\ 
                  & L' & 2012 March 23 & 9 & HD136754 & 7.2 & western nucleus & 12 & 0 \\
IRAS 14348$-$1447 & K  & 2012 April 15 & 9 & FS132 & 11.8 & 0749-0288838 & 15 & 27 \\ 
                  & L' & 2012 April 15 & 13.5 & HD106965 & 7.3 & 0749-0288838 & 15 & 27 \\
Arp 220           & K  & 2012 April 15 & 9 & P272D & 11.2 & nucleus & 8 & 0 \\ 
                  & L' & 2012 April 15 & 13.5 & HD136754 & 7.2 & nucleus & 8 & 0 \\
IRAS 16468$+$5200 & K  & 2011 August 23 & 12 & p138-c & 11.1 & 1419-0298775 & 18 & 37 \\  
                  & L' & 2011 August 23 & 13.5 & HD162208 & 7.1 & 1419-0298775 & 18 & 37 \\
IRAS 16474$+$3430 & K  & 2011 July 30 & 8 & FS 139 & 12.1 & 1244-0244739 & 15 & 17 \\ 
                  & L' & 2012 May 20 & 18 & HD136754 & 7.2 & 1244-0244739 & 15 & 17 \\
IRAS 16487$+$5447 & K  & 2011 August 23 & 11.3 & p138-c & 11.1 & 1447-0253801 & 13 & 54 \\ 
                  & L' & 2011 August 23 & 13.5 & HD162208 & 7.1 & 1447-0253801 & 13 & 54 \\
IRAS 17044$+$6720 & K  & 2012 April 15 & 9 & P272D & 11.2 & 1572-0201777 & 14 & 62 \\  
                  & L' & 2012 April 15 & 4.5 & HD136754 & 7.2 & 1572-0201777 & 14 & 62 \\
IRAS 21208$-$0519 & K  & 2011 August 23 & 15 & S813D & 11.1 & 0848-0617074 & 15 & 17 \\ 
                  & L' & 2011 Augsut 23 & 22.5 & GL811.1 & 6.7 & 0848-0617074 & 15 & 17 \\
IRAS 23233$+$2817 & K  & 2012 October 16 & 9 & FS155 & 9.4 & 1185-0599614 & 14 & 63 \\ 
                  & L' & 2012 October 16 & 13.5 & HD203856 & 6.9 & 1185-0599614 & 14 & 63 \\
IRAS 23234$+$0946 & K  & 2011 August 22 & 11.3 & FS154 & 11.1 & 1000-0612015 & 18 & 61 \\ 
                  & L' & 2011 August 22 & 9 & HD1160 & 7.1 & 1000-0612015 & 18 & 61 \\
IRAS 23327$+$2913 & K  & 2011 August 23 & 11.3 & FS155 & 9.4 & 1194-0587875 & 18 & 36 \\ 
                  & L' & 2012 October 16 & 18 & HD203856 &  6.9 & 1194-0587875 & 18 & 36 \\
IRAS 23389$+$0300 & K  & 2011 August 22 & 11.3 & FS154 & 11.1 & 0932-0712575 & 16 & 52 \\ 
                  & L' & 2011 August 22 & 13.5 & HD1160 & 7.1 & 0932-0712575 & 16 & 52 \\
IRAS 23498$+$2423 & K  & 2012 October 16 & 9 & FS155 & 9.4 & north-western nucleus & 17 & 0 \\ 
                  & L' & 2012 October 16 & 9 & HD203856 & 6.9 & north-western nucleus & 17 & 0 \\
UGC 5101 & K  & 2012 April 15 & 3.75 & FS125 & 10.4 & 1513-0178261 & 15 & 15 \\  
         & L' & 2012 April 15 & 9 & HD84800 & 7.5 & 1513-0178261 & 15 & 15 \\
Mrk 463  & K  & 2012 March 23 & 9 & P272D & 11.2 & 1083-0240150 & 17 & 65 \\  
         & L' & 2012 March 23 & 9 & HD136754 & 7.2 & 1083-0240150 & 17 & 65 \\
Mrk 739  & K  & 2012 April 15 & 9 & FS125 & 10.4 & 1116-0211017 & 17 & 31 \\  
         & L' & 2012 April 15 & 13.5 & HD84800 & 7.5 & 1116-0211017 & 17 & 31 \\
NGC 3393 & K  & 2012 May 20 & 9 & FS132 & 11.8 & 0648-0254969 & 16 & 55 \\  
         & L' & 2012 May 20 & 9 & HD106965 & 7.3 & 0648-0254969 & 16 & 55 \\
NGC 6240 & K  & 2011 June 20 & 48 & FS137 & 11.8 & 0924-0386013 \tablenotemark{a}& 12 & 37 \\  
         & L' & 2011 June 20 & 5 & HD129655 & 6.7 & 0924-0386013 \tablenotemark{a}& 12 & 37 \\
IRAS 20210$+$1121 & K  & 2011 June 20 & 7.5 & FS149 & 10.1 & 1015-0589702 \tablenotemark{a}&13 & 47  \\  
         & L' & 2011 June 20 & 6.3 & HD201941 & 6.6 & 1015-0589702 \tablenotemark{a}& 13 & 47 \\
\enddata

\tablenotetext{a}{NGS-AO guide star}

\tablecomments{
Col.(1): Object name.  
Col.(2): Observed band. $K$- or $L'$-band.
Col.(3): Observing date in UT.
Col.(4): Net on-source exposure time in min.
Col.(5): Standard star's name.
Col.(6): Standard star's magnitude in the $K$- or $L'$-band.
Col.(7): Guide star name (USNO number) used for the LGS-AO tip-tilt
correction or NGS-AO correction.
Col.(8): Guide star's optical $R$-band magnitude.
Col.(9): Separation between the target object and guide star in arcsec.
}
\end{deluxetable}

\begin{deluxetable}{lcrcrrc}
\tabletypesize{\scriptsize}
\tablecaption{
Nuclear Photometry and Estimated AGN Component \label{tbl-3}}
\tablewidth{0pt}
\tablehead{
\colhead{Object} & \colhead{$K$(0$\farcs$5)} &  
\colhead{$L'$(0$\farcs$5)}  & \colhead{$K-L'$(0$\farcs$5)} &
\colhead{AGN} &  \colhead{L$_{\rm AGN}$} & \colhead{WISE(3.4$\mu$m)} \\ 
\colhead{} & \colhead{(mag)} & \colhead{(mag)} & \colhead{(mag)} &
\colhead{(\%)} & \colhead{(10$^{44}$ ergs s$^{-1}$)} & \colhead{(mag)}\\  
\colhead{(1)} & \colhead{(2)} & \colhead{(3)} & \colhead{(4)} 
& \colhead{(5)} & \colhead{(6)} & \colhead{(7)} 
}
\startdata
IRAS 00091$-$0738 S & 15.7 & 14.7 & 1.0$\pm$0.2 & 49 & 0.045 & 13.7\tablenotemark{A} \\
IRAS 00091$-$0738 N & 16.5 & $>$15.3 & $<$1.2 & $<$64 & $<$0.034 & 13.7\tablenotemark{A} \\
IRAS 00188$-$0856   & 13.9 & 12.7 & 1.2$\pm$0.2 & 63 & 0.44 &12.3\\
IRAS 05024$-$1941 E & 15.1 & 13.9 & 1.2$\pm$0.2 & 63 & 0.35& 13.3\tablenotemark{A}\\
IRAS 05024$-$1941 W & 15.8 & $>$15.3 & $<$0.5 & 0 & 0 &13.3\tablenotemark{A}\\
IRAS 05189$-$2524   &  10.9 & 8.6 & 2.3$\pm$0.2 & 100 & 2.9 & 8.7\\
IRAS 08572$+$3915 NW & 13.8 & 9.3\tablenotemark{B} & 4.5$\pm$0.2 & 100 & 3.0 &10.3\\
IRAS 08572$+$3915 SE & 16.5 & $>$14.8\tablenotemark{B} & $<$1.7 & $<$90 & $<$0.017 &\nodata\\ 
IRAS 12127$-$1412 NE & 13.5 & 10.7 & 2.8$\pm$0.2 & 100 & 4.8 & 10.9\\ 
IRAS 12127$-$1412 SW & 18.3 & $>$15.2 & $<$3.1 & $<$100 & $<$0.076 &\nodata\\
Mrk 231 & 9.2 & 7.2\tablenotemark{C} & 2.0$\pm$0.2 & 100 & 10 & 7.4\\
IRAS 13335$-$2612 N & 15.6 & 14.8 & 0.8$\pm$0.2 & 32 & 0.030 &13.0\tablenotemark{A}\\
IRAS 13335$-$2612 S & 15.3 & 14.2 & 1.1$\pm$0.2 & 57 & 0.094 &13.0\tablenotemark{A}\\
IRAS 13335$-$2612 SE & 16.0 & $>$15.4 & $<$0.6 & $<$12 & $<$0.0066&13.0\tablenotemark{A}\\
Mrk 273 SW & 13.6 & 11.3\tablenotemark{D} & 2.3$\pm$0.2 & 100 & 0.19 &10.4\tablenotemark{A}\\
Mrk 273 NE & 13.1 & 11.5\tablenotemark{D} & 1.6$\pm$0.2 & 85 & 0.14 &10.4\tablenotemark{A}\\
IRAS 13443$+$0802 NE & 14.6 & 14.1 & 0.5$\pm$0.2 & 0 & 0 &12.7\\
IRAS 13443$+$0802 E  & 16.1 & 14.6 & 1.5$\pm$0.2 & 80 & 0.11 &13.8\\
IRAS 13443$+$0802 SW & 16.8 & $>$15.3 & $<$1.5 & $<$81 & $<$0.058 &14.8\\
PKS 1345+12 W & 14.4 & 11.4  & 3.0$\pm$0.2 & 100 & 2.1 &11.4\tablenotemark{A}\\ 
PKS 1345+12 E & 15.6 & $>$14.6 & $<$1.0 & $<$50 & $<$0.054 &11.4\tablenotemark{A}\\ 
IRAS 14348$-$1447 SW & 14.3 & 13.4 & 0.9$\pm$0.2 & 41 & 0.059 & 12.1\tablenotemark{A}\\ 
IRAS 14348$-$1447 NE & 15.0 & 13.7 & 1.3$\pm$0.2 & 70 & 0.076 & 12.1\tablenotemark{A}\\ 
Arp 220 W & 13.0 & 11.7 & 1.3$\pm$0.2 & 70 & 0.021 & 10.1\tablenotemark{A}\\
Arp 220 E & 13.2 & 12.1 & 1.1$\pm$0.2 & 57 & 0.012 & 10.1\tablenotemark{A}\\
IRAS 16468$+$5200 E & 16.5 & 15.0 & 1.5$\pm$0.2 & 80 & 0.094 &14.4\tablenotemark{A}\\
IRAS 16468$+$5200 W & 16.7 & $>$15.6 & $<$1.1 & $<$57 & $<$0.039 & 14.4\tablenotemark{A}\\
IRAS 16474$+$3430 S & 14.6 & 13.6 & 1.0$\pm$0.2 & 49 & 0.11 &12.3\tablenotemark{A}\\ 
IRAS 16474$+$3430 M \tablenotemark{E} & 14.8 & 13.7 & 1.1$\pm$0.2 & 57 & 0.12 &12.3\tablenotemark{A}\\
IRAS 16487$+$5447 SW & 15.7 & 14.6 & 1.1$\pm$0.2 & 57 & 0.044 &13.1\tablenotemark{A}\\ 
IRAS 16487$+$5447 NE & 16.0 & $>$15.2 & $<$0.8 & $<$33 & $<$0.015 &13.1\tablenotemark{A}\\ 
IRAS 17044$+$6720    & 14.5 & 11.9 & 2.6$\pm$0.2 & 100 & 1.6 &12.0\\
IRAS 21208$-$0519 NE & 15.1 & 14.5 & 0.6$\pm$0.2 & 12 & 0.016 &13.2\\
IRAS 21208$-$0519 SW & 16.1 & $>$16.0 & $<$0.1 & 0 & 0 &\nodata\\
IRAS 23233$+$2817 N  & 14.6 & 12.7 & 1.9$\pm$0.2 & 97 & 0.52 &12.5\tablenotemark{A}\\
IRAS 23233$+$2817 S  & 16.3 & $>$15.0 & $<$1.3 & $<$70 & $<$0.046&12.5\tablenotemark{A} \\
IRAS 23234$+$0946 NW & 15.2 & 13.7 & 1.5$\pm$0.2 & 80 & 0.22 & 13.4\tablenotemark{A}\\
IRAS 23234$+$0946 SE & 16.7 & $>$14.3 & $<$2.4 & $<$100 & $<$0.16 &13.4\tablenotemark{A}\\
IRAS 23327$+$2913 S  & 14.5 & 13.2 & 1.3$\pm$0.2 & 70 & 0.21 & 12.8\\
IRAS 23327$+$2913 N  & 15.7 & $>$14.9 & $<$0.8 & $<$33 & $<$0.020 &13.8\\
IRAS 23389$+$0300 N  & 15.0 & 14.0 & 1.0$\pm$0.2 & 49 & 0.13 & 13.7\tablenotemark{A}\\
IRAS 23389$+$0300 S  & 17.1 & $>$15.1 & $<$2.0 & $<$100 & $<$0.10 &13.7\tablenotemark{A}\\
IRAS 23498$+$2423 NW & 13.7 & 11.4 & 2.3$\pm$0.2 & 100 & 7.0 & 11.5\tablenotemark{A}\\
IRAS 23498$+$2423 SE & 17.4 & $>$15.4 & $<$2.0 & $<$100 & $<$0.18 & 11.5\tablenotemark{A}\\
UGC 5101  & 12.0 & 9.5 & 2.5$\pm$0.2 & 100 & 1.1 & 10.0\\
Mrk 463 E & 10.9 & 8.3 & 2.6$\pm$0.2 & 100 & 5.5 &8.5\tablenotemark{A}\\
Mrk 463 W & 14.1 & 13.3 & 0.8$\pm$0.2 & 32 & 0.017 & 8.5\tablenotemark{A} \\
Mrk 739 E & 12.2 & 10.2 & 2.0$\pm$0.2 & 100 & 0.33 &9.9 \\
Mrk 739 W & 14.2 & 13.8 & 0.4$\pm$0.2 & 0 & 0 &\nodata \\
NGC 3393  & 13.0 & 11.9 & 1.1$\pm$0.2 & 57 & 0.0072 &10.0\\
NGC 6240 S \tablenotemark{F} & 11.8 & 10.5\tablenotemark{G} & 1.3$\pm$0.2 & 70 & 0.11 &9.3\tablenotemark{A}\\
NGC 6240 N \tablenotemark{F} & 13.0 & 11.9\tablenotemark{G} & 1.1$\pm$0.2 & 57 & 0.025 &9.3\tablenotemark{A}\\
IRAS 20210$+$1121 S & 13.5 & 11.0 & 2.5$\pm$0.2 & 100 & 0.57 &11.2\\
IRAS 20210$+$1121 N & 14.6 & 13.9 & 0.7$\pm$0.2 & 22 & 0.0087 &\nodata\\
\enddata

\tablenotetext{A}{
More than one nucleus combined, due to 
low-spatial-resolution data.}

\tablenotetext{B}{
\citet{zho93} estimated 10.0 mag and $>$15.0 mag at 3.4 $\mu$m using a
2$\farcs$5 aperture for the NW and SE nuclei, respectively.
For the NW nucleus, our $L'$-band (3.8 $\mu$m) magnitude, measured with a
smaller aperture, is $\sim$0.7 mag brighter, which can naturally be
explained by a very red, steeply rising continuum flux from 3.4 $\mu$m
to 3.8 $\mu$m \citep{ima08}.}

\tablenotetext{C}{
\citet{zho93} estimated 7.5 mag at 3.4 $\mu$m using a 2$\farcs$5
aperture.
Our measurement at $L'$ with a smaller aperture is 0.3 mag brighter.
This galaxy shows a red, rising continuum flux from 3.4 $\mu$m 
to 3.8 $\mu$m \citep{ima10b} and possesses a Seyfert 1 nucleus, 
in which case, a flux time variation may be present on a $\sim$20-year time 
scale.}

\tablenotetext{D}{
\citet{zho93} estimated 10.9 mag at 3.4 $\mu$m using a
2$\farcs$5 aperture for the SW and NE nuclei combined.}

\tablenotetext{E}{
This nucleus is first spatially-resolved in our AO images, and is
different from the northern nucleus defined by \citet{kim02}.
We call as M (= middle nucleus). 
A third faint emission component may be present between the M and S
nuclei in the $L'$-band image in Figure 1.
Photometry of this component was not attempted because 
its signal largely overlaps with the signal of the brighter M nucleus 
in the chosen $\sim$0$\farcs$5 aperture.}

\tablenotetext{F}{
The nuclear separation in the $L'$-band is estimated to be
$\sim$1$\farcs$59$\pm$0$\farcs$05, where the uncertainties of 
both peak position coordinate determination and pixel scale 
(20.57$\pm$0.04 mas pixel$^{-1}$) are taken into account. 
This is comparable to the values shown by \citet{max07} 
(1$\farcs$542$\pm$0$\farcs$007 at $L'$), estimated in the 2--10 keV
X-ray (1$\farcs$5$\pm$0$\farcs$2) \citep{kom03,max07} and in the 
radio 1.4--5 GHz (6--21 cm) (1$\farcs$5--1$\farcs$52) \citep{bes01,gal04}.
}

\tablenotetext{G}{
\citet{zho93} estimated 10.1 mag and 10.8 mag at 3.4 $\mu$m using a
1$\farcs$7 aperture for the S and N nuclei, respectively.}

\tablecomments{
Col.(1): Object name.
Col.(2): $K$-band (2.2 $\mu$m) magnitude within the central
$\sim$0$\farcs$5-radius aperture (10 pixels for 52.77 mas
pixel$^{-1}$ data). 
For Mrk 231, the 0$\farcs$5 aperture is set from 25 pixels 
$\times$ 20.57 mas pixel$^{-1}$. 
The possible uncertainty for spatially unresolved compact emission 
is taken as $<$0.3 mag (see $\S$4.2).
Col.(3): $L'$-band (3.8 $\mu$m) magnitude within the central
$\sim$0$\farcs$5 radius aperture (25 pixels for 20.57 mas
pixel$^{-1}$ data). 
The possible uncertainty for spatially unresolved compact emission 
is taken as $<$0.2 mag (see $\S$4.2).
For undetected $L'$-band nuclei, the upper limits are derived 
at the $K$-band peak position based on 3$\sigma$ 
of sky fluctuation. 
Col.(4): $K-L'$ color magnitude within the 
central $\sim$0$\farcs$5-radius aperture.  
The possible uncertainty for spatially unresolved compact emission 
is regarded as $<$0.2 mag (see $\S$4.2).
Col.(5): AGN fraction in the $L'$-band in \%. 
Col.(6): AGN luminosity in 10$^{44}$ ergs s$^{-1}$ from AGN-origin
$\nu$F$_{\nu}$ value at $L'$ after removing the starburst contribution.
The possible uncertainty is $<$0.2 mag in the sense that the AGN 
luminosity may be underestimated. 
Col.(7): WISE 3.4 $\mu$m photometric magnitude with 6$\farcs$1 angular
resolution, for comparison \citep{wri10}.}

\end{deluxetable}

\begin{deluxetable}{lccc}
\tabletypesize{\scriptsize}
\tablecaption{Comparison of our $K$-band (2.2 $\mu$m) Photometry with 2.2 $\mu$m 
Photometry by \citet{sco00}  \label{tbl-4}}
\tablewidth{0pt}
\tablehead{
\colhead{Object} & \colhead{2.2 $\mu$m mag}  & \colhead{$K$-mag} &
\colhead{Difference} \\
\colhead{} & \colhead{\citet{sco00}} & \colhead{our data} &
\colhead{} \\
\colhead{(1)} & \colhead{(2)} & \colhead{(3)} & \colhead{(4)} 
}
\startdata
IRAS 05189$-$2524   & 10.33 & 10.9 & 0.57 \\
IRAS 08572$+$3915 N & 13.53 & 13.8 & 0.27 \\
IRAS 08572$+$3915 S & 15.76 & 16.5 & 0.74 \\
Mrk 273 S  & 13.29 & 13.6 & 0.31 \\
Mrk 273 N  & 12.91 & 13.1 & 0.19 \\
PKS 1345$+$12 W & 13.96 & 14.4 & 0.44 \\
PKS 1345$+$12 E & 15.31 & 15.6 & 0.29 \\
IRAS 14348$-$1447 S & 14.19 & 14.3 & 0.11 \\
IRAS 14348$-$1447 N & 14.92 & 15.0 & 0.08 \\
Arp 220 W  & 12.79 & 13.0 & 0.21 \\
Arp 220 E  & 13.13 & 13.2 & 0.07 \\
UGC 5101   & 11.77 & 12.0 & 0.23 \\
NGC 6240 S & 11.36 & 11.8 & 0.44 \\
NGC 6240 N & 12.72 & 13.0 & 0.28 \\
\enddata

\tablecomments{
Col.(1): Object name.
Col.(2): Nuclear 1$\farcs$1-diameter magnitude at 2.2 $\mu$m derived by
\citet{sco00}. 
Col.(3): Nuclear 0$\farcs$5-radius (1$\farcs$0 diameter) $K$-band (2.2
$\mu$m) magnitude based on our data.
Col.(4): Difference in the photometry between \citet{sco00} and our
data.
For all sources, our data tend to show slightly fainter photometric 
magnitudes than computed by
\citet{sco00} due to the smaller aperture size and the
signal spread into the seeing-sized halo outside the AO core.}

\end{deluxetable}

\begin{deluxetable}{lccccc}
\tabletypesize{\scriptsize}
\tablecaption{Luminosity Ratio and Nuclear Separation in 
Seeing-based Multiple Nuclei (U)LIRGs 
\label{tbl-5}}
\tablewidth{0pt}
\tablehead{
\colhead{Object} & \colhead{K(stellar) ratio}  & \colhead{L$'$ ratio} &
\colhead{L$' $(AGN) ratio} & \colhead{Separation} &
\colhead{Separation} \\ 
\colhead{} & \colhead{} & \colhead{} & \colhead{} & \colhead{(arcsec)} &
\colhead{(kpc)} \\
\colhead{(1)} & \colhead{(2)} & \colhead{(3)} & \colhead{(4)} 
& \colhead{(5)} & \colhead{(6)} 
}
\startdata
IRAS 00091$-$0738 N, S & 1.35 (15.34, 15.67) & $<$0.58 & $<$0.76 & 1.2 & 2.6\\
IRAS 08572$+$3915 NW, SE & 6.37 (13.66, 15.67) & $>$158 & $>$176 &5.5&6.1\\
IRAS 12127$-$1412 NE, SW & 75.9 (13.63, 18.33) & $>$63 & $>$63 &10.4&24.2 \\ 
IRAS 13335$-$2612 S, N &  1.20 (15.20, 15.40) \tablenotemark{A} & 1.7 & 3.1&1.6&3.5 \\
IRAS 13443$+$0802 NE, E  & 4.02 (14.52, 16.03) & 1.6 & 0 &4.8&11.3 \\
IRAS 13443$+$0802 E, SW & 1.80 (16.03, 16.67) & $>$1.9 & $>$1.8 & 13.4&31.7 \\
PKS 1345+12 W, E & 2.31 (14.54, 15.45) \tablenotemark{A} & $>$19 & $>$38&2.0&4.4\\ 
IRAS 14348$-$1447 SW, NE & 1.64 (13.98, 14.52) \tablenotemark{A} & 1.3 & 0.78&3.4&5.2 \\ 
IRAS 16468$+$5200 E, W & 1.19 (16.64, 16.83) & $>$1.7 & $>$2.4&3.2&8.3 \\
IRAS 16487$+$5447 SW, NE & 1.53 (14.89, 15.35) & $>$1.7 & $>$2.9&3.1&5.9  \\ 
IRAS 21208$-$0519 NE, SW & 2.15 (15.16, 15.99) & $>$3.9 & $\infty$&6.6&15.2 \\
IRAS 23234$+$0946 NW, SE & 3.70 (15.32, 16.74) \tablenotemark{A} & $>$1.7 & $>$1.3&3.5&8.0 \\
IRAS 23327$+$2913 S, N  &  2.54 (14.29, 15.30) & $>$4.7 & $>$10&12.5&24.1 \\
IRAS 23389$+$0300 N, S  &  6.49 (15.08, 17.11) & $>$2.7 & $>$1.3&2.3&5.7 \\
IRAS 23498$+$2423 NW, SE & 32.8 (14.08, 17.87) & $>$39 & $>$38&4.1&13.9 \\
Mrk 463 E, W & 1.74 (10.49, 11.09) & 100 & 324&3.9&3.8 \\
Mrk 739 E, W & 3.53 (11.37, 12.74) & 28 & $\infty$&6.1&3.6 \\
IRAS 20210$+$1121 S, N & 2.25 (12.76, 13.64) & 15 & 66 &12.3&13.2\\
\enddata

\tablenotetext{A}{Based on the velocity dispersion obtained through
near-infrared spectroscopy, SMBH mass ratios are computed to be 
2.45, 1.72, 0.61, and 3.29 for IRAS 13335$-$2612 (S, N), 
PKS 1345+12 (W, E), IRAS 14348$-$1447 (SW, NE), and IRAS 23234$+$0946
(NW, SE), respectively \citep{das06a,das06b}.}


\tablecomments{
Col.(1): Object name.
Col.(2): $K$-band flux ratio, measured with the same aperture size
between two nuclei, as an approximation of stellar emission luminosity ratio. 
$K$-band photometric values in individual nuclei are shown in
parentheses; the first value is for the nucleus shown first in
column 1.   
The 4-kpc aperture $K$-band photometry by \citet{kim02} is basically adopted, 
but for the last three sources (Mrk 463, Mrk 739, and IRAS 20210$+$1121), 
photometry from the 2MASS point source catalog (4$''$ aperture) is employed. 
Col.(3): Nuclear $L'$-band (3.8 $\mu$m) luminosity ratio based on our
photometry (Table 3, column 3).
Col.(4): Nuclear AGN-origin $L'$-band (3.8 $\mu$m) luminosity ratio 
after the subtraction of stellar emission component (Table 3, column 6).
In Cols. (2)--(4), the luminosity at the nucleus listed first in column
1 is divided by that listed second in column 1.
Col.(5): Apparent nuclear separation in arcsec calculated from our Subaru
LGS/NGS-AO $K$-band images. 
Col.(6): Apparent nuclear separation in kpc calculated using H$_{0}$ $=$
71 km s$^{-1}$ Mpc$^{-1}$, $\Omega_{\rm M}$ = 0.27, and 
$\Omega_{\rm \Lambda}$ = 0.73 \citep{kom09}. 
}

\end{deluxetable}

\clearpage

\begin{figure}
\includegraphics[angle=0,scale=.6]{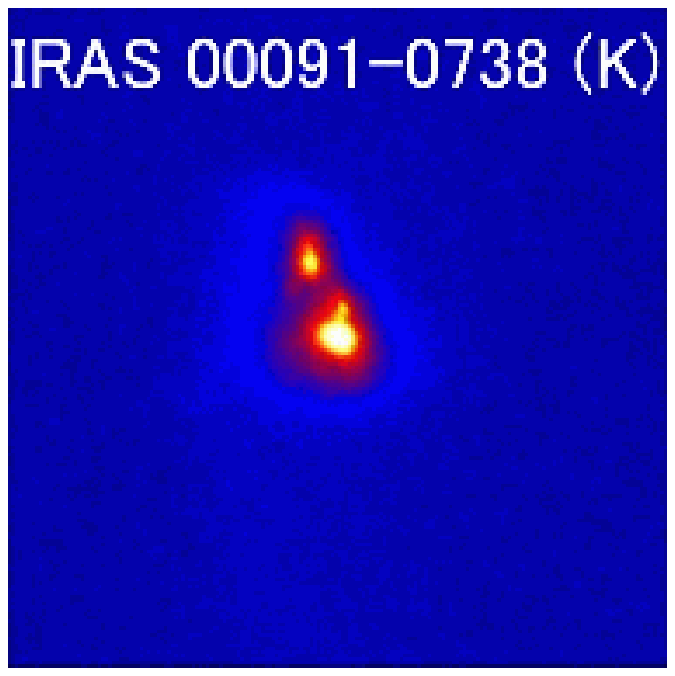} 
\includegraphics[angle=0,scale=.47]{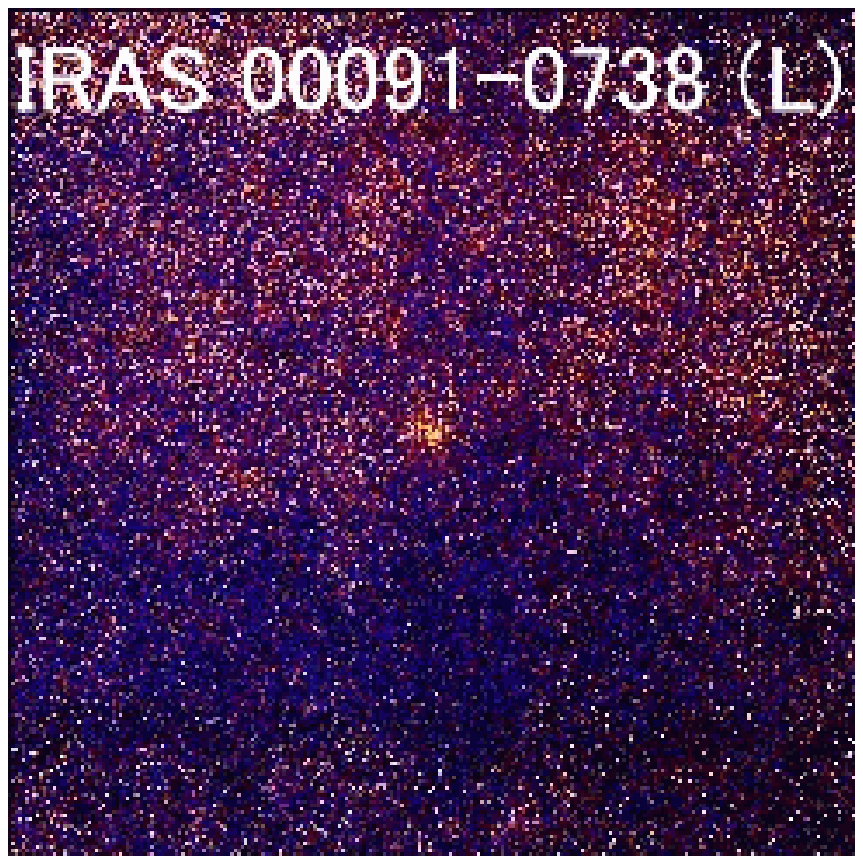} 
\includegraphics[angle=0,scale=.6]{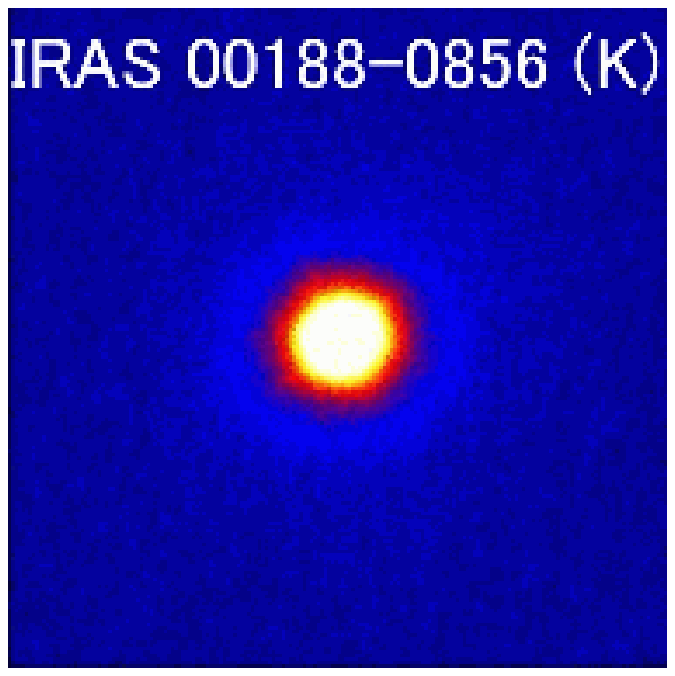} 
\includegraphics[angle=0,scale=.47]{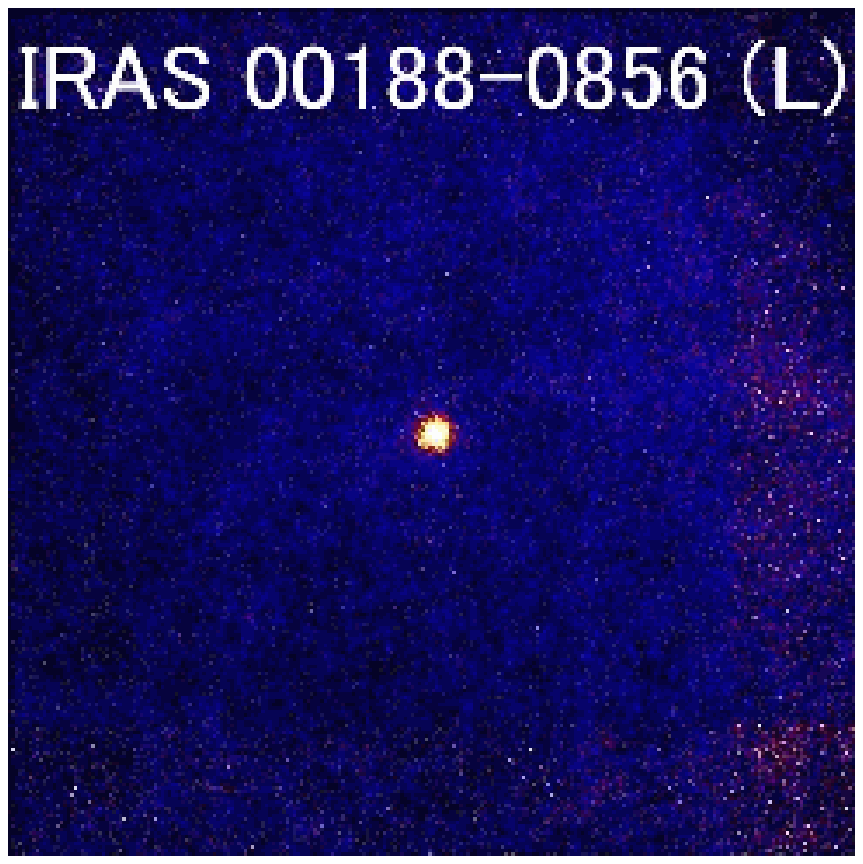} \\
\includegraphics[angle=0,scale=.6]{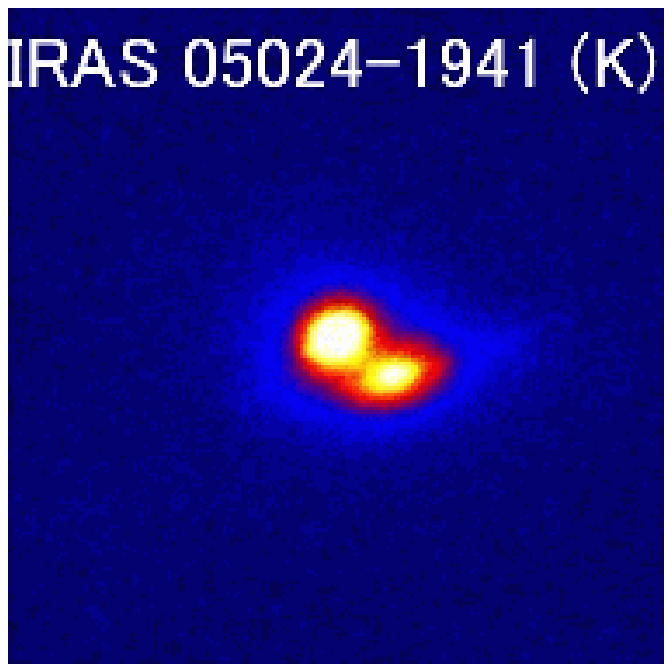} 
\includegraphics[angle=0,scale=.47]{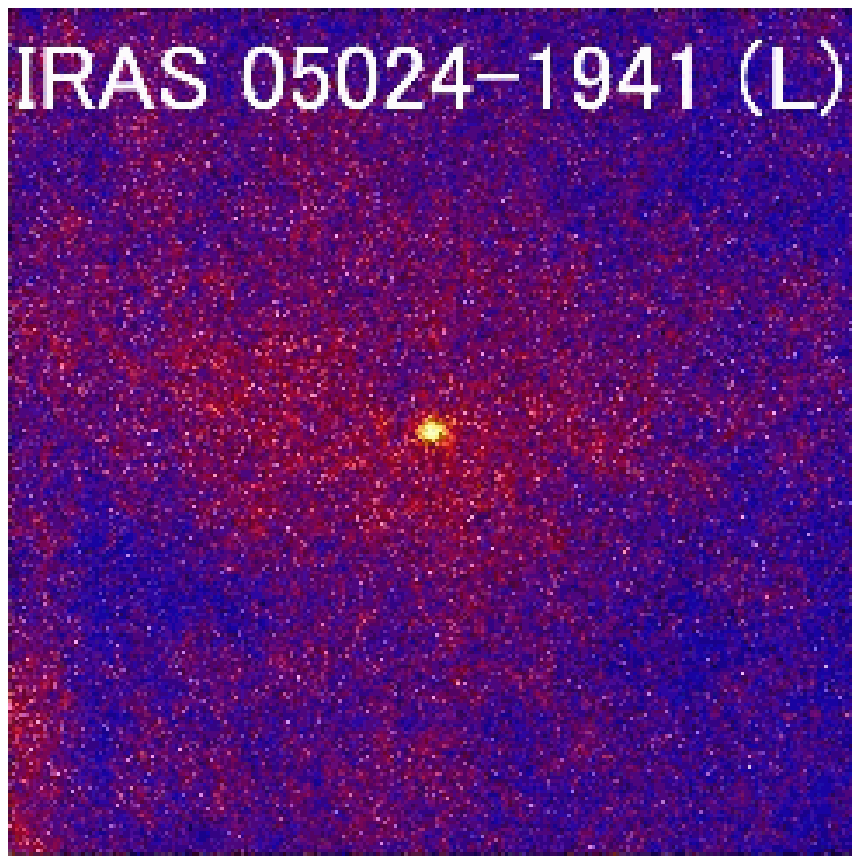} 
\includegraphics[angle=0,scale=.6]{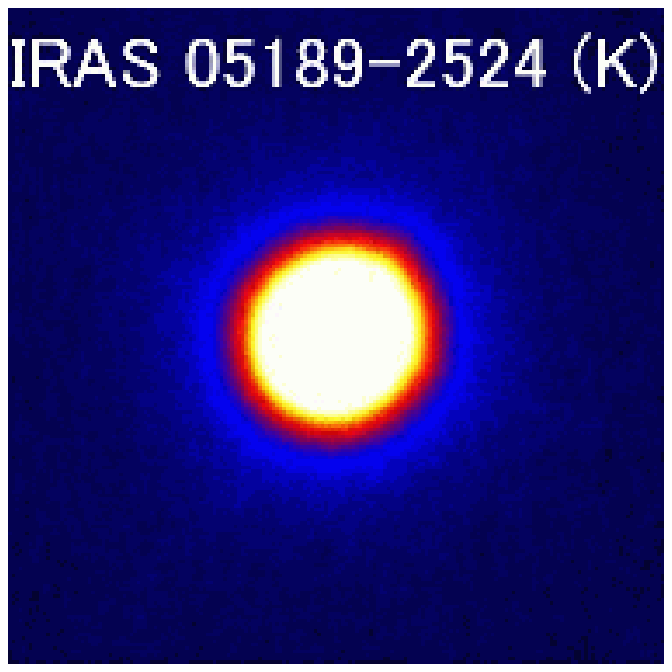} 
\includegraphics[angle=0,scale=.47]{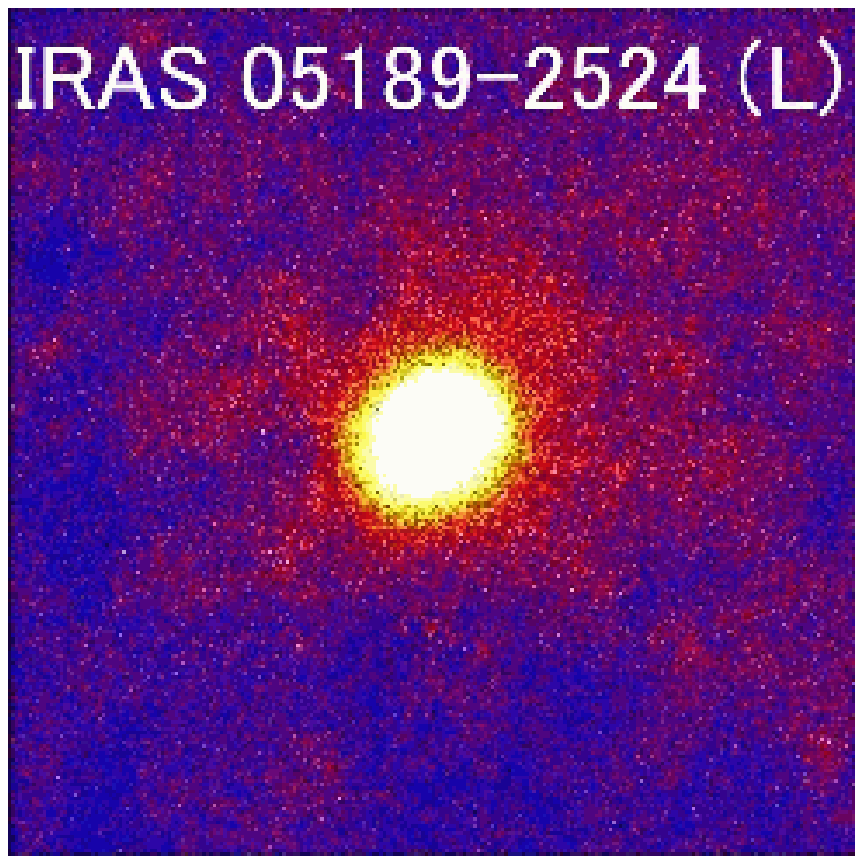} \\
\includegraphics[angle=0,scale=.6]{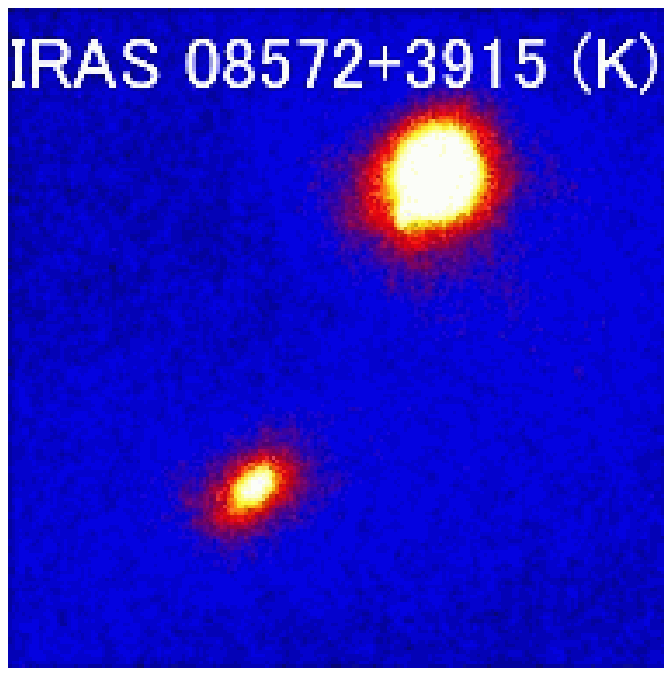} 
\includegraphics[angle=0,scale=.47]{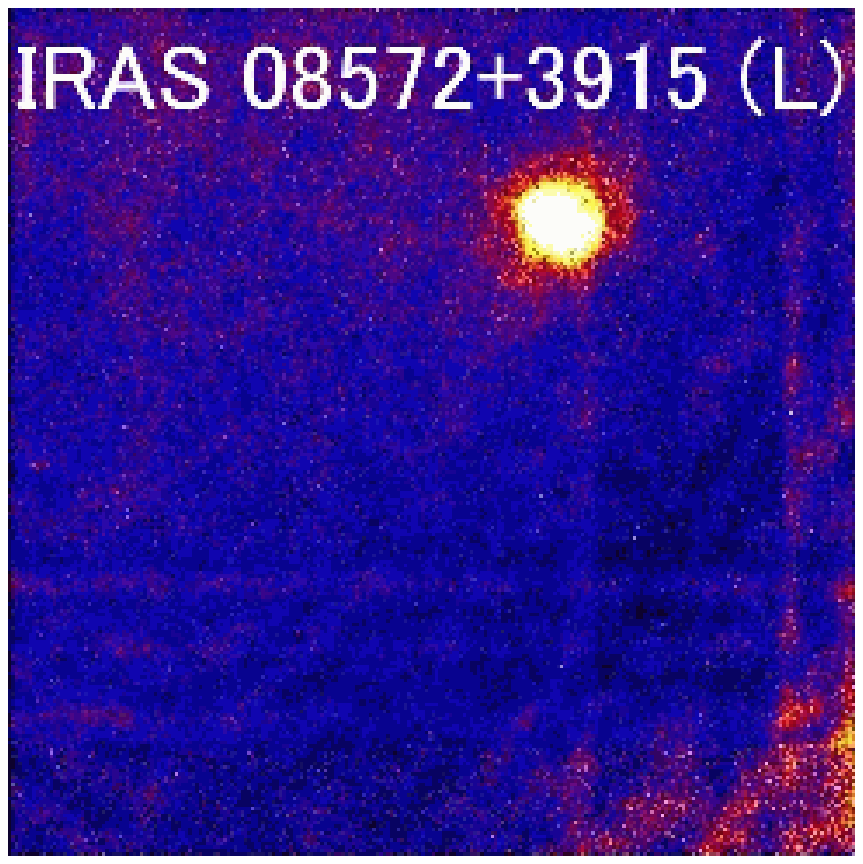} 
\includegraphics[angle=0,scale=.6]{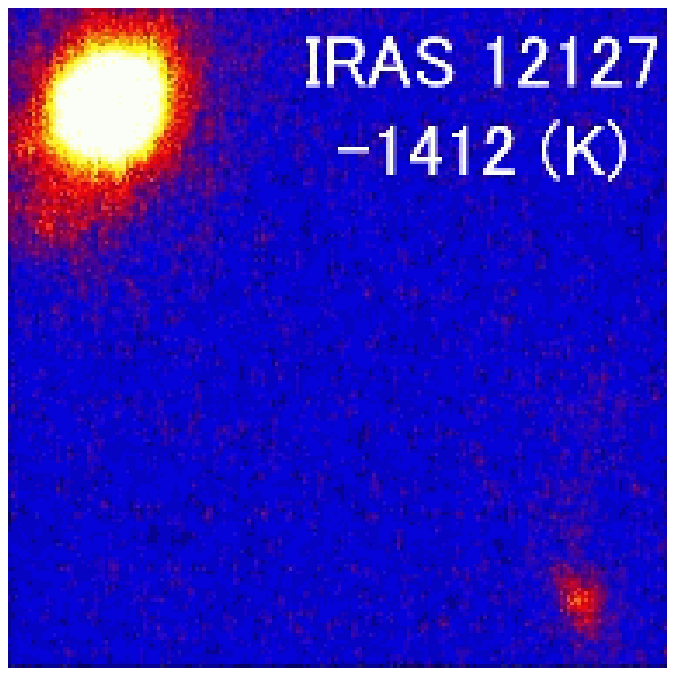} 
\includegraphics[angle=0,scale=.47]{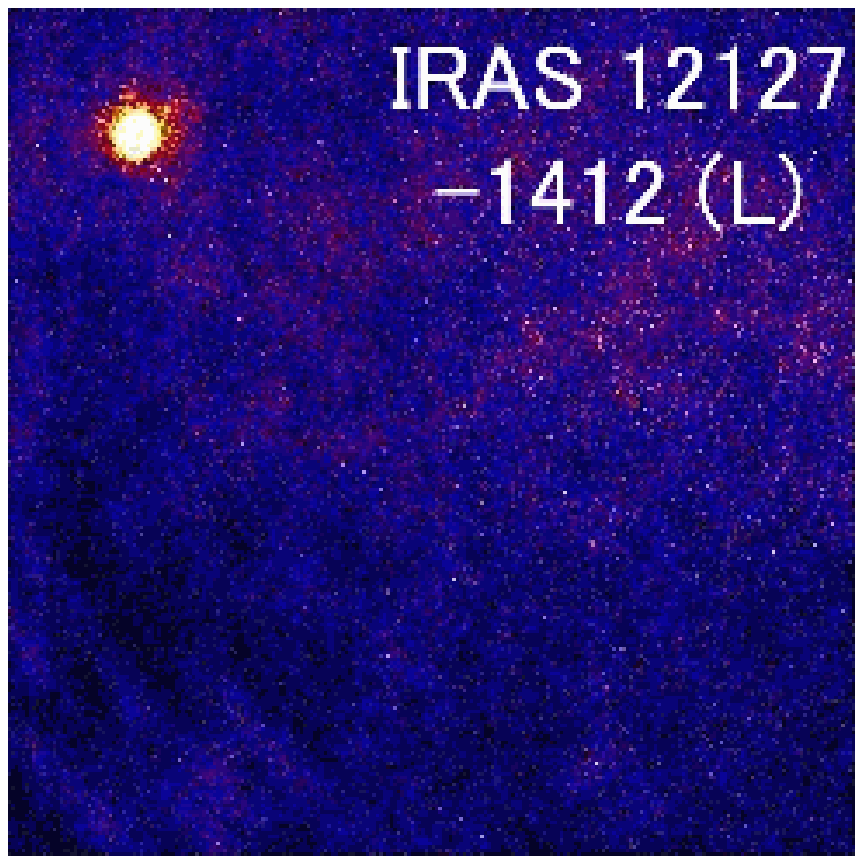} \\
\includegraphics[angle=0,scale=.47]{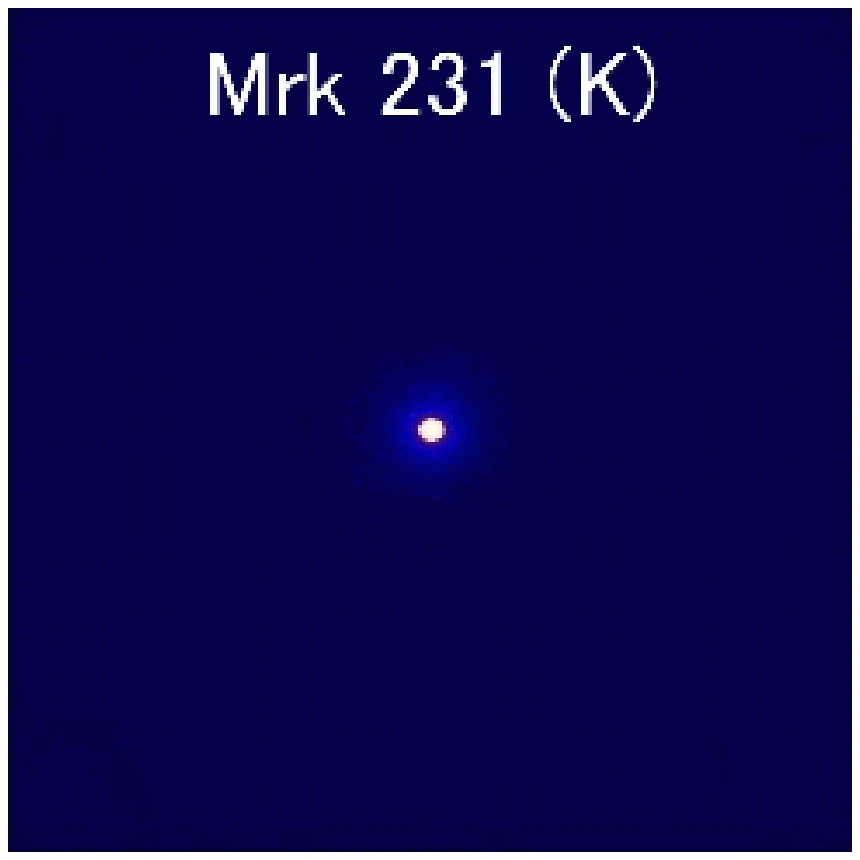} 
\includegraphics[angle=0,scale=.47]{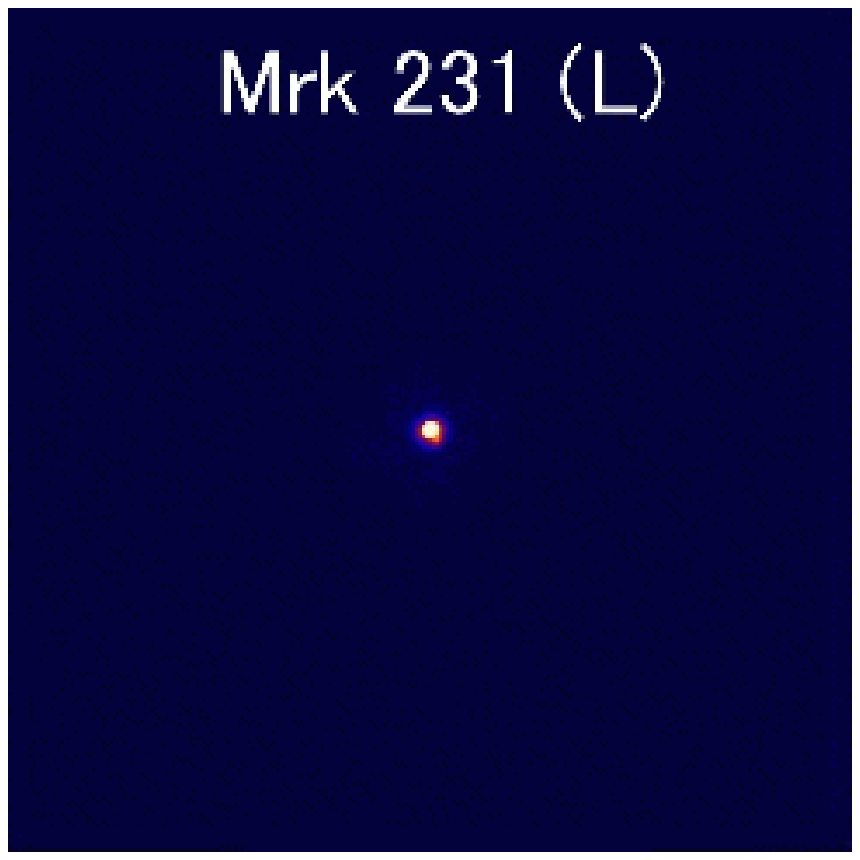} 
\includegraphics[angle=0,scale=.6]{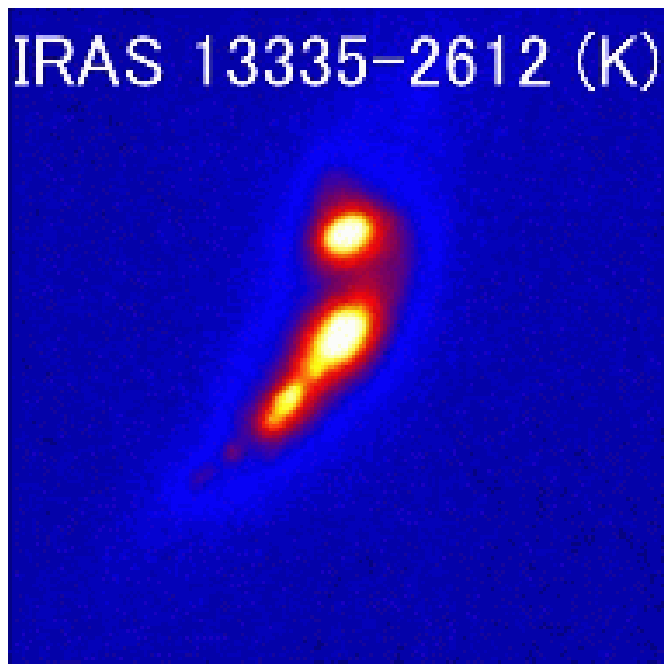} 
\includegraphics[angle=0,scale=.47]{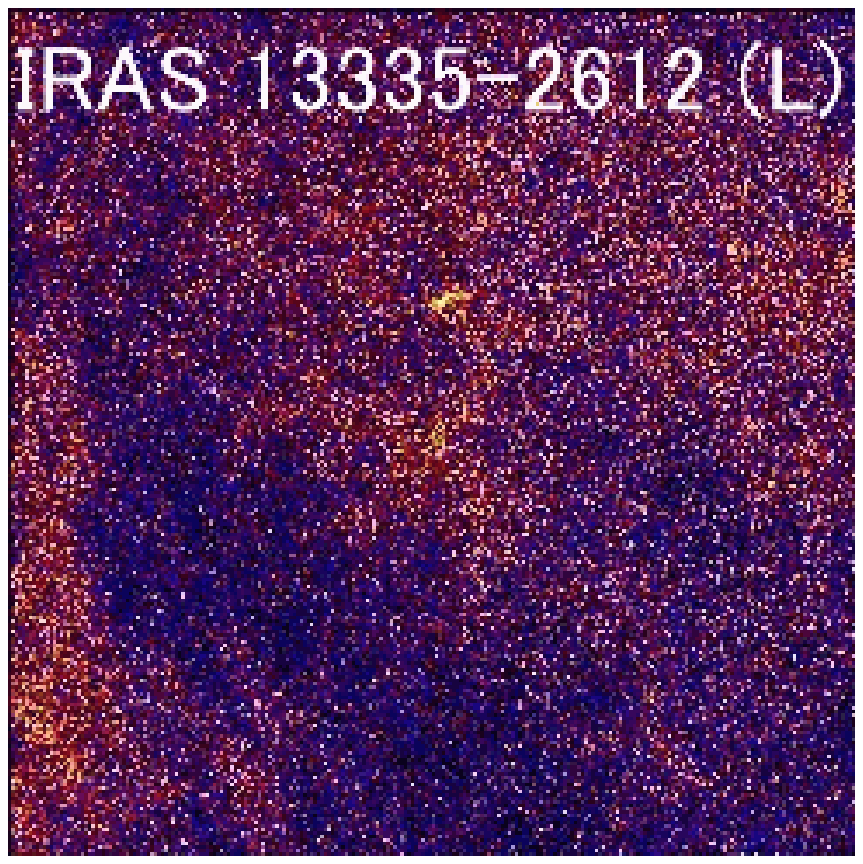} \\
\includegraphics[angle=0,scale=.6]{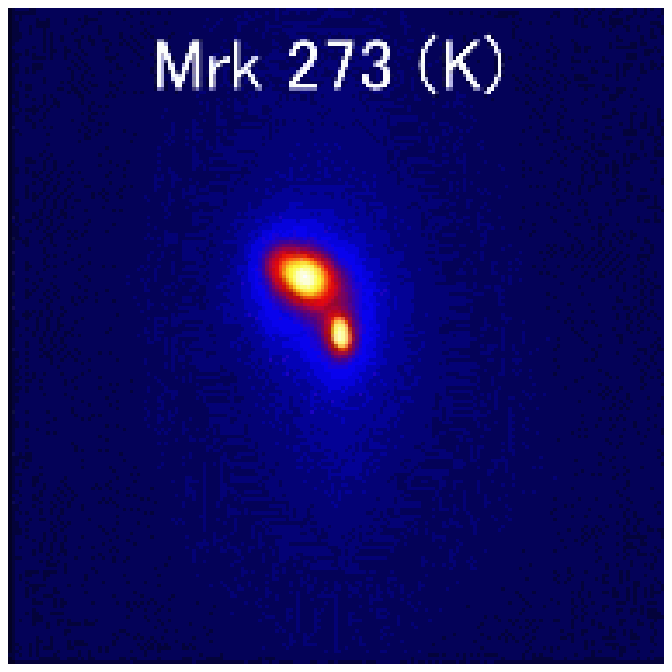} 
\includegraphics[angle=0,scale=.47]{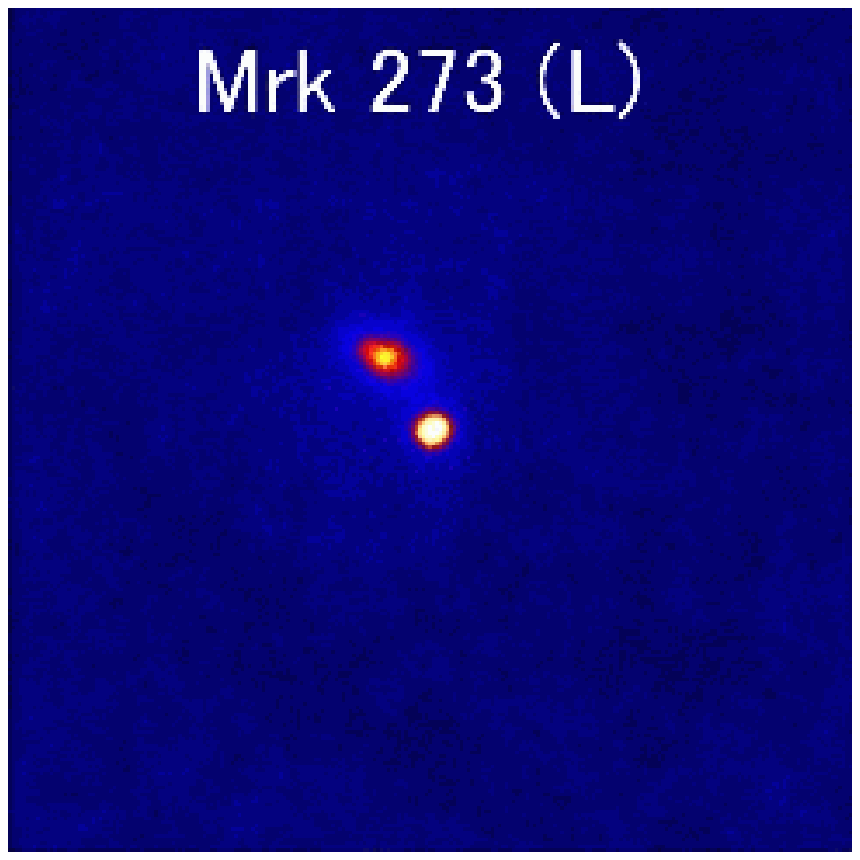} 
\includegraphics[angle=0,scale=.6]{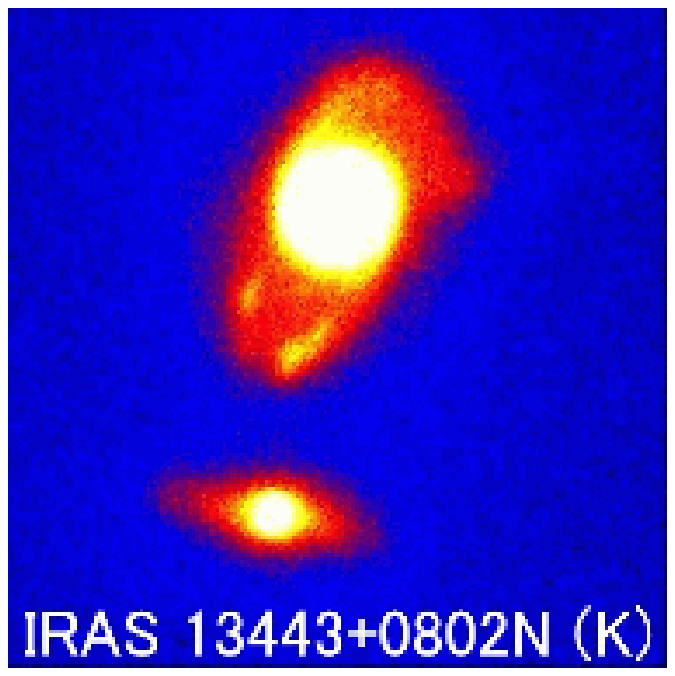} 
\includegraphics[angle=0,scale=.47]{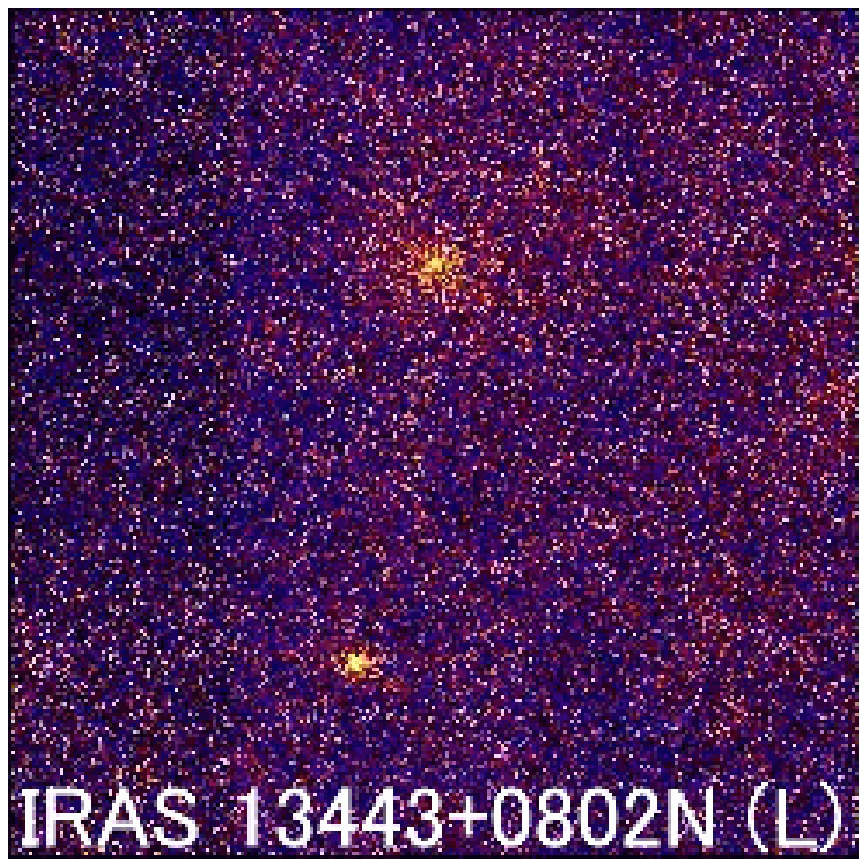} \\
\end{figure}

\clearpage
\begin{figure}
\includegraphics[angle=0,scale=.6]{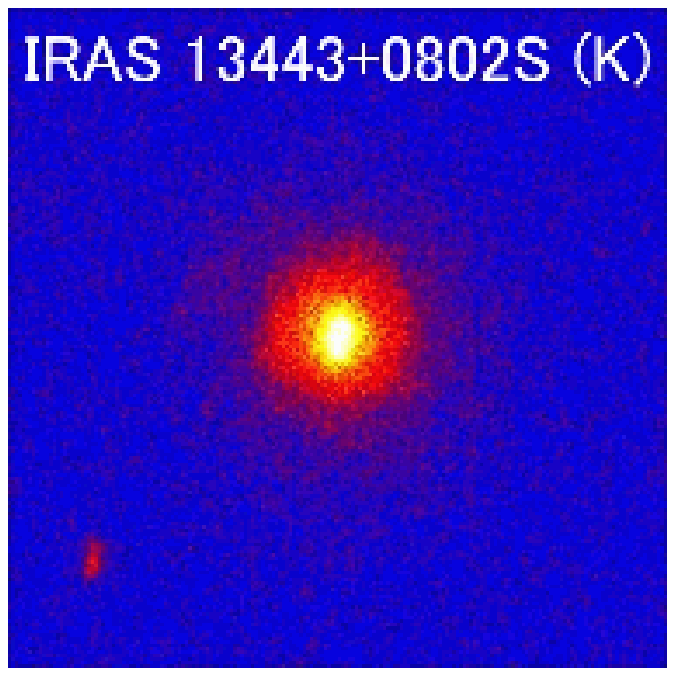} 
\includegraphics[angle=0,scale=.47]{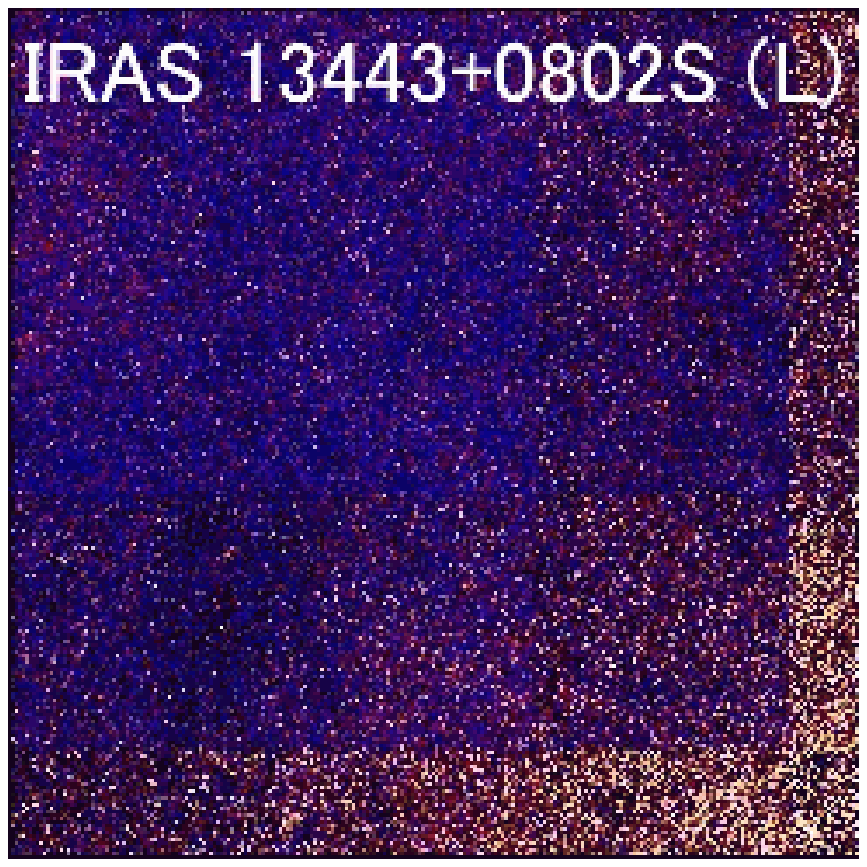} 
\includegraphics[angle=0,scale=.6]{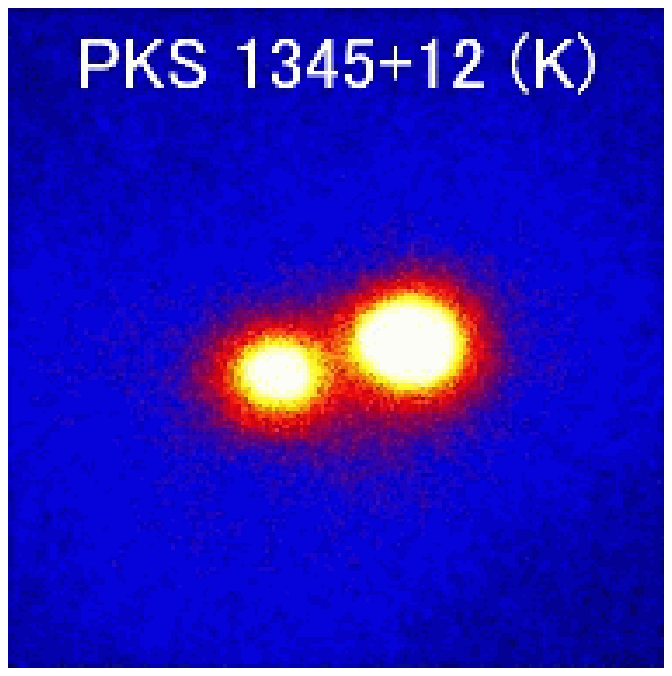} 
\includegraphics[angle=0,scale=.47]{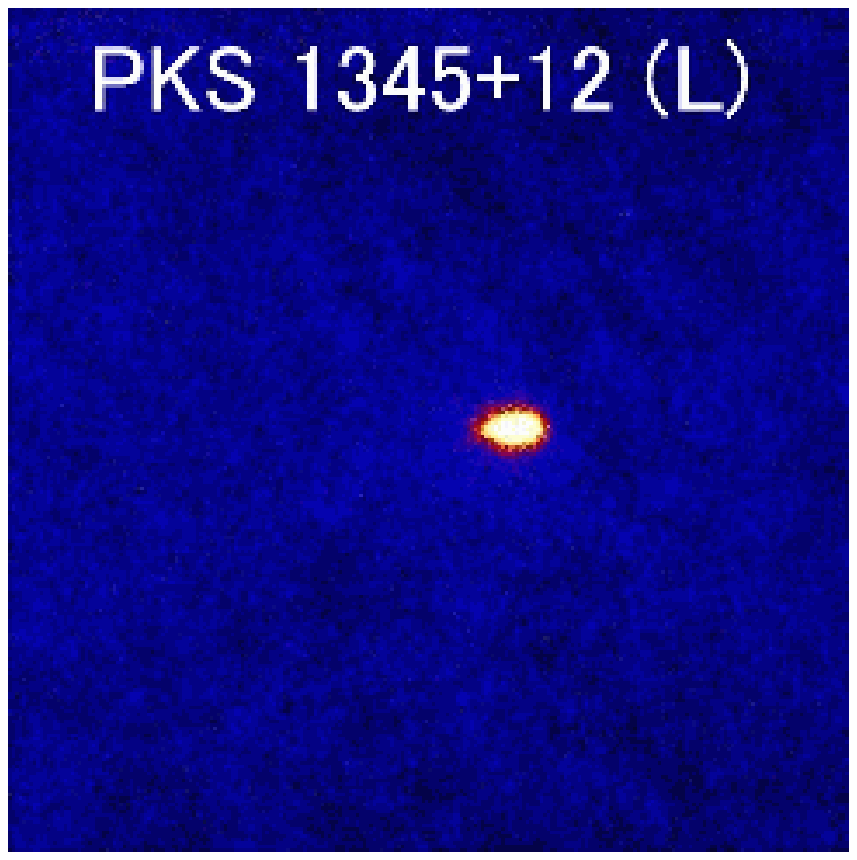} \\ 
\includegraphics[angle=0,scale=.6]{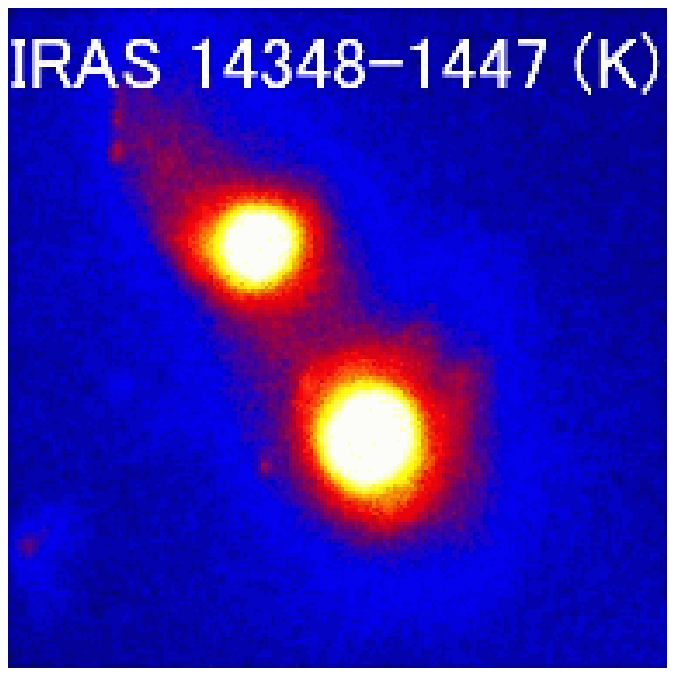} 
\includegraphics[angle=0,scale=.47]{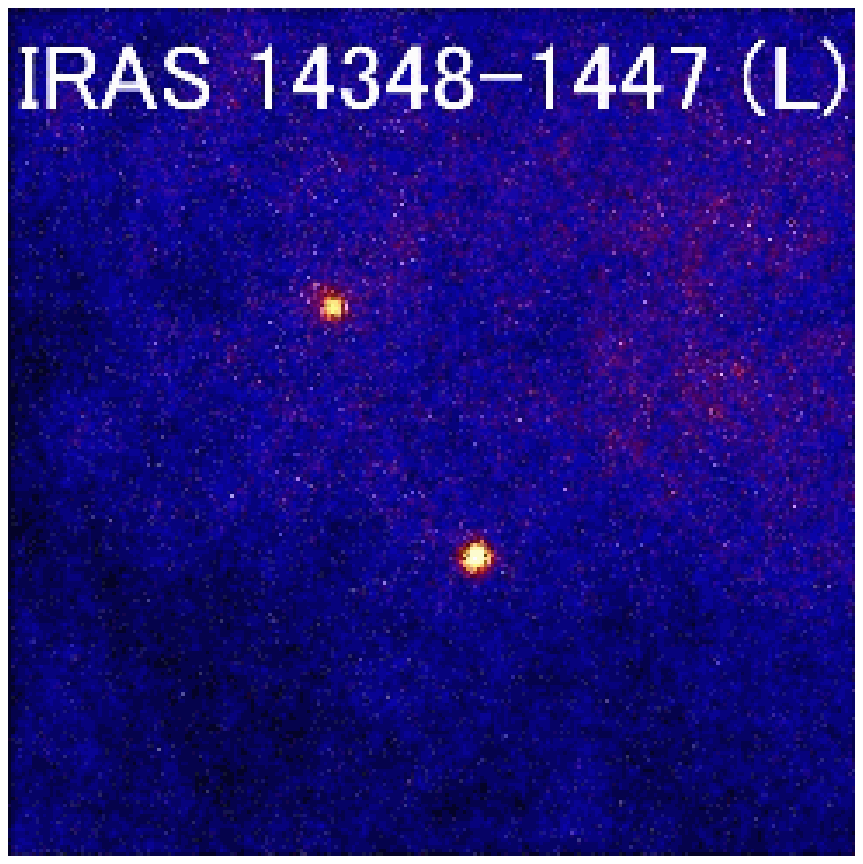} 
\includegraphics[angle=0,scale=.6]{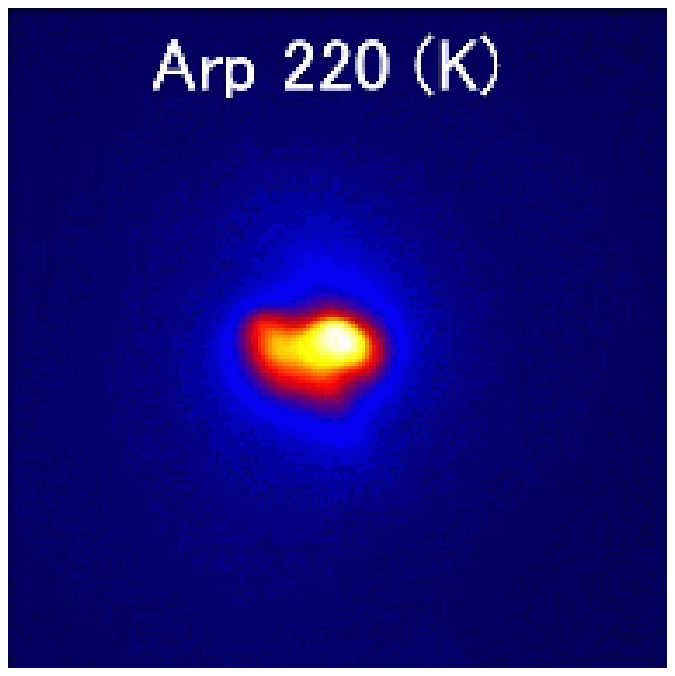} 
\includegraphics[angle=0,scale=.47]{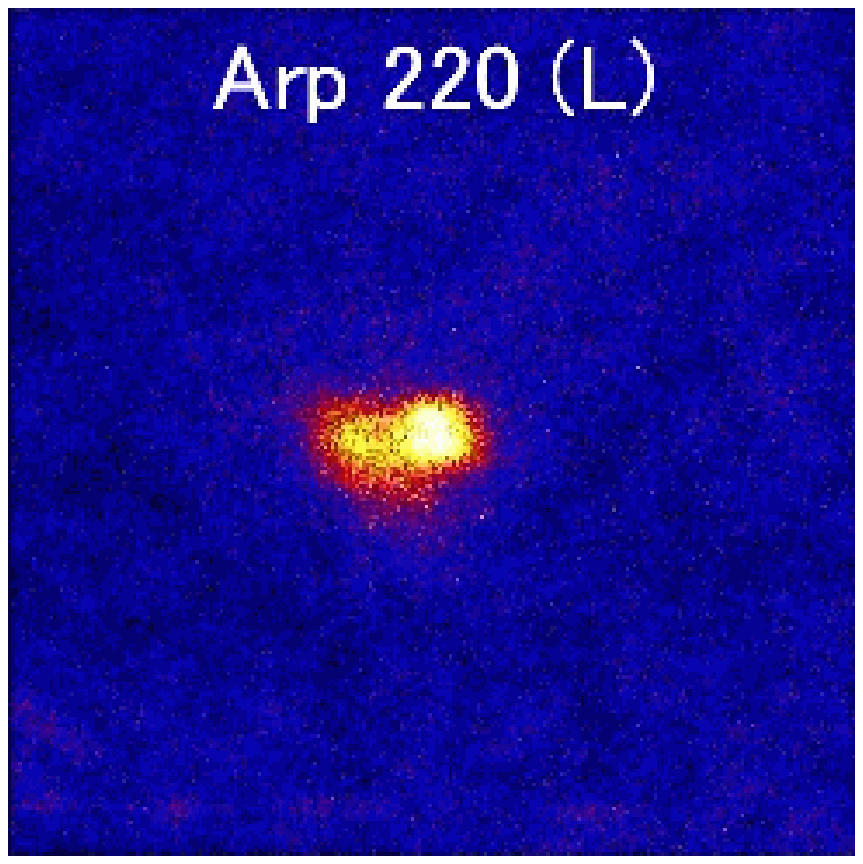} \\
\includegraphics[angle=0,scale=.6]{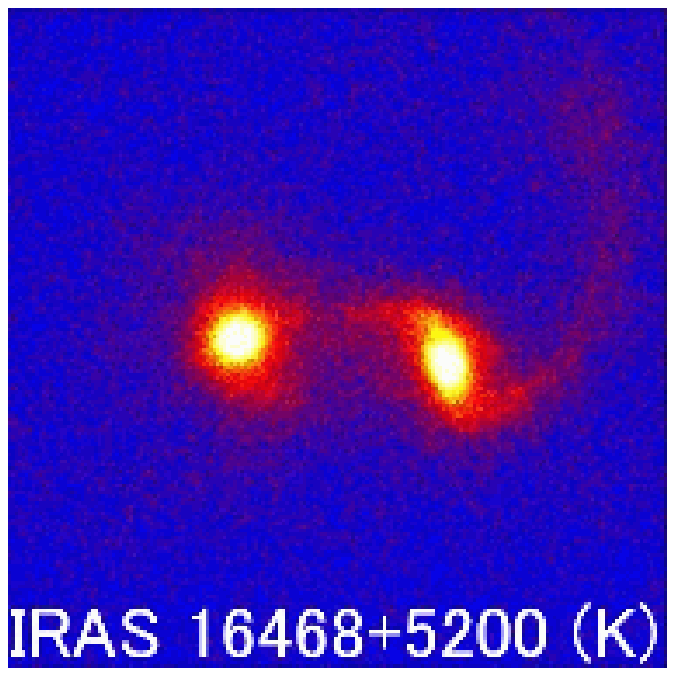} 
\includegraphics[angle=0,scale=.47]{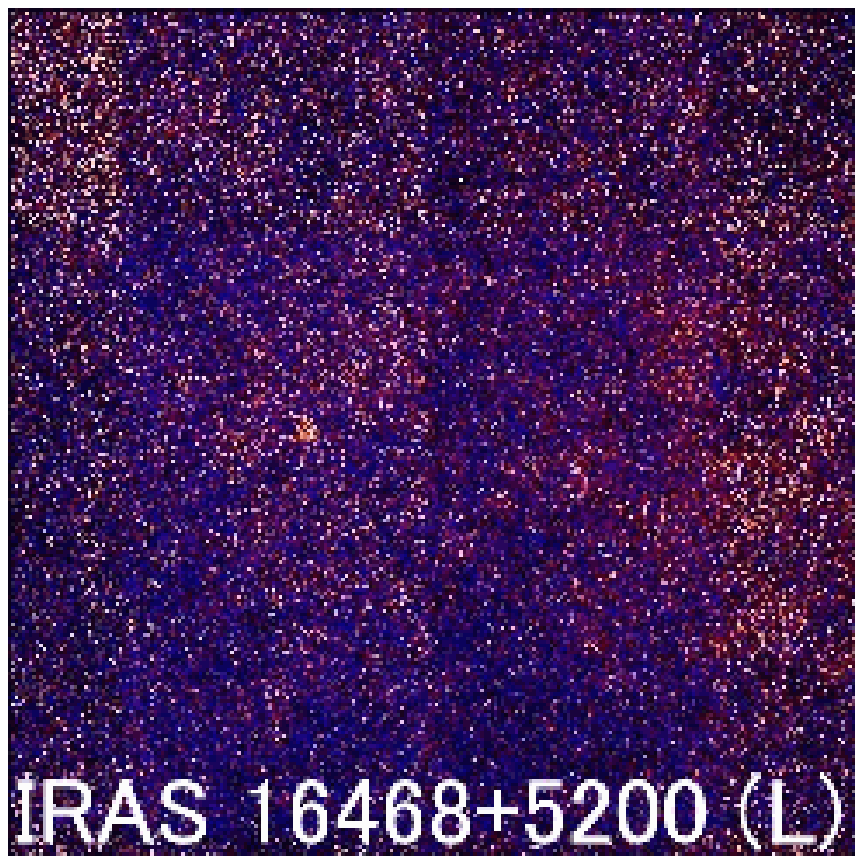} 
\includegraphics[angle=0,scale=.6]{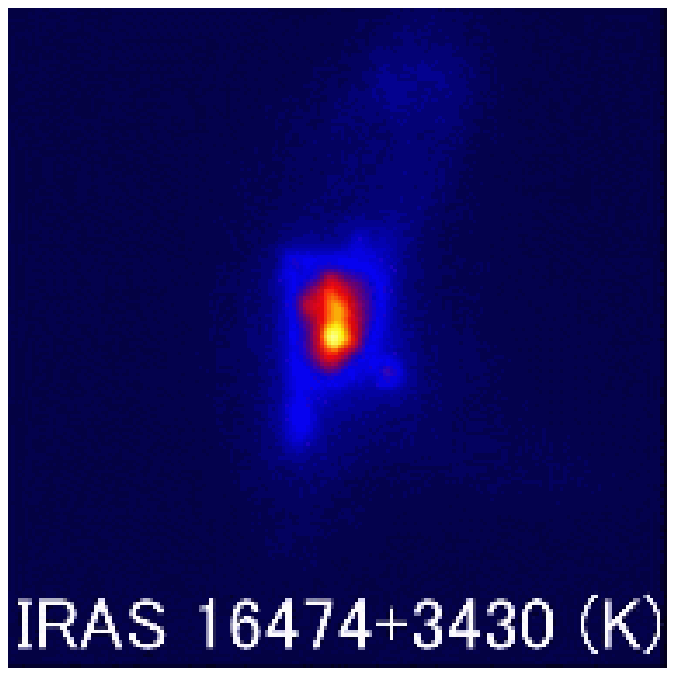} 
\includegraphics[angle=0,scale=.47]{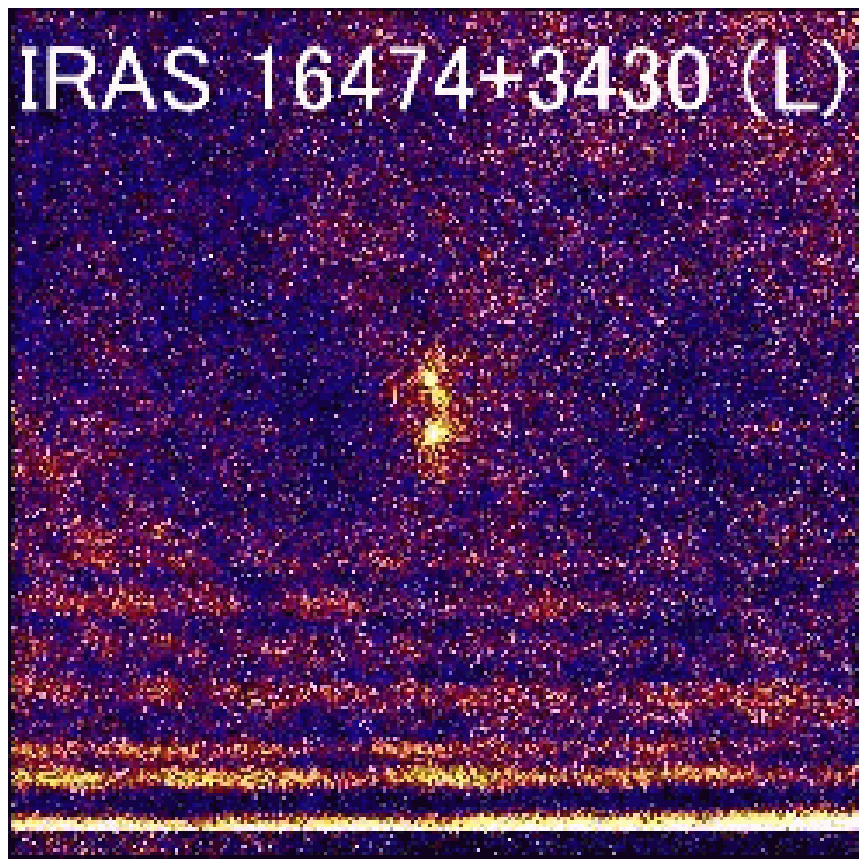} \\ 
\includegraphics[angle=0,scale=.6]{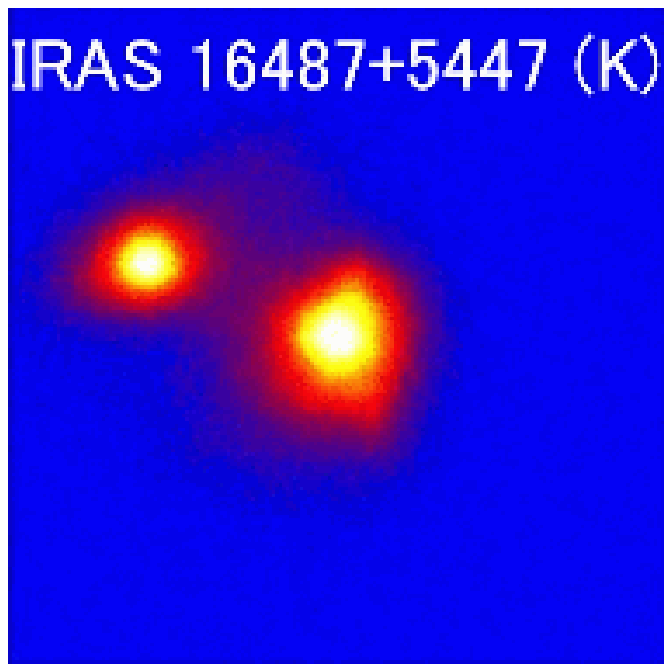} 
\includegraphics[angle=0,scale=.47]{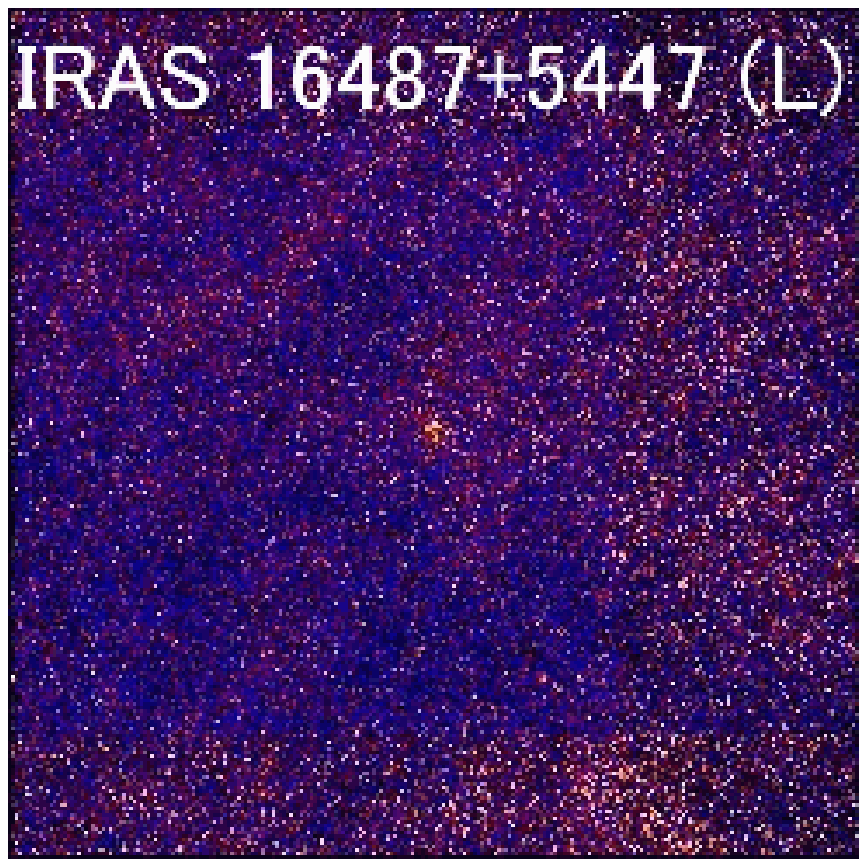} 
\includegraphics[angle=0,scale=.6]{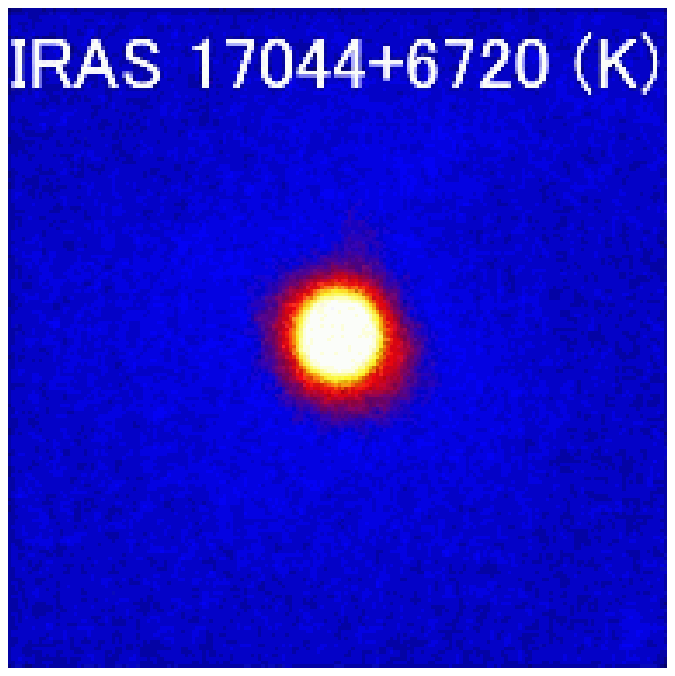} 
\includegraphics[angle=0,scale=.47]{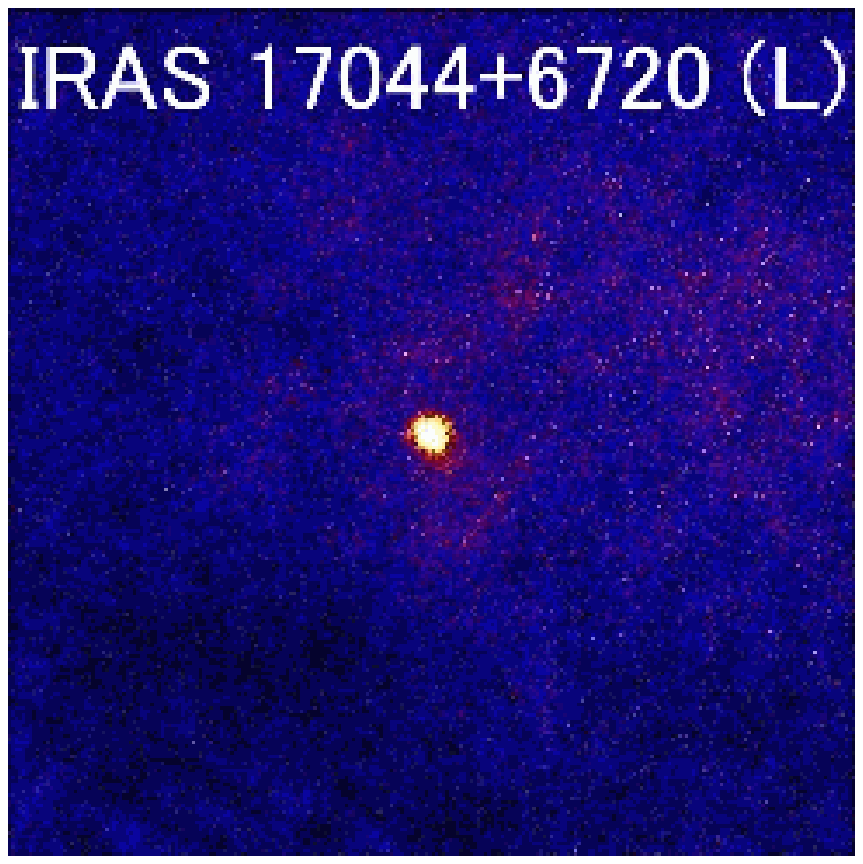} \\
\includegraphics[angle=0,scale=.6]{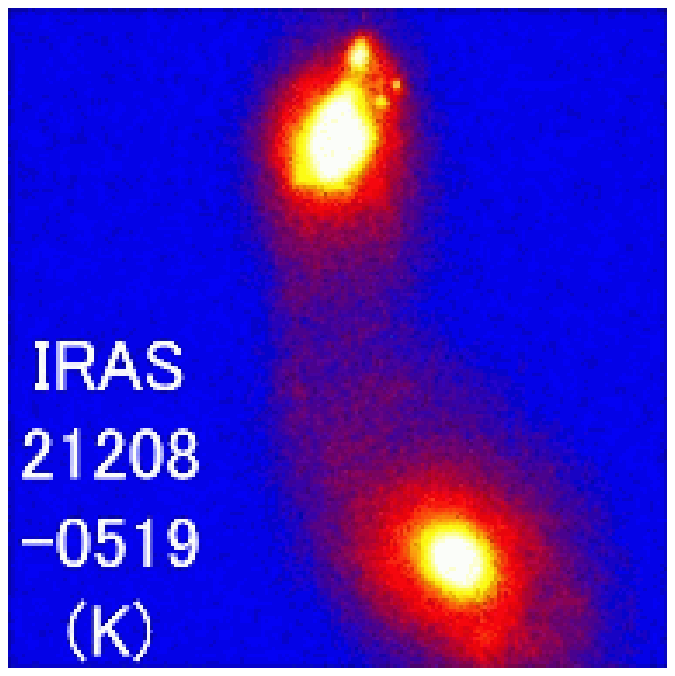} 
\includegraphics[angle=0,scale=.47]{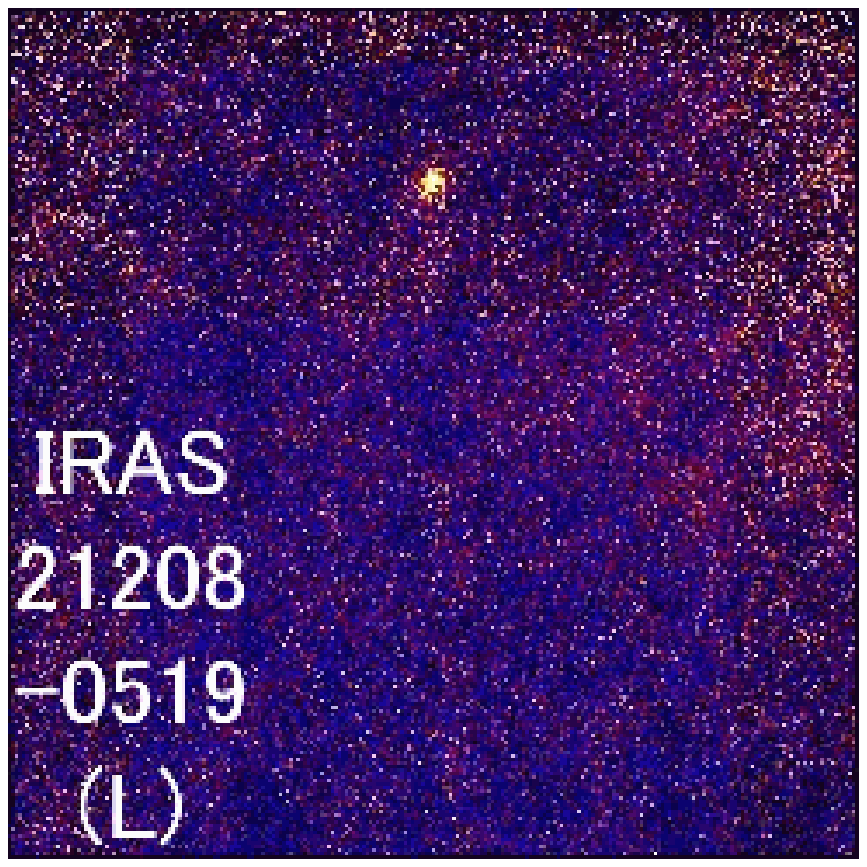} 
\includegraphics[angle=0,scale=.6]{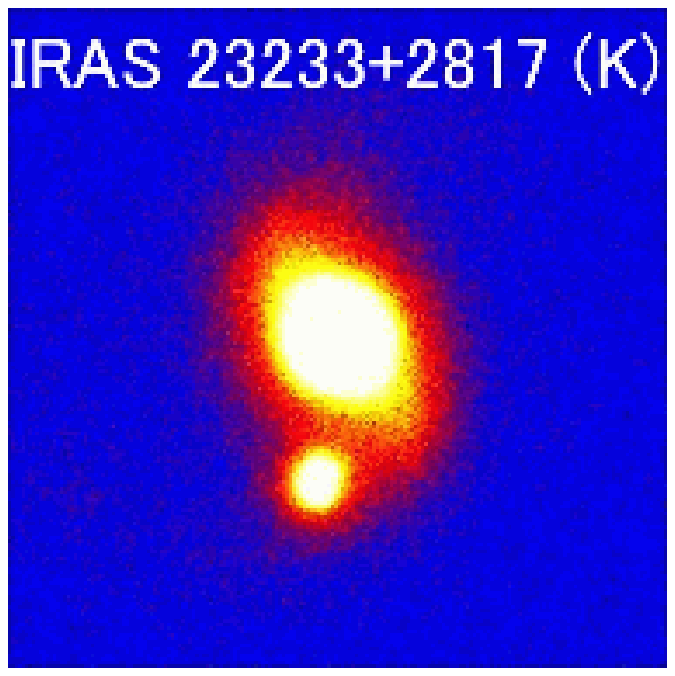} 
\includegraphics[angle=0,scale=.47]{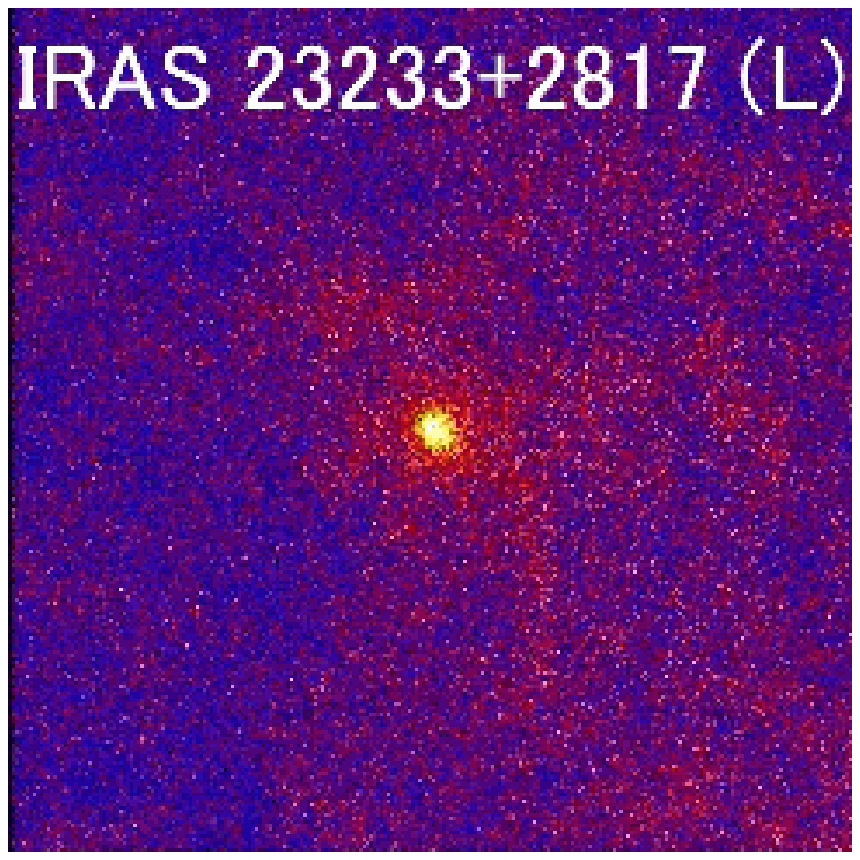} \\
\end{figure}

\clearpage
\begin{figure}
\includegraphics[angle=0,scale=.6]{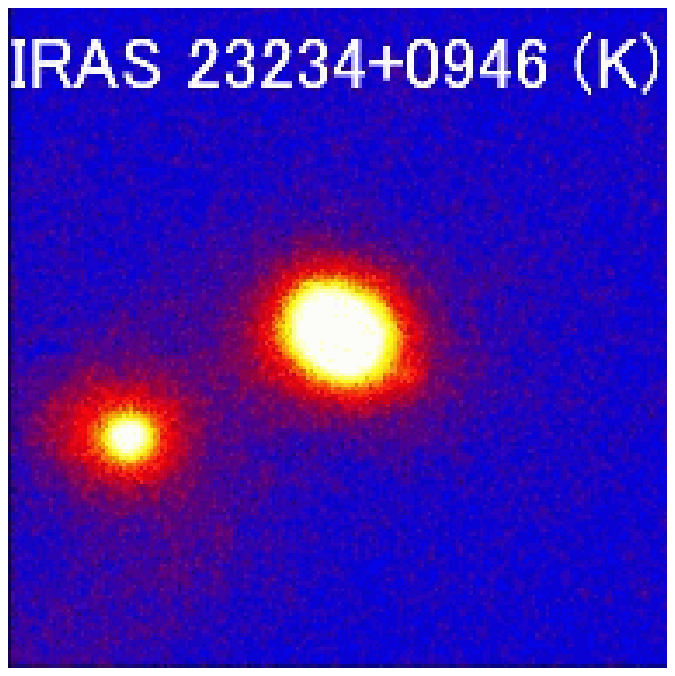} 
\includegraphics[angle=0,scale=.47]{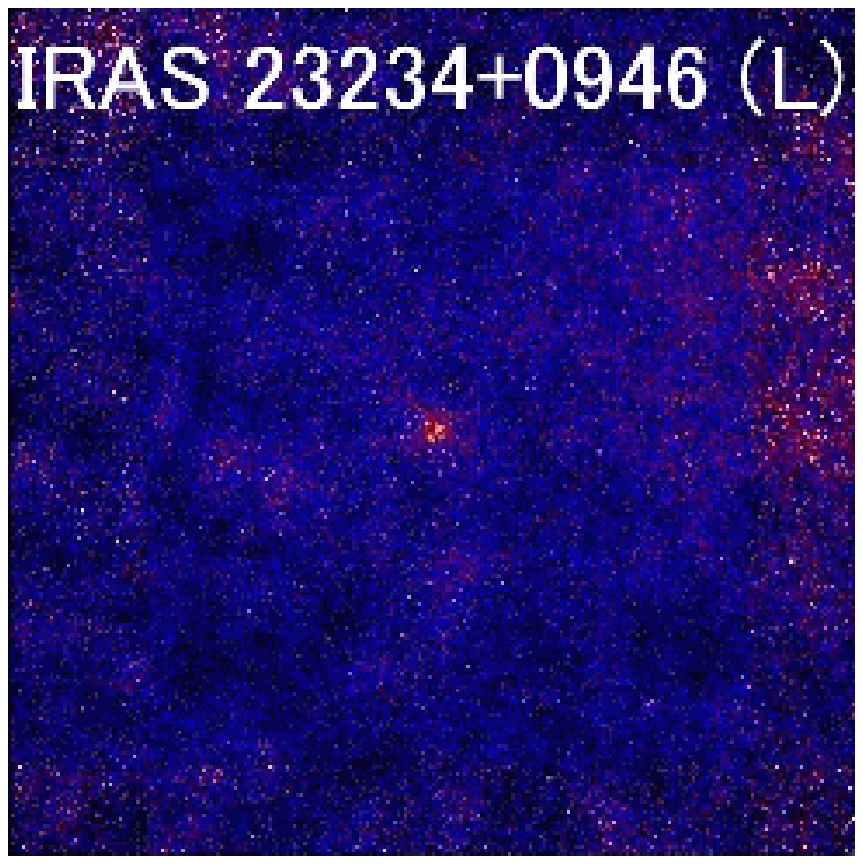}
\includegraphics[angle=0,scale=.6]{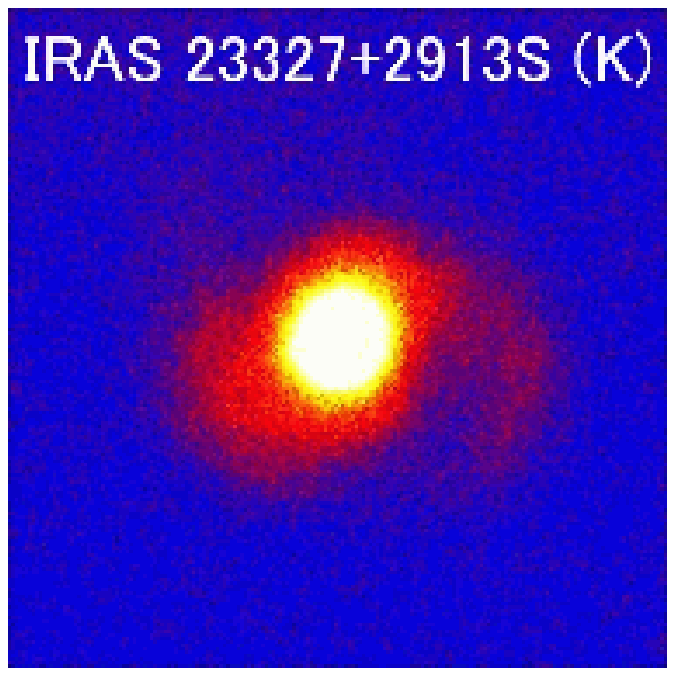} 
\includegraphics[angle=0,scale=.47]{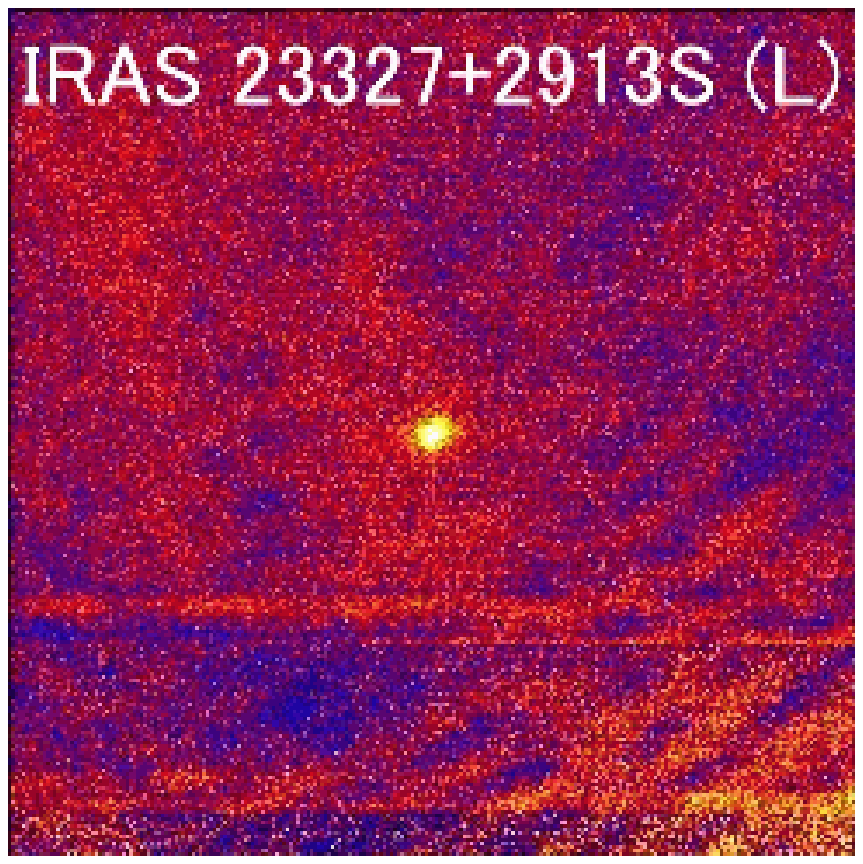} \\
\includegraphics[angle=0,scale=.6]{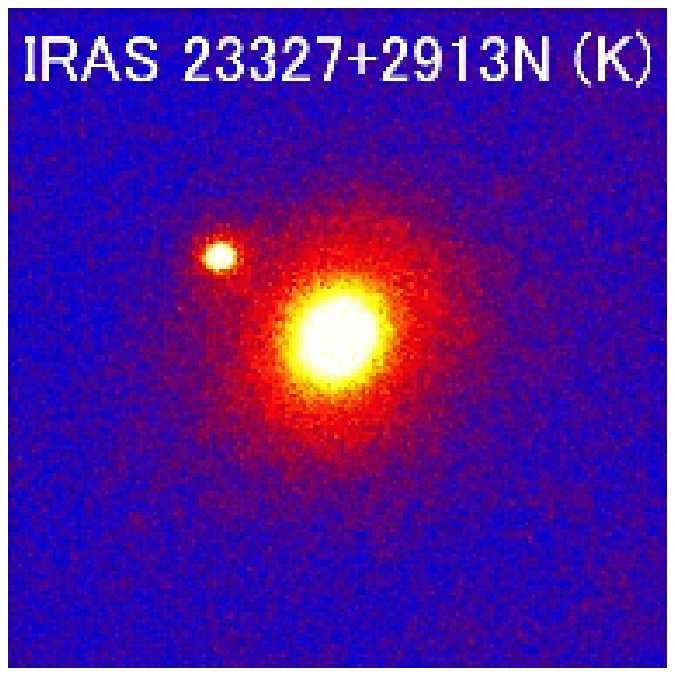} 
\includegraphics[angle=0,scale=.47]{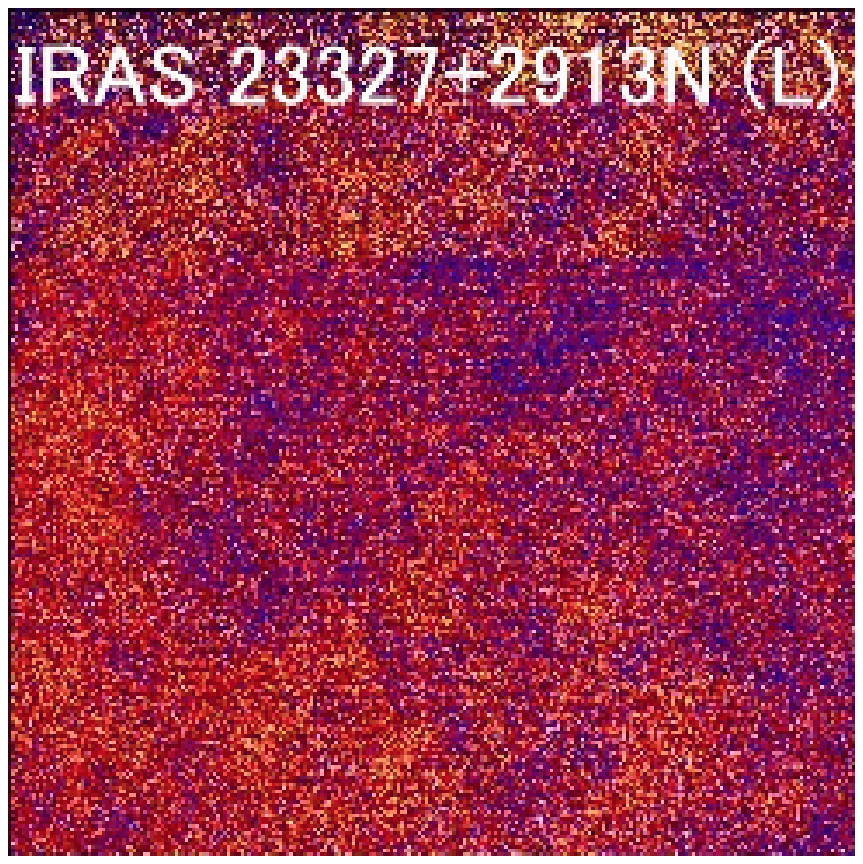} 
\includegraphics[angle=0,scale=.6]{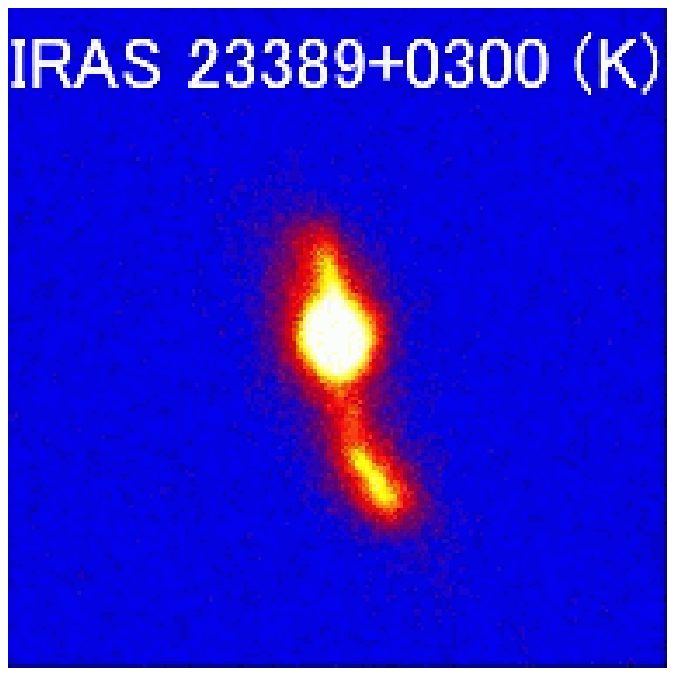} 
\includegraphics[angle=0,scale=.47]{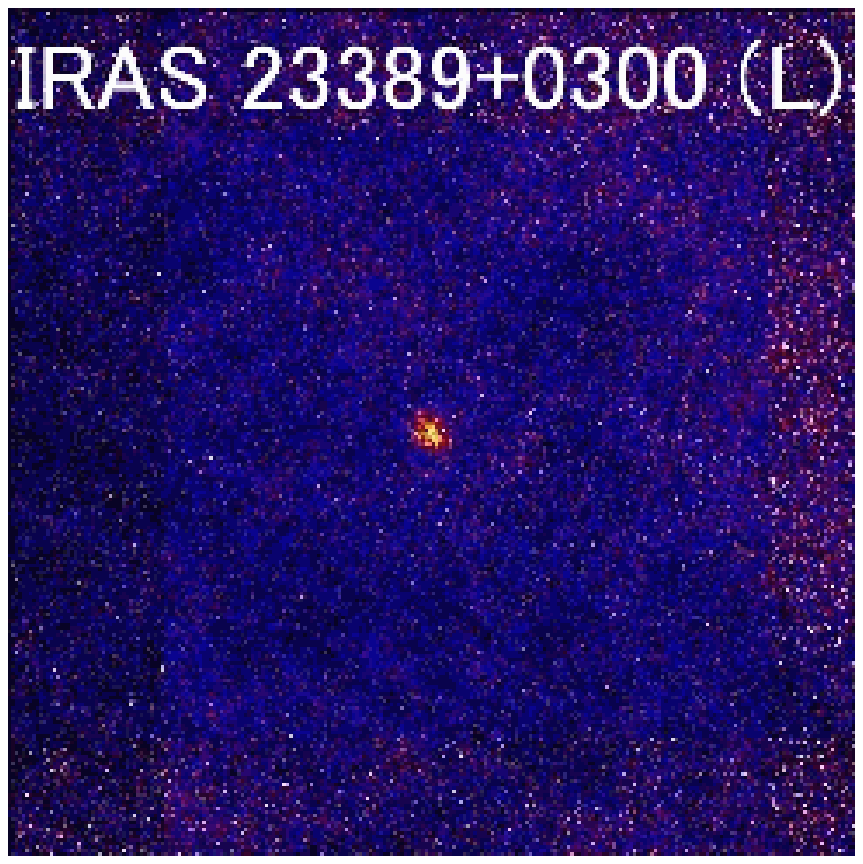} \\ 
\includegraphics[angle=0,scale=.6]{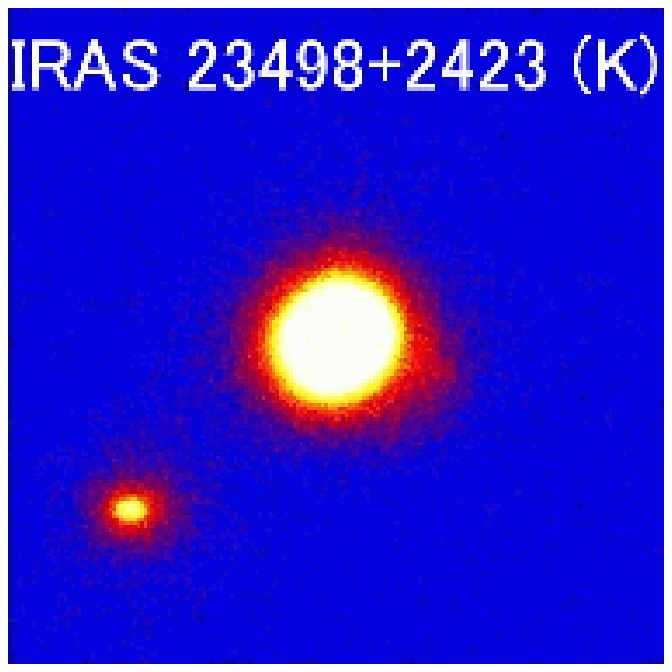} 
\includegraphics[angle=0,scale=.47]{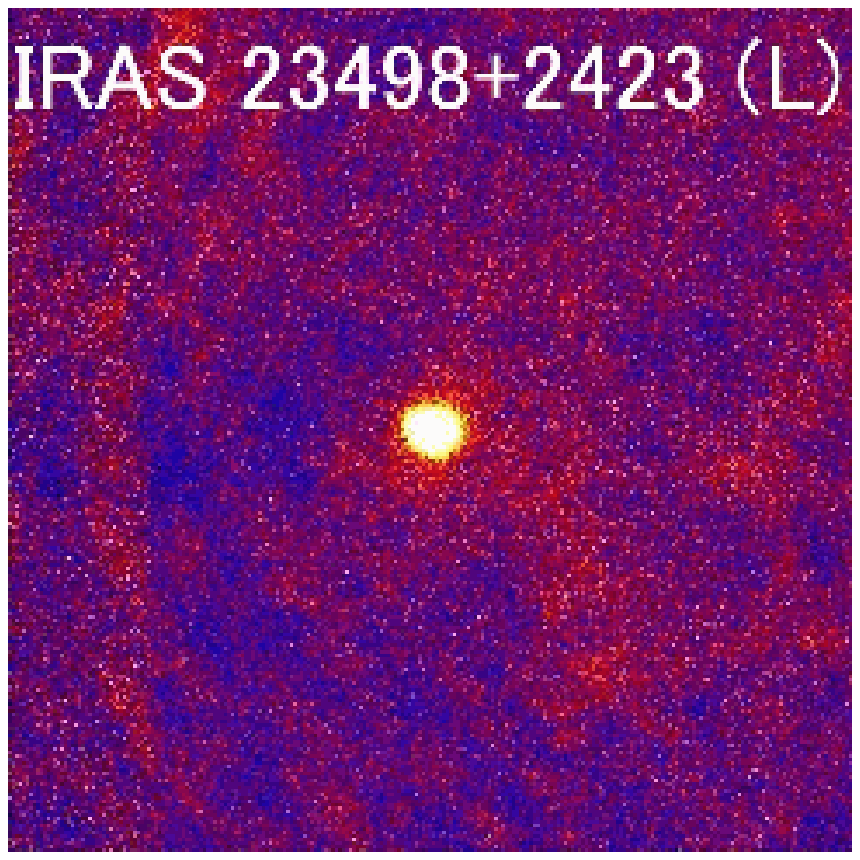} 
\includegraphics[angle=0,scale=.6]{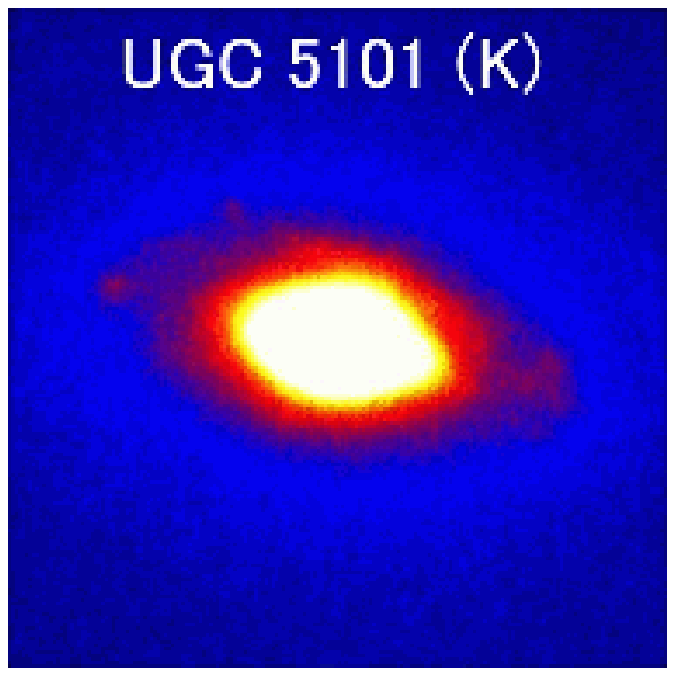} 
\includegraphics[angle=0,scale=.47]{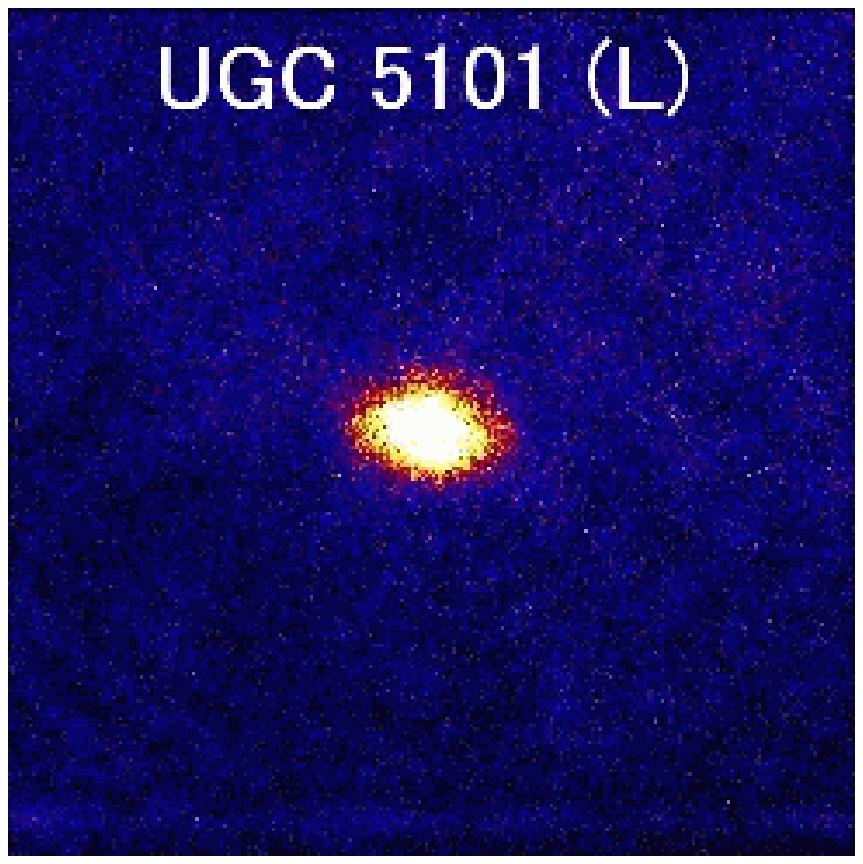} \\
\includegraphics[angle=0,scale=.6]{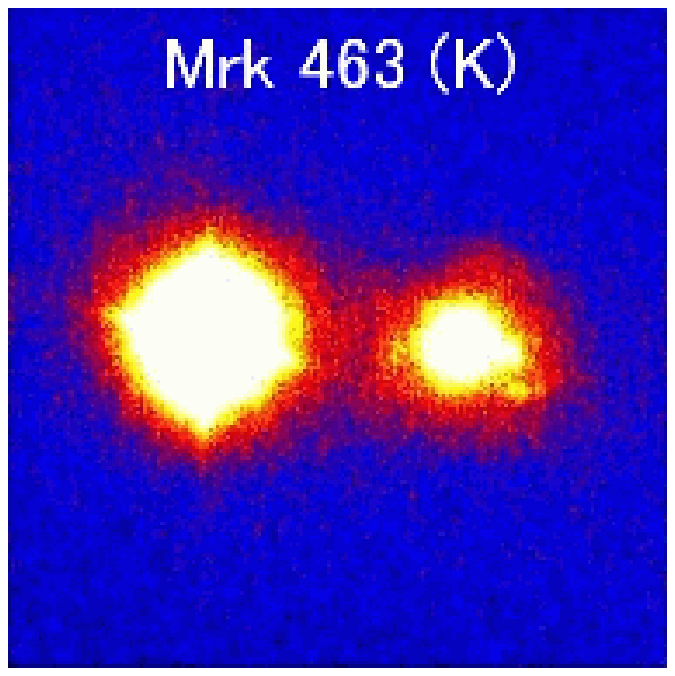} 
\includegraphics[angle=0,scale=.47]{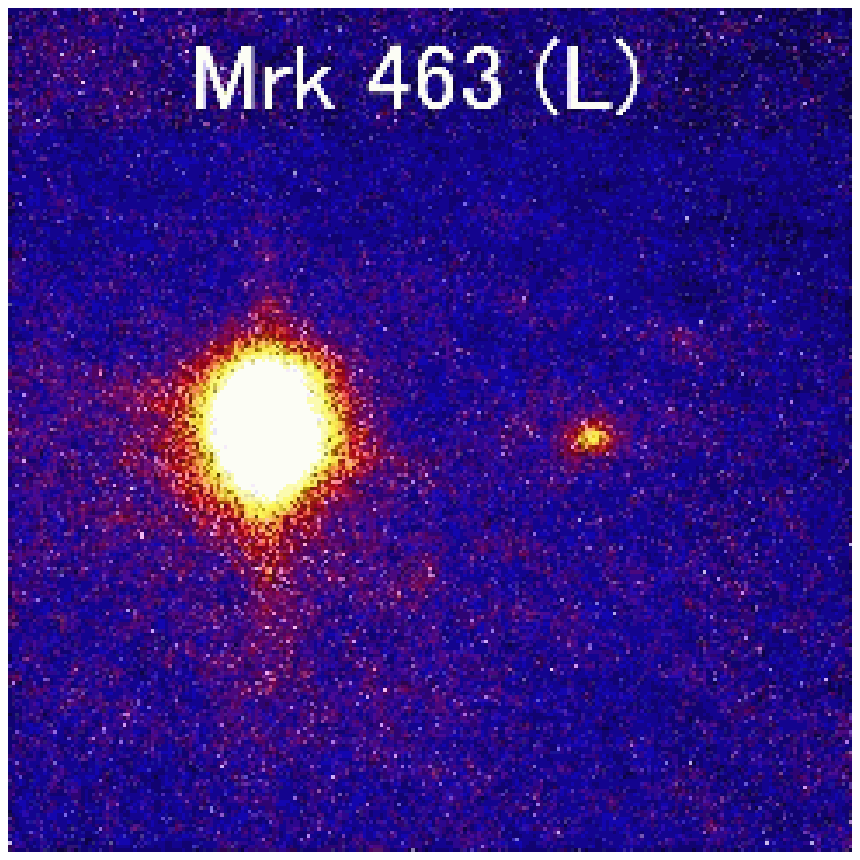} 
\includegraphics[angle=0,scale=.6]{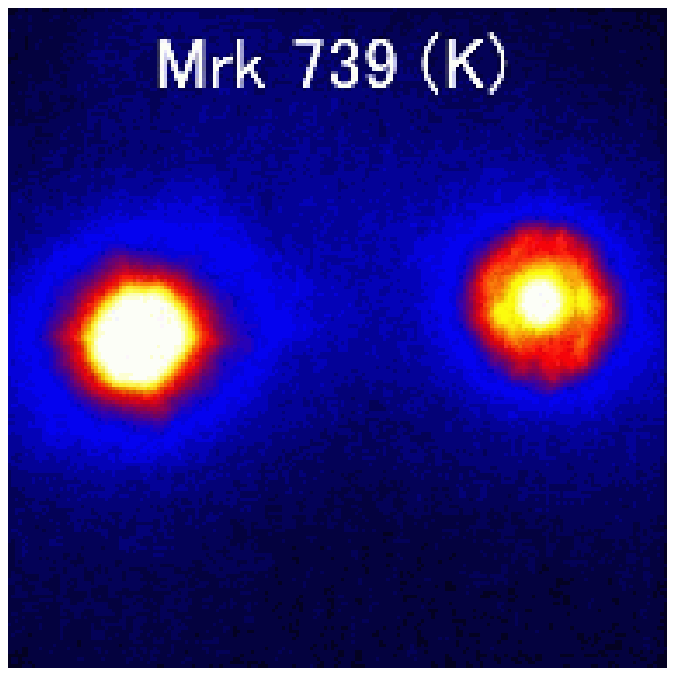} 
\includegraphics[angle=0,scale=.47]{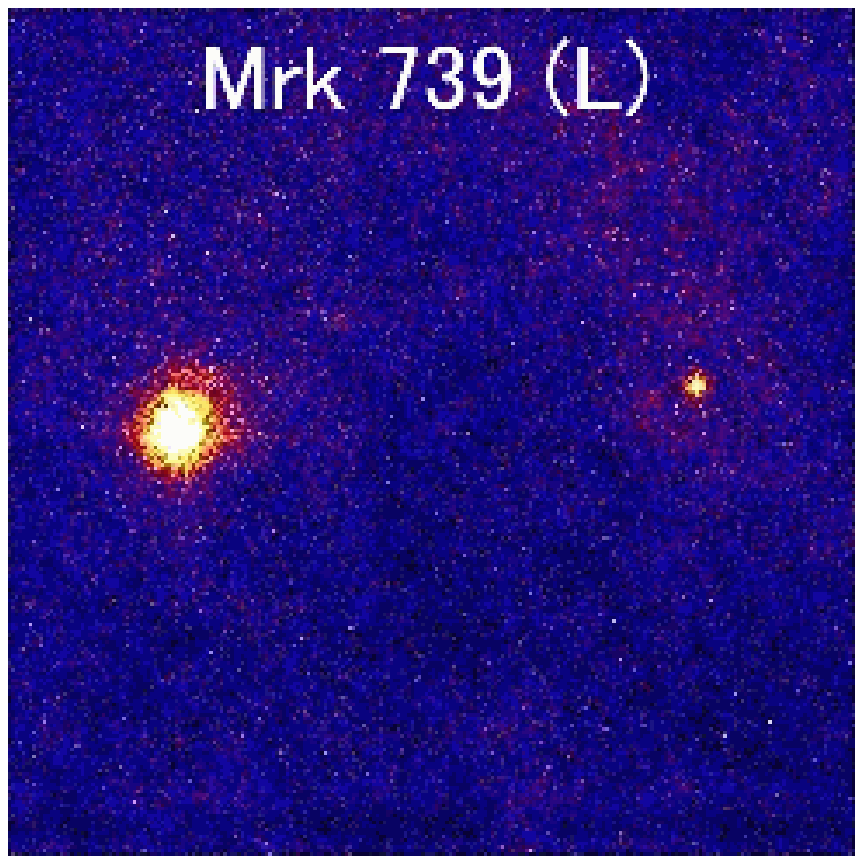} \\
\includegraphics[angle=0,scale=.6]{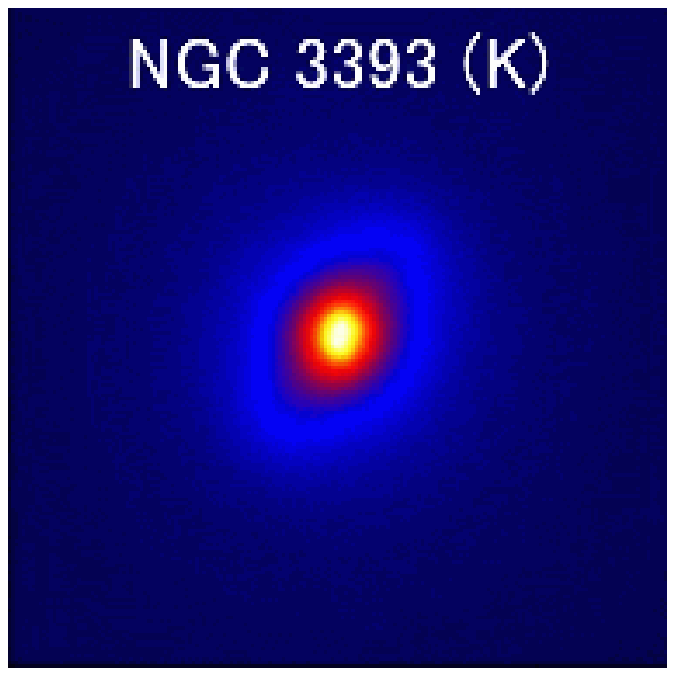} 
\includegraphics[angle=0,scale=.47]{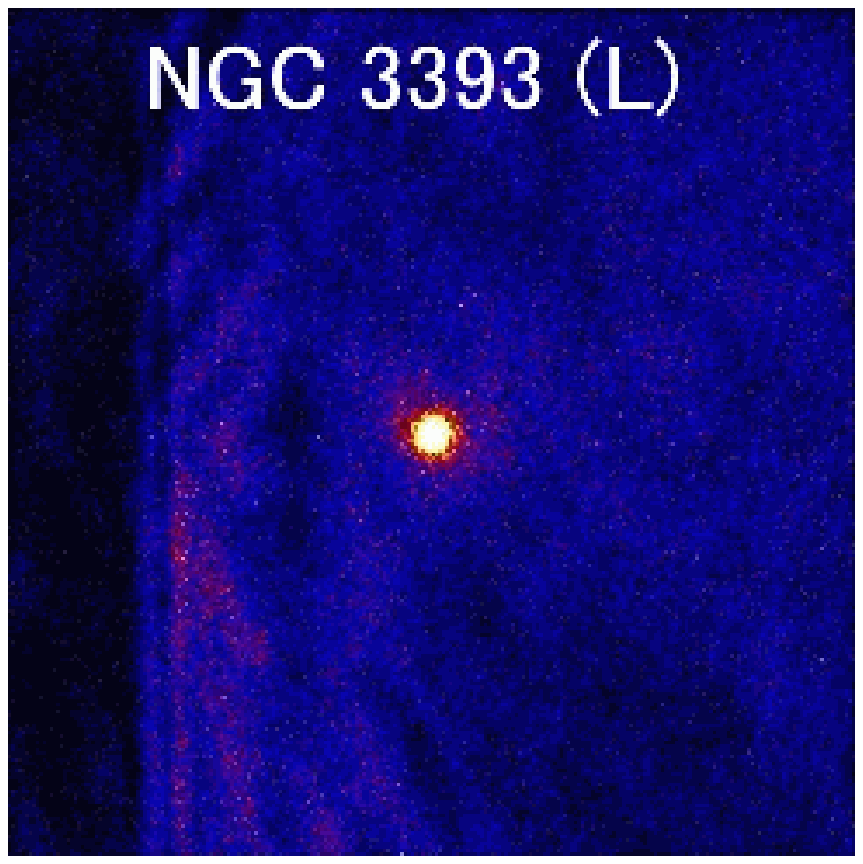} 
\includegraphics[angle=0,scale=.6]{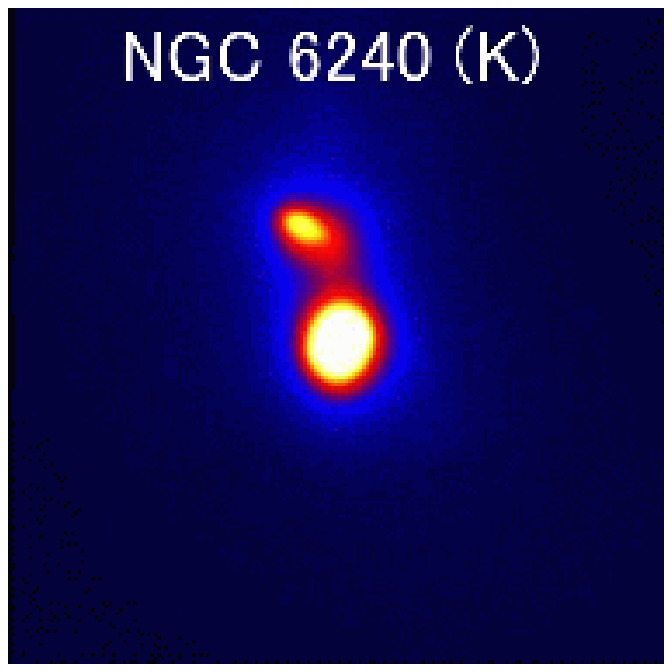} 
\includegraphics[angle=0,scale=.47]{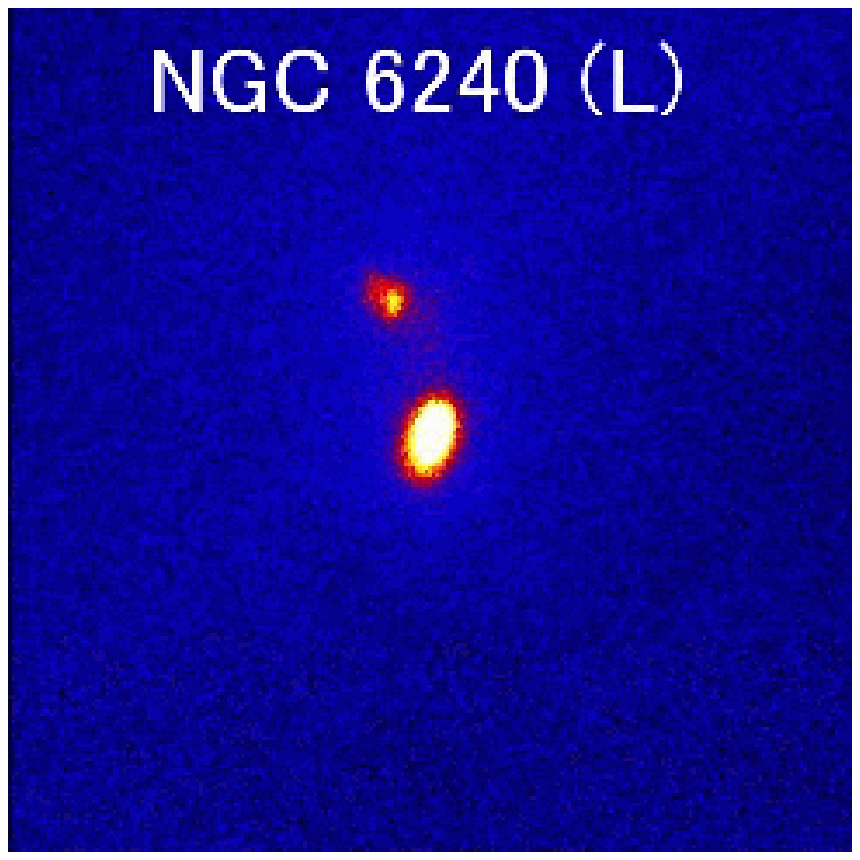} \\ 
\end{figure}

\clearpage
\begin{figure}
\includegraphics[angle=0,scale=.6]{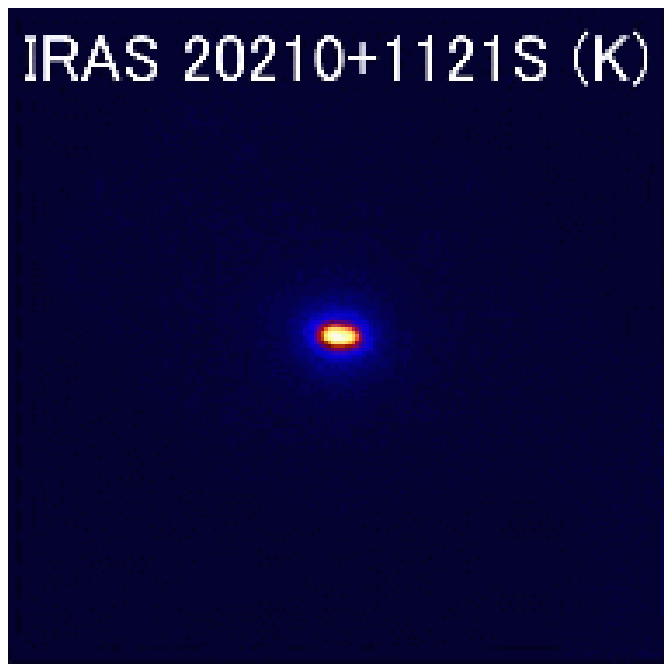} 
\includegraphics[angle=0,scale=.47]{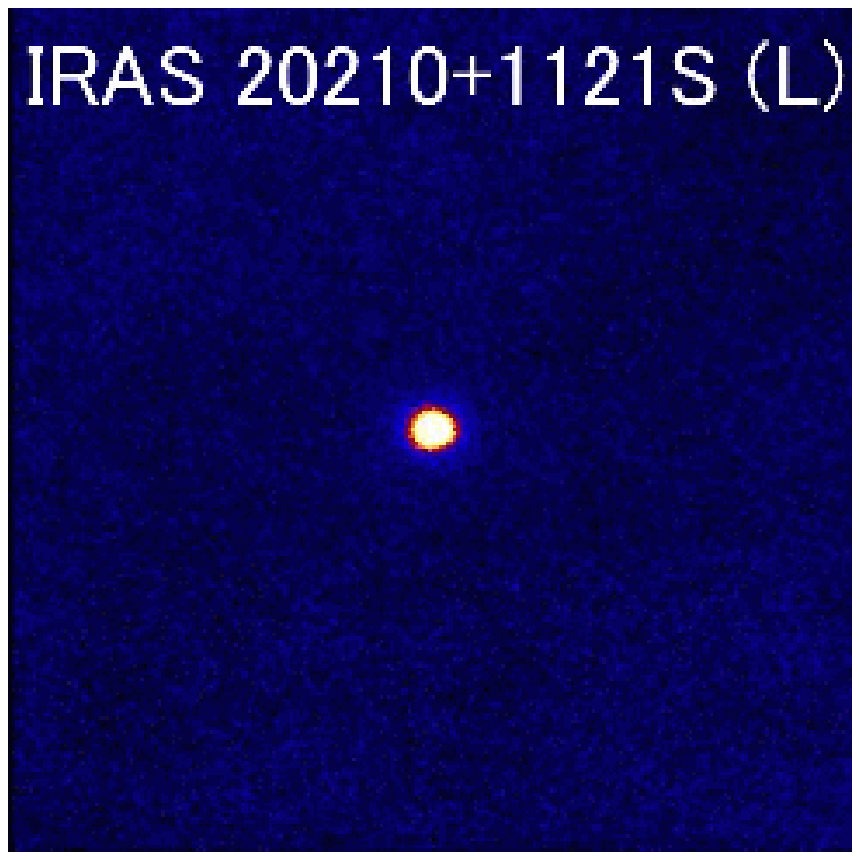}  
\includegraphics[angle=0,scale=.6]{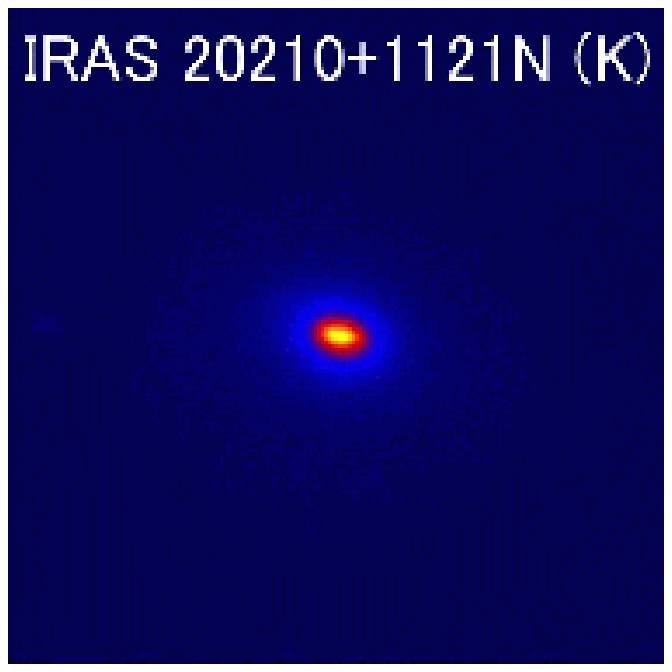} 
\includegraphics[angle=0,scale=.47]{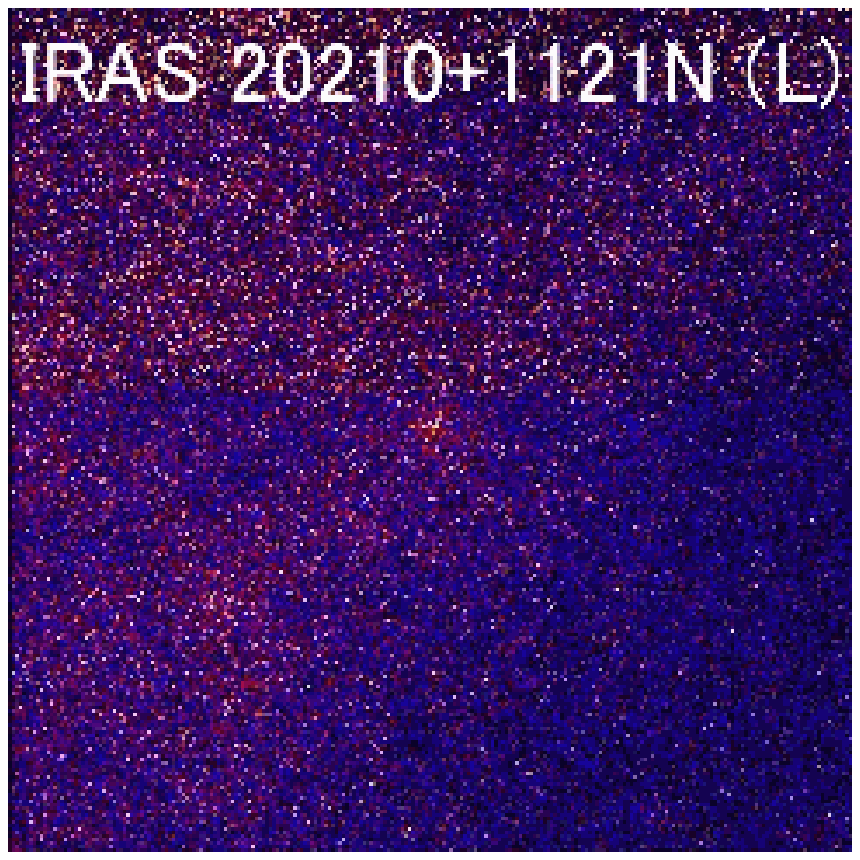} \\ 
\caption{
Our Subaru AO $K$- (2.2 $\mu$m) and $L'$-band (3.8 $\mu$m) images of
observed (U)LIRG nuclei. 
The field of view (FOV) is 10$''$ x 10$''$. 
North is up, and east is to the left. 
For IRAS 13443+0802, 23327+2913, and 20210+1121, two separate images are
shown because the largest separations of multiple nuclei are $>$10$''$
\citep{kim02,pic10}.}
\label{fig:fig-1}
\end{figure}

\begin{figure}
\includegraphics[angle=0,scale=.464]{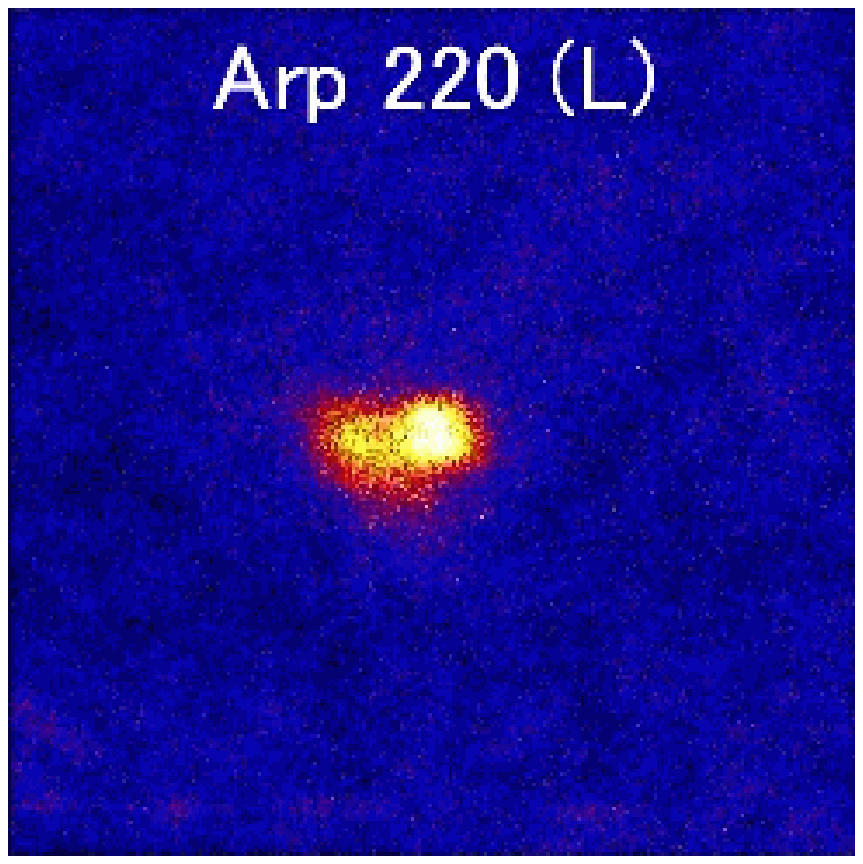} 
\includegraphics[angle=0,scale=0.79]{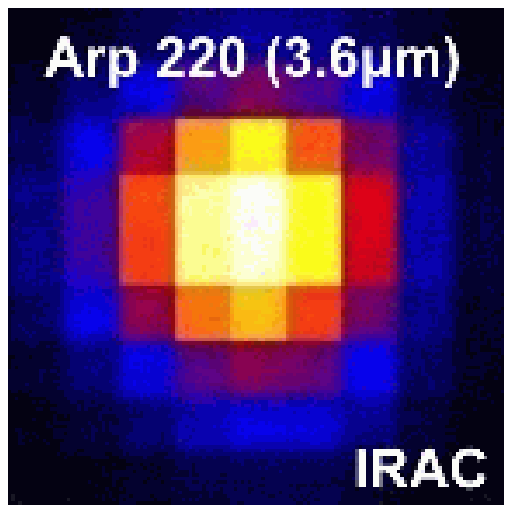} 
\includegraphics[angle=0,scale=.464]{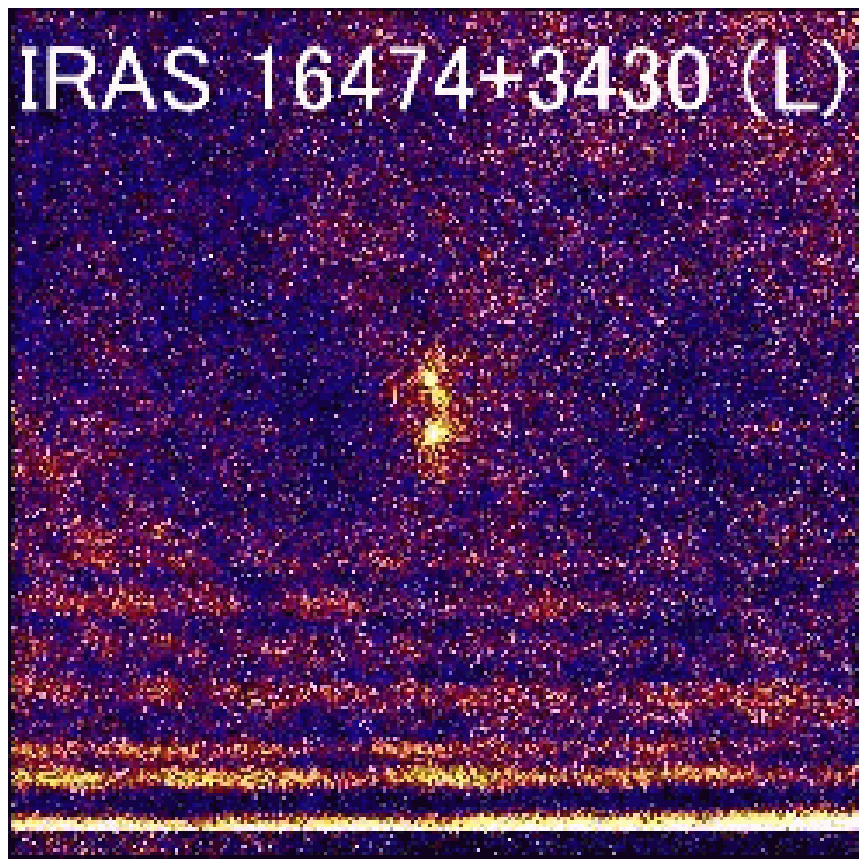} 
\includegraphics[angle=0,scale=0.79]{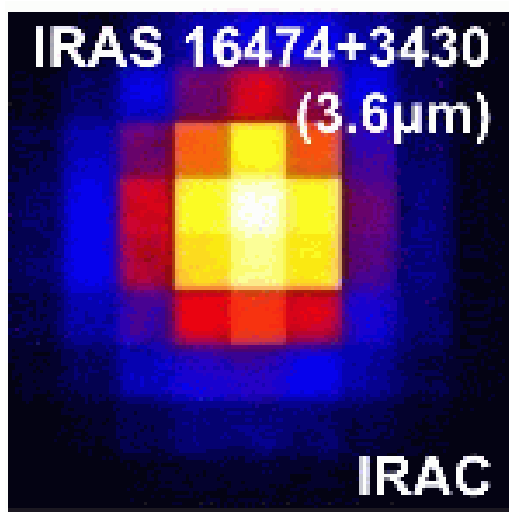} 
\caption{
Comparison of our Subaru AO $L'$-band (3.8 $\mu$m) images with Spitzer
IRAC 3.6 $\mu$m images of two ULIRGs (Arp 220 and IRAS 16474$+$3430).  
The FOV is 10$''$ $\times$ 10$''$. North is up, and east is to the left.}
\label{fig:fig-2}
\end{figure}

\begin{figure}
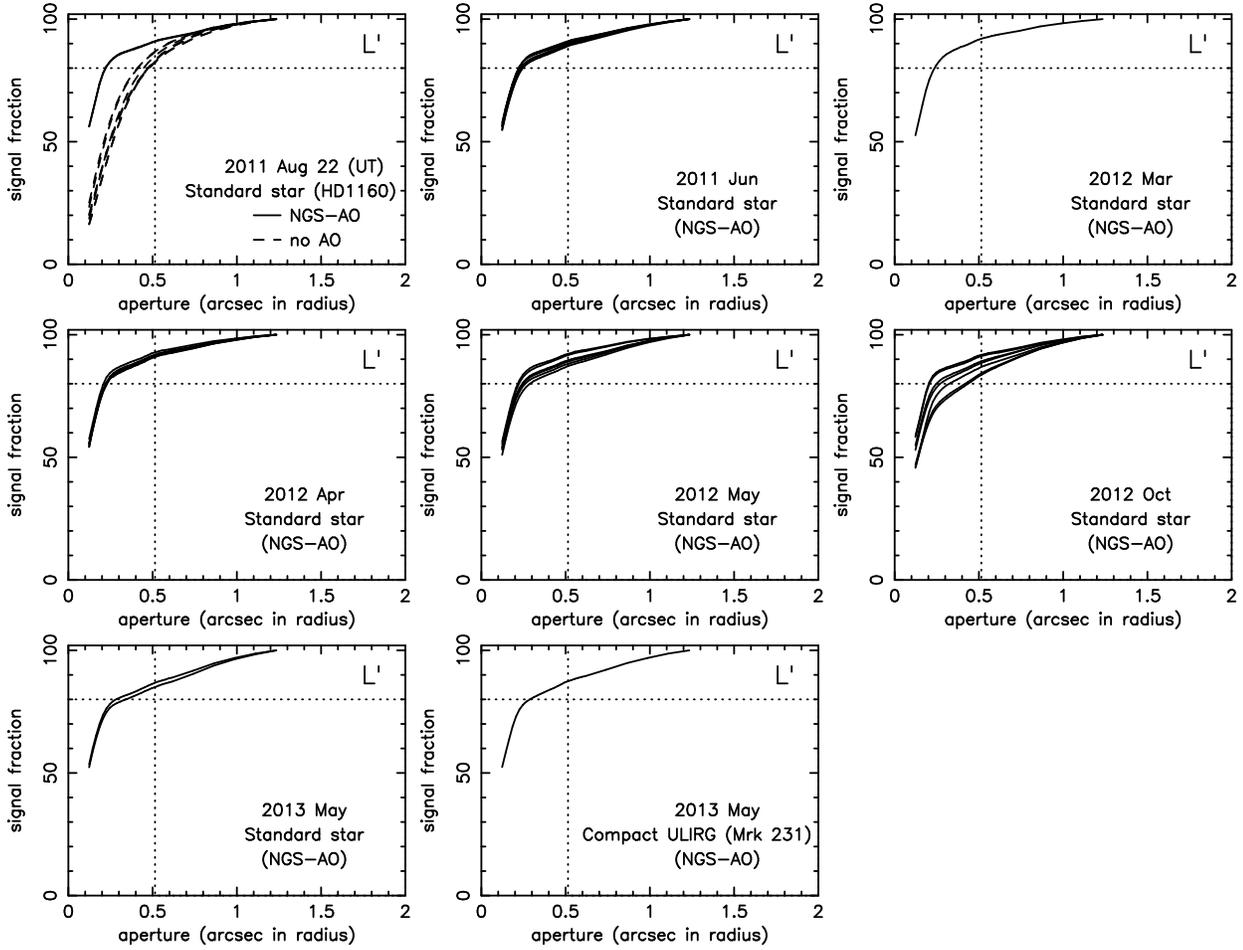

\includegraphics[angle=-90,scale=.24]{f3a.eps} 
\includegraphics[angle=-90,scale=.24]{f3b.eps} 
\includegraphics[angle=-90,scale=.24]{f3c.eps} \\
\includegraphics[angle=-90,scale=.24]{f3d.eps} 
\includegraphics[angle=-90,scale=.24]{f3e.eps} 
\includegraphics[angle=-90,scale=.24]{f3f.eps} \\
\includegraphics[angle=-90,scale=.24]{f3g.eps} 
\includegraphics[angle=-90,scale=.24]{f3h.eps} 
\caption{
The growth of the curve of the encircled signal in the $L'$-band for
standard stars observed with NGS-AO on various nights (Table 2) using
the standard stars themselves as AO guide stars. 
The growth of the signal of a ULIRG dominated by spatially unresolved 
compact emission (Mrk 231), observed with NGS-AO, is also plotted.  
For the upper left plot, the growth of the curves of a standard star (HD 1160) 
observed with NGS-AO (two data sets) and without AO (seven data sets) 
on the same night (2011 August 22 UT) are shown and compared.  
NGS-AO data (solid lines) show a higher central concentration of signals 
than non-AO data (dashed lines),
demonstrating the power of Subaru AO for improved photometry of
spatially unresolved compact sources using a small aperture.   
The horizontal and vertical dotted lines indicate an 80\% signal fraction 
and employed $\sim$0$\farcs$5-radius aperture size, respectively.
It is seen that the $\sim$0$\farcs$5-radius aperture (25 pixels $\times$
20.57 mas pixel$^{-1}$) consistently contains 83--93\% of
spatially unresolved compact source signals in the Subaru $L'$-band
NGS-AO images.}
\label{fig:fig-3}
\end{figure}

\clearpage

\begin{figure}
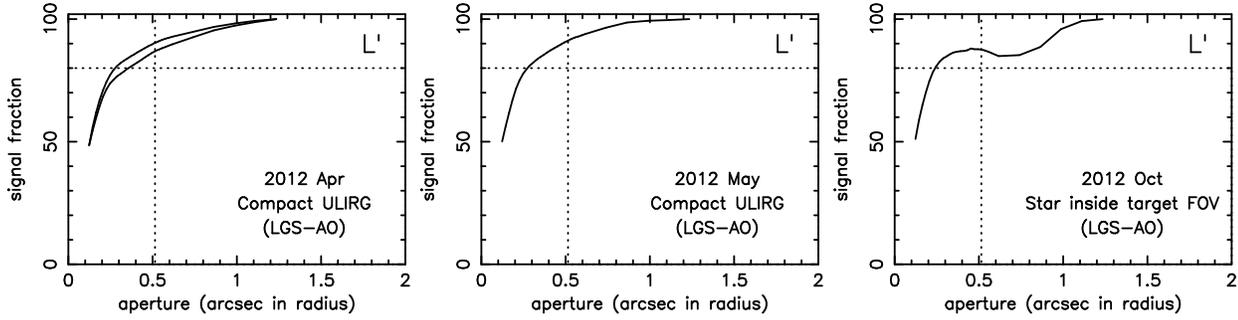

\includegraphics[angle=-90,scale=.24]{f4a.eps} 
\includegraphics[angle=-90,scale=.24]{f4b.eps} 
\includegraphics[angle=-90,scale=.24]{f4c.eps} 
\caption{
The growth of the curve of the encircled signal in the $L'$-band
observed with LGS-AO.  
{\it (Left)}: 
Compact merging nuclei with no discernible extended emission component 
(Mrk 739 eastern nucleus and IRAS 08572$+$3915 north-western nucleus),  
observed in 2012 April.
In the $L'$-band images of merging galaxies, it is harder to find bright 
stars with high detection significance inside frames than in the $K$-band
because of the smaller field of view and larger background noise.  
For these reasons, compact merging nuclei are used to estimate the
signal growth of the curve for LGS-AO at $L'$. As these merging nuclei
could contain spatially extended host galaxy emission components, the 
signal fraction at each aperture size is a lower limit on 
the spatially unresolved emission components.
{\it (Middle)}: Compact merging nuclei observed in 2012 May 
(IRAS 12127$-$1412 north-eastern nucleus). 
{\it (Right)}: A bright star inside the field of view of IRAS
05024$-$1941 (USNO 0703-0054437) observed in 2012 October. 
This star is located at the edge of the $L'$-band image, where a 
significant noise pattern is recognizable. The growth of the curve may be
affected by this noise.  
The horizontal and vertical dotted lines indicate the 80\% signal fraction 
and the $\sim$0$\farcs$5 radius aperture size, respectively.
In all plots, 85--93\% of the signals from 
spatially unresolved compact sources are recovered using 
the $\sim$0$\farcs$5-radius aperture 
(25 pixels $\times$ 20.57 mas pixel$^{-1}$) in the Subaru $L'$-band
LGS-AO images.
This signal fraction is comparable to that of Subaru $L'$-band   
NGS-AO images (Figure 3).}
\label{fig:fig-4}
\end{figure}

\clearpage

\begin{figure}
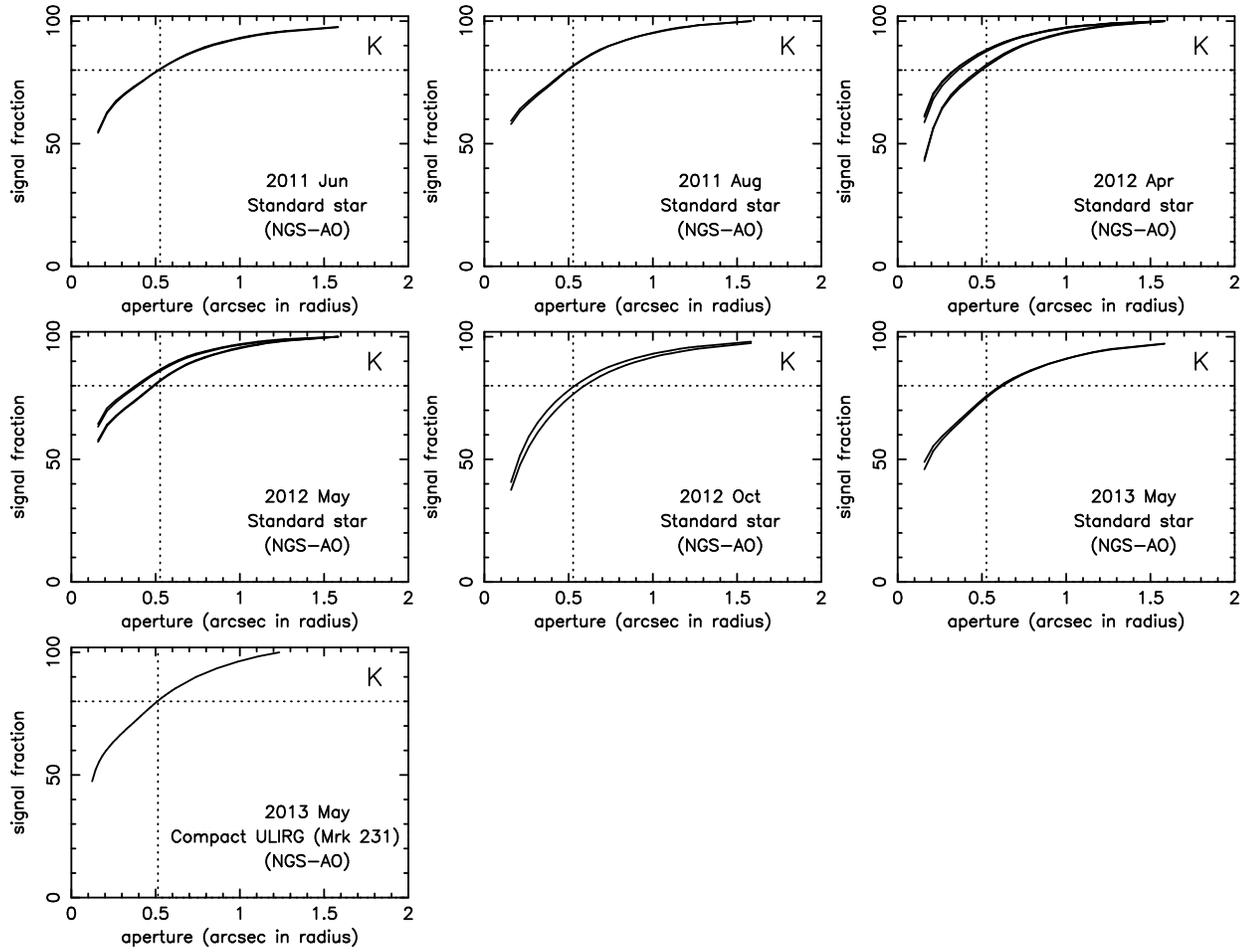

\includegraphics[angle=-90,scale=.24]{f5a.eps} 
\includegraphics[angle=-90,scale=.24]{f5b.eps}  
\includegraphics[angle=-90,scale=.24]{f5c.eps} \\
\includegraphics[angle=-90,scale=.24]{f5d.eps} 
\includegraphics[angle=-90,scale=.24]{f5e.eps} 
\includegraphics[angle=-90,scale=.24]{f5f.eps} \\
\includegraphics[angle=-90,scale=.24]{f5g.eps} 
\caption{
The growth of the curve of the encircled signal in the $K$-band for
standard stars observed with NGS-AO during various observation nights
(Table 2) using the standard stars themselves as AO guide stars.
A plot of a ULIRG dominated by spatially unresolved compact emission
(Mrk 231), observed with NGS-AO, is also shown.  
The horizontal and vertical dotted lines indicate the 80\% signal fraction 
and employed $\sim$0$\farcs$5 radius aperture size, respectively.
In all plots, 75--90\% of signals from 
spatially unresolved compact source emission are recovered using
a $\sim$0$\farcs$5-radius aperture (10 pixels $\times$ 52.77 mas
pixel$^{-1}$) for Subaru $K$-band NGS-AO data.  
For Mrk 231, the $\sim$0$\farcs$5 radius aperture is set from 25 pixels
$\times$ 20.57 mas pixel$^{-1}$ (see $\S$ 3).}
\label{fig:fig-5}
\end{figure}

\clearpage

\begin{figure}
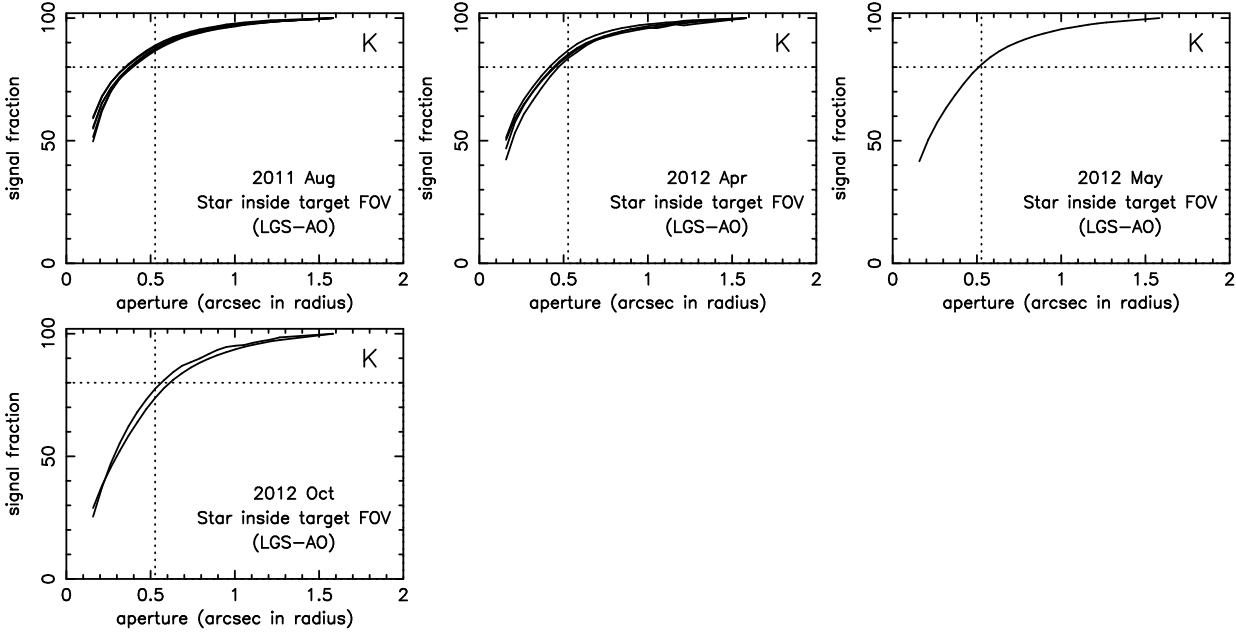

\includegraphics[angle=-90,scale=.24]{f6a.eps} 
\includegraphics[angle=-90,scale=.24]{f6b.eps} 
\includegraphics[angle=-90,scale=.24]{f6c.eps} \\
\includegraphics[angle=-90,scale=.24]{f6d.eps} 
\caption{
The growth of the curve of the encircled signal in the $K$-band,
estimated using bright stars inside the field of view of the target
(U)LIRGs, observed with LGS-AO on various nights. 
{\it (Upper Left)}: Three stars in the IRAS 00188$-$0856 LGS-AO image 
(USNO 0813-0003716, 0813-0003719, 0813-0003721) and
three stars in the IRAS 21208$-$0519 LGS-AO image (USNO 0848-0617068,
0848-0617074, 0848-0617085), taken in 2011 August are used.    
{\it (Upper Middle)}: A bright star in the IRAS 14388$-$1447 LGS-AO
image (USNO 0749-0288838), 
a star in the IRAS 17044$+$6720 LGS-AO image (USNO 1572-0201749), and
two stars in the Mrk 739 LGS-AO image (USNO 1116-0211017, 1116-0211021)
obtained in 2012 April are used.  
{\it (Upper Right)}: A bright star in the IRAS 12127$-$1412 LGS-AO image 
(USNO 0755-0260793), taken in 2012 May, is used.  
{\it (Lower Left)}: A bright star in the IRAS 05024$-$1941 LGS-AO
image (USNO 0703-0054437) and a bright star $\sim$8$''$ west of
IRAS 23498+2423 (seen in the $K$-band image 
of \citet{kim02}) taken in 2012 October are used. 
The horizontal and vertical dotted lines indicate the 80\% signal fraction 
and $\sim$0$\farcs$5-radius aperture size, respectively.
In all plots, the chosen $\sim$0$\farcs$5 radius aperture recovered 
75--90\% of the spatially unresolved compact source signals in the 
Subaru $K$-band LGS-AO images.}
\label{fig:fig-6}
\end{figure}

\begin{figure}
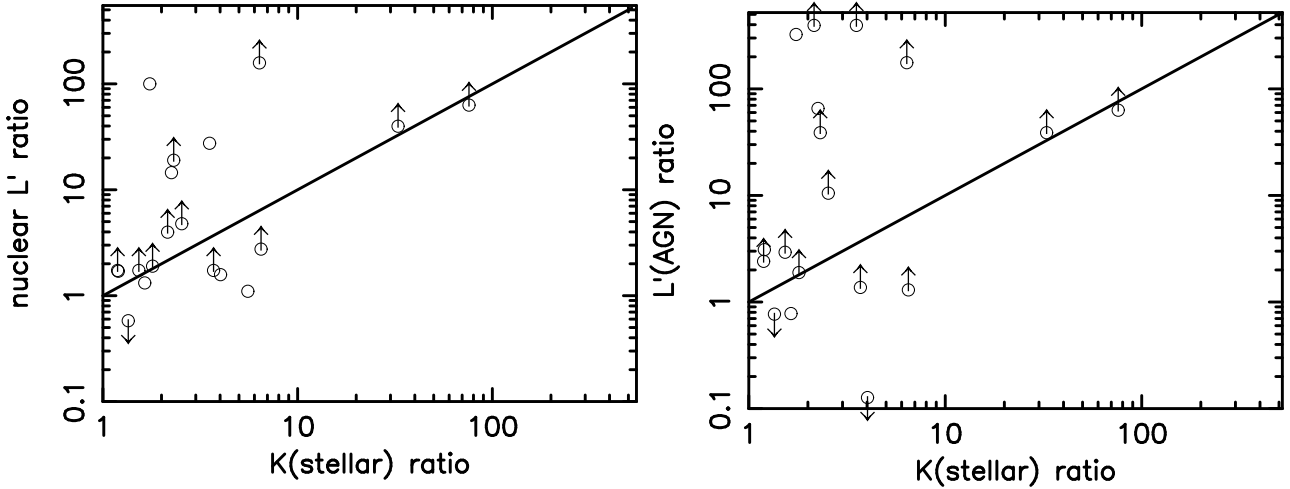

\includegraphics[angle=-90,scale=.38]{f7a.eps} 
\includegraphics[angle=-90,scale=.38]{f7b.eps} 
\caption{
{\it (Left)}: Luminosity ratio among multiple nuclei for (U)LIRGs listed
in Table 5.
The abscissa is the $K$-band photometry, including host galaxy emission.
The ordinate is the nuclear $L'$-band ($\sim$0$\farcs$5-radius aperture)
photometry (Table 3, column 3). 
The abscissa is taken as the stellar luminosity ratio, which is
converted to the SMBH mass ratio. 
If AGN-origin $L'$-band luminosity excess compared with stellar
emission is different for multiple nuclei, then the source will deviate
from the solid line.
{\it (Right)}: The ordinate is the AGN-origin $L'$-band luminosity
ratio after subtracting stellar emission (Table 3, column 6).  
In the ordinate of both plots, the $L'$-band luminosity at the brighter 
nucleus in the $K$-band stellar emission is divided by that at the
$K$-band fainter nucleus. 
Sources appearing around the upper-left (lower-right) of the solid
line suggest that larger-mass SMBHs show higher (lower) mass
accretion rates when normalized to SMBH mass than do smaller-mass SMBHs.
The upper (down) arrows in the ordinate indicate sources
whose $L'$-band emission was not detected ($<$3$\sigma$) at the $K$-band 
fainter (brighter) nuclei.
}
\label{fig:fig-7}
\end{figure}

\begin{figure}
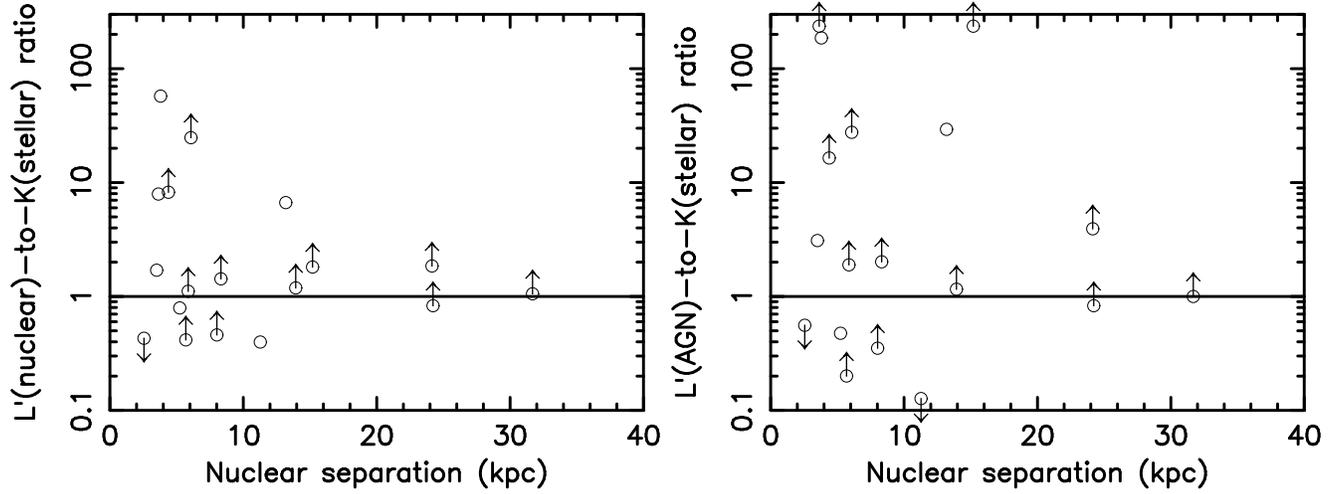

\includegraphics[angle=-90,scale=.38]{f8a.eps} 
\includegraphics[angle=-90,scale=.38]{f8b.eps} 
\caption{
{\it (Left)}: The abscissa is apparent nuclear separation in kpc 
for (U)LIRGs in Table 5.
The ordinate is the comparison of the nuclear $L'$-band to 
galaxy-wide $K$-band stellar luminosity ratio
between two nuclei.  
{\it (Right)}: Same as the left panel, but the ordinate is the
comparison of the AGN-origin $L'$-band to $K$-band 
stellar luminosity ratio between two nuclei.
In both plots, the ratio at the nucleus with brighter $K$-band stellar 
emission is divided by that at the nucleus with fainter $K$-band 
stellar emission.
Since the $K$-band stellar luminosity is taken to be
proportional to the SMBH mass, the ordinate is proportional to the
Eddington ratio (= mass accretion rate per SMBH mass).
Sources above the solid horizontal lines mean that the Eddington ratios
are higher at the $K$-band brighter nuclei (= nuclei with larger-mass
SMBHs).  
The upper (down) arrows in the ordinate indicate sources
whose $L'$-band emission was not detected ($<$3$\sigma$) at the $K$-band 
fainter (brighter) nuclei.
}
\label{fig:fig-8}
\end{figure}

\end{document}